\documentclass[11pt,a4paper]{article}
\pdfoutput=1 % ensure pdflatex processing
\usepackage[utf8]{inputenc} % source file character encoding
\usepackage[T1]{fontenc} % output font encoding
\usepackage[UKenglish]{babel} % hyphenation etc.
\usepackage{microtype} % improves word spacing etc.
\usepackage{mathtools} % amsmath extension package
\usepackage{xcolor} % named colors etc.
\usepackage{booktabs} % better looking tables
\usepackage{subcaption} % subfigure-environment
\usepackage{bbm} % blackboard math (for CM font)
\usepackage{bm} % bolded math symbols
\usepackage{siunitx} % typeset quantities with units easily
\usepackage{slashed} % Feynman slash notation

\usepackage{jheppub} % includes hyperref as last package

\usepackage{cleveref} % easier cross-refs using \cref (load after hyperref!)
    \crefname{equation}{}{}
    \crefname{figure}{}{} % example: \crefformat{figure}{#figure~#1#3}
    \crefname{table}{}{}
    \crefname{section}{}{} % example: \crefformat{section}{#2section~#1#3}
    \crefname{appendix}{}{}
    \crefname{footnote}{}{}
    
    \crefalias{subequation}{equation}

\newcommand{\widebar}[1]{\mkern 1.3mu\overline{\mkern-1.3mu#1\mkern-1.3mu}\mkern 1.3mu}
\renewcommand{\bar}{\widebar}

\newcommand{\im}{{\mathrm{i}}} % imaginary unit
\newcommand{\e}{\mathop{}\!\mathrm{e}} % Euler's number (Napier's constant), use for \e^x too

\newcommand{\evec}[1]{{\bm{#1}}} % Euclidean vector
\newcommand{\Uevec}[1]{{\hat{\bm{#1}}}} % Euclidean unit vector

\newcommand{\idmat}{\bm{1}} % identity matrix; variants: \mathbbm{1}, \mathds{1}
\newcommand{\transp}{{\mathrm{T}}} % transpose; variants: \mathsf{T}, \intercal, \top

\newcommand{\dd}{\mathop{}\!\mathrm{d}} % differential d

\newcommand{\sfrac}[2]{{\textstyle{\frac{#1}{#2}}}} % small fraction
\newcommand{\shalf}{{\textstyle \frac{1}{2}}} % small half
\newcommand{\ihalf}{{\textstyle \frac{\im}{2}}}

\newcommand{\convol}{\mathbin{\ast}} % convolution
\newcommand{\nconvol}{\mathbin{\!\convol\!}} % variant with narrower spacing

\newcommand{\lt}{{\scriptscriptstyle <}} % for super- or subscripts
\newcommand{\gt}{{\scriptscriptstyle >}} % for super- or subscripts
\newcommand{\smallless}{{\scriptstyle <}} % for text mode
\newcommand{\smallgreater}{{\scriptstyle >}} % for text mode

\newcommand{\tin}{{t_{\mathrm{in}}}} % initial time
\newcommand{\tf}{{t_{\mathrm{f}}}} % final time
\newcommand{\cw}{c_{\mathrm{w}}} % c_w, the weak isospin multiplicity factor

\newcommand{\eg}{e.g.}
\newcommand{\ie}{i.e.}
\newcommand{\cf}{cf.}

\DeclareMathOperator{\sgn}{sgn} % sign function
\DeclareMathOperator{\tr}{tr} % trace
\DeclareMathOperator{\Tr}{Tr} % Trace
\DeclareMathOperator{\PV}{PV} % principal value
\DeclareMathOperator{\Res}{Res} % residue
\DeclareMathOperator{\diag}{diag} % diagonal matrix

\renewcommand{\Re}{\operatorname{Re}} % real part
\renewcommand{\Im}{\operatorname{Im}} % imaginary part

\newcommand{\SU}{\operatorname{SU}} % special unitary group

\DeclarePairedDelimiter{\abs}{\lvert}{\rvert} % absolute value
\DeclarePairedDelimiter{\comm}{[}{]} % commutator
\DeclarePairedDelimiter{\anticomm}{\{}{\}} % anti-commutator
\DeclarePairedDelimiter{\expectval}{\langle}{\rangle} % expectation value

\DeclareOldFontCommand{\rm}{\normalfont\rmfamily}{\mathrm}

\let\oldepsilon\epsilon % dummy variable
\let\epsilon\varepsilon
\let\varepsilon\oldepsilon

\title{Flavour mixing transport theory and resonant leptogenesis}

\author[a,b]{Henri Jukkala,}
\author[a,b]{Kimmo Kainulainen}
\author[a,b]{and Pyry M. Rahkila}

\affiliation[a]{Department of Physics, University of Jyväskylä,\\ P.O.~Box 35 (YFL), FI-40014 Jyväskylä, Finland}
\affiliation[b]{Helsinki Institute of Physics, University of Helsinki,\\ P.O.~Box 64, FI-00014 Helsinki, Finland}

\emailAdd{henri.a.jukkala@jyu.fi}
\emailAdd{kimmo.kainulainen@jyu.fi}
\emailAdd{pyry.m.rahkila@jyu.fi}

\abstract{We derive non-equilibrium quantum transport equations for flavour-mixing fermions. We develop the formalism mostly in the context of resonant leptogenesis with two mixing Majorana fermions and one lepton flavour, but our master equations are valid more generally in homogeneous and isotropic systems. We give a hierarchy of quantum kinetic equations, valid at different approximations, that can accommodate helicity and arbitrary mass differences. In the mass-degenerate limit the equations take the familiar form of density matrix equations. We also derive the semiclassical Boltzmann limit of our equations, including the CP-violating source, whose regulator corresponds to the flavour coherence damping rate. Boltzmann equations are accurate and insensitive to the particular form of the regulator in the weakly resonant case $\Delta m \gg \Gamma$, but for $\Delta m \lesssim \Gamma$ they are qualitatively correct at best, and their accuracy crucially depends on the form of the CP-violating source.}


\arxivnumber{2104.03998}

\begin{document}
\maketitle
\flushbottom

%%%%%%%%%%%%%%%%%%%%%%%%%%%%%%%%%%%%%%%%%%%%%%%%%%%%%%%%%%%%%%%%%%%%%%%%%%%%%%%
%
\section{Introduction}
\label{sec:introduction}
%
%%%%%%%%%%%%%%%%%%%%%%%%%%%%%%%%%%%%%%%%%%%%%%%%%%%%%%%%%%%%%%%%%%%%%%%%%%%%%%%

Quantum coherence, and the related mixing and oscillation of quantum states, plays an important role in many interesting phenomena in particle physics and in the early universe. Examples include neutrino mixing and oscillations~\cite{Bilenky:1987ty,Barbieri:1989ti,Enqvist:1990ad,Kainulainen:1990ds,Enqvist:1990ek,Barbieri:1990vx,Enqvist:1991qj,Sigl:1992fn,Vlasenko:2013fja}, particle scattering from phase transition walls in baryogenesis~\cite{Joyce:1994zn,Cline:1997vk,Kainulainen:2001cn,Kainulainen:2002th,Cline:2020jre,Nelson:1991ab,Huet:1995mm,Riotto:1997gu,Postma:2019scv}, mixing of Majorana neutrinos in leptogenesis~\cite{Fukugita:1986hr,Davidson:2008bu,Blanchet:2012bk}, and particle production after inflation or in phase transitions~\cite{Chung:1999ve,Kofman:1986wm,Fairbairn:2018bsw}. In these applications a quantum field theoretical treatment of coherence is essential. Several different types of coherence may be relevant depending on the problem: particle production is driven by the particle-antiparticle coherence, and coherent mixing of left- and right-moving particles can be the engine for creating the particle-antiparticle asymmetry during the electroweak phase transition. Finally, coherence between different flavour states powers the familiar phenomenon of neutrino oscillations as well as the particle-antiparticle asymmetry generation in leptogenesis, which is the main topic of this paper.

The primary goal of this work is to develop a general and transparent formalism for treating problems involving quantum flavour coherence. While our results are more general, a large part of the development will be done in the context of (resonant) leptogenesis. We will develop a series of implementations of quantum transport equations with different levels of sophistication. At the most general level our formalism includes also the particle-antiparticle coherences. From there we move by a well defined reduction process to equations fully describing the flavour coherence separately in the particle and antiparticle sectors. These equations are valid for arbitrary helicities and neutrino masses. We also show how these equations can be further reduced to a helicity-symmetric density matrix equation and eventually to the Boltzmann limit with a soundly motivated form for the CP-violating source in the leptogenesis application. This hierarchy of implementations serves both to compare our results against the existing literature, and provides a ``library of methods'' from where one can choose the one that is best suitable for the given application at hand.

In the leptogenesis scenario~\cite{Fukugita:1986hr} an initial lepton asymmetry $n_L$ gets converted to the baryon asymmetry via the $B + L$-violating sphaleron processes~\cite{Kuzmin:1985mm} present in the standard model (SM)~\cite{tHooft:1976snw,Klinkhamer:1984di,Arnold:1987mh,Arnold:1987zg}. Several versions of the leptogenesis mechanism exist~\cite{Davidson:2008bu,Blanchet:2012bk}. In standard thermal leptogenesis $n_L$ is generated by CP-violating out-of-equilibrium decays of heavy Majorana neutrinos and the same basic mechanism carries over to resonant leptogenesis. In the latter the neutrino masses are nearly degenerate however, which leads to an enhanced efficiency. The enhancement is maximal when the timescale $\sim 1/\Delta m$ of the flavour oscillations is comparable to the timescale $\sim 1/\Gamma$ of the change of the neutrino abundances. In this case the flavour mixing has been found to be the dominant source of the lepton asymmetry~\cite{Davidson:2008bu,Blanchet:2012bk}. Thermal leptogenesis can be well treated using standard kinetic equations, but the resonant case needs a more refined treatment to correctly account for the flavour mixing.

We will use the Schwinger--Keldysh closed time path (CTP) formalism~\cite{Schwinger:1960qe,Keldysh:1964ud} to derive Kadanoff--Baym evolution equations~\cite{Baym:1961zz,Kadanoff:1962book} for the two-point correlation function of the mixing fermions from first principles, using the two-particle irreducible effective action (2PIEA) method~\cite{Cornwall:1974vz,Calzetta:1986cq}. We specialize to spatially homogeneous and isotropic systems and include the expansion of the universe, relevant for the leptogenesis application. The central object for our study is the \emph{local} Wightman function $S^\lt_{\evec{k}}(t,t)$, which encodes the statistical properties of the system including flavour diagonal and off-diagonal correlations. The key element of our approach is finding a closed equation for $S^\lt_{\evec{k}}(t,t)$. Our method does not rely on restricted forms of the correlation function, such as the Kadanoff--Baym or quasiparticle ansätze~\cite{Kadanoff:1962book,Greiner:1998vd}. Rather, it is based on the identification of a proper background solution and a judicious approximation to compute the collision terms. This makes it well suited for a study of dynamical mixing as all components of the correlation function are treated on an equal footing. We point out that leptogenesis has already been studied extensively using first principles methods~\cite{Buchmuller:2000nd,Hohenegger:2008zk,Anisimov:2008dz,Garny:2009rv,Anisimov:2010aq,Beneke:2010wd,Garny:2010nz,Beneke:2010dz,Garbrecht:2010sz,Anisimov:2010dk,Garbrecht:2012qv,Drewes:2012ma,Garbrecht:2012pq,Frossard:2012pc,Garbrecht:2013gd,Garbrecht:2013urw,Frossard:2013bra,Garbrecht:2013iga,Dev:2017trv,Garbrecht:2019zaa} and resonant leptogenesis in particular~\cite{DeSimone:2007gkc,DeSimone:2007edo,Cirigliano:2007hb,Garny:2009qn,Garbrecht:2011aw,Garny:2011hg,Iso:2013lba,Iso:2014afa,Hohenegger:2014cpa,Garbrecht:2014aga,Dev:2014wsa,Kartavtsev:2015vto,Drewes:2016gmt,Dev:2017wwc,Garbrecht:2018mrp}. Technically our approach is closest to that of ref.~\cite{Garbrecht:2011aw}.

Our formalism is a generalisation of the coherent quasiparticle approximation (cQPA) first developed in~\cite{Herranen:2008hi,Herranen:2008hu,Herranen:2008yg,Herranen:2008di,Herranen:2010mh,Fidler:2011yq,Herranen:2011zg} and further studied in~\cite{Jukkala:2019slc}. The cQPA is a two-step approximation where the structure of the Wightman function is solved first in a collisionless approximation in the Wigner representation. This results in a spectral shell structure, including particle and coherence shells, which is then used to solve the full dynamical equation. The present method is not restricted to spectral structures as the equations are solved directly in the two-time representation, which allows taking into account a finite width of the pole propagators. If the width is neglected however, our equations are equivalent to cQPA equations.

Our main results include the quantum transport equations~\cref{eq:delta-f-equation} and~\cref{eq:delta-f-m-equation}, the helicity symmetric equations~\cref{eq:offd_equation_approx,eq:diag2_equation_approx,eq:diag1_equation_approx} and the Boltzmann equation source term~\cref{eq:SCP-approx2} with the CP-violating parameter~\cref{eq:sum_epsilon}. We provide an explicit implementation for a benchmark model with two Majorana neutrinos and one lepton flavour including decay and inverse decay interactions, but generalising to more complicated fermion sectors and scattering processes would be straightforward. Numerically all approaches are in good agreement in the weakly resonant case, $m \gg \Delta m \gg \Gamma$, and the helicity-symmetric equation is in good agreement with the full master equation for all parameter values. However, when $\Delta m \lesssim \Gamma$ the Boltzmann equation results depend strongly on the choice of the CP-violating source, the precise form of which has been debated in the literature~\cite{Garny:2011hg,Garbrecht:2014aga,Dev:2017trv,Dev:2017wwc}. Our result agrees with ref.~\cite{Garny:2011hg}. We show that the effective width that defines this source corresponds to the off-diagonal damping of flavour coherence in the density matrix equations. We also find that the Boltzmann equations equipped with this source are in best numerical agreement with the full master equation results.

This paper is organised as follows: in section~\cref{sec:kb-equations} we review the underlying CTP formalism and examine the general structure of solutions. In section~\cref{sec:quantum-kinetic-equation} we derive the transport equation for the local correlation function $S^\lt_{\evec{k}}(t,t)$ and the key approximation leading to a closure is introduced in section~\cref{sec:local-approximation}. In section~\cref{sec:leptogenesis} we introduce the leptogenesis model, compute the self-energy functions for the decay processes, find the adiabatic background solutions and adapt the transport equations of section~\cref{sec:quantum-kinetic-equation} to the leptogenesis case. We conclude the section with renormalised master equations for leptogenesis, including an equation for the lepton asymmetry and explicit forms for the source and washout terms. In section~\cref{sec:noneq_distribution_functions} we project the neutrino master equation onto different frequency, helicity and flavour quantum states, recasting it as a generalised density matrix equation, which we then further simplify by averaging over the particle-antiparticle oscillations. In section~\cref{sec:expansion-of-universe} we generalise to the case of an expanding universe and in section~\cref{sec:Numerical_results} we give detailed numerical results for the lepton asymmetry in some benchmark cases. In section~\cref{sec:helicity_symmetric_equations} we introduce further approximations, first by dropping the helicity dependence and then reducing the master equation to a density matrix equation in the quasidegenerate case and finally into a Boltzmann equation in the decoupling limit, including a semiclassical source term. In section~\cref{sec:comparison_and_discussion} we present detailed comparisons to earlier work in the literature. Further details of the derivation are presented in several appendices. Finally, section~\cref{sec:conclusions} contains our conclusions and outlook.

%%%%%%%%%%%%%%%%%%%%%%%%%%%%%%%%%%%%%%%%%%%%%%%%%%%%%%%%%%%%%%%%%%%%%%%%%%%%%%%
%
\section{Kadanoff-Baym equations}
\label{sec:kb-equations}
%
%%%%%%%%%%%%%%%%%%%%%%%%%%%%%%%%%%%%%%%%%%%%%%%%%%%%%%%%%%%%%%%%%%%%%%%%%%%%%%%

\begin{figure}[t!]
    \centering
    \includegraphics{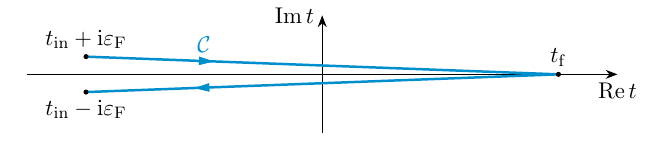}
    \caption{Shown is the Schwinger--Keldysh closed time path $\mathcal{C}$. Arrows show the direction of increasing contour-time.}
    \label{fig:schwinger-keldysh-path}
\end{figure}

For completeness and to introduce the notations, we start with a brief review of the CTP formalism. The basic quantity of interest in the CTP quantum transport theory is the contour-time ordered two-point correlation function. For flavoured fermions it is defined by
\begin{equation}
    \im S_{ij,\alpha\beta}(u,v) \equiv
    \expectval[\big]{\mathcal{T}_\mathcal{C}\bigl[\psi_{i,\alpha}(u) \bar\psi_{j,\beta}(v)\bigr]}
    \text{,} \label{eq:def-fermion-ctp-propagator}
\end{equation}
where $\mathcal{C}$ is a complex time contour and the expectation value is defined as a trace weighted by the non-equilibrium density operator of the system. We will usually suppress the Dirac ($\alpha$, $\beta$) and flavour ($i$, $j$) indices when possible, as they follow the spacetime coordinates $u$ and $v$ of the fermion field $\psi$. All products involving the two-point and self-energy functions are thus implicitly matrix products in Dirac as well as flavour indices. In this paper we consider the Schwinger--Keldysh path shown in figure~\cref{fig:schwinger-keldysh-path} and parametrise the contour function~\cref{eq:def-fermion-ctp-propagator} in terms of four real-time valued correlation functions: the Wightman functions
\begin{subequations}
    \label{eq:wightman-functions-def}
    \begin{alignat}{2}
        &\im S^\lt(u,v) &&\equiv \expectval{\bar\psi(v) \psi(u)}
        \text{,} \\*
        &\im S^\gt(u,v) &&\equiv \expectval{\psi(u) \bar\psi(v)}
        \text{,}
    \end{alignat}
\end{subequations}
and the retarded and advanced pole propagators%
\footnote{A related parametrisation is $S^{ab}$, where $a,b=\pm$ denote the CTP branches of the time arguments $u^0$ and $v^0$: $+$ referring to the upper and $-$ to the lower branch in figure~\cref{fig:schwinger-keldysh-path}. Then $S^\lt =-S^{+-}$ and $S^\gt = S^{-+}$, while $S^{++} = S^{\rm T}$ and $S^{--} = S^{\bar{\rm T}}$ are the time-ordered and reverse time-ordered propagators. Note that our definition of $S^\lt$ has an additional minus sign for fermions compared to what is often used elsewhere in the literature.}
\begin{subequations}
    \label{eq:ret-adv-propagators-def}
    \begin{alignat}{2}
        &\im S^r(u,v) &&\equiv \hphantom{-}2\theta(u^0 - v^0) \mathcal{A}(u,v)
        \text{,} \\*
        &\im S^a(u,v) &&\equiv           - 2\theta(v^0 - u^0) \mathcal{A}(u,v)
        \text{.}
    \end{alignat}
\end{subequations}
Here $\mathcal{A}(u,v) \equiv \frac{1}{2} \expectval{\anticomm{\psi(u), \bar\psi(v)}}$ is the spectral function, which satisfies $\mathcal{A} =\ihalf(S^\gt + S^\lt) = \ihalf(S^r - S^a)$. We also denote $\mathcal{A} = S^{\mathcal{A}}$ below. We also need the Hermitian part of the pole propagators $S^{\rm H} \equiv \shalf(S^r + S^a)$. The fermion self-energy $\Sigma$ can be divided into real-time components similarly and the various self-energy functions $\Sigma^{\lt,\gt}$, $\Sigma^{r,a}$, $\Sigma^{\mathcal{A},{\rm H}}$ satisfy analogous relations.

%------------------------------------------------------------------------------
%
%
%------------------------------------------------------------------------------

With the leptogenesis application in mind, we now specialize to spatially homogeneous and isotropic systems where the correlation and self-energy functions $F \in \{S, \Sigma\}$ have spatial translational invariance: $F(u,v) = F(u^0, v^0; \evec{u} - \evec{v})$. In this case it is convenient to use the two-time representation
\begin{equation}
    F_\evec{k}(t_1,t_2) \equiv
    \int \dd^3 \evec{r} \e^{-\im\evec{k} \cdot \evec{r}} F(t_1, t_2; \evec{r})
    \text{,} \label{eq:two-time-def}
\end{equation}
which is just the spatial Fourier transform and $\evec{k}$ is the Euclidean three-momentum vector.

The Schwinger--Dyson equation obeyed by the two-point function~\cref{eq:def-fermion-ctp-propagator} can be cast into four real-time Kadanoff--Baym (KB) equations for the real-time correlation and self-energy functions. In the homogeneous and isotropic case and in the two-time representation~\cref{eq:two-time-def}
they are given by
\begin{subequations}
    \label{eq:KB-two-time}
    \begin{alignat}{2}
        & \bigl[
            (S_{0,\evec{k}}^{-1} - \Sigma^p_\evec{k}) \convol S^p_\evec{k}
        \bigr](t_1, t_2) &&= \delta(t_1 - t_2)
        \text{,} \label{eq:KB-two-time-A}
        \\*
        & \bigl[
            (S_{0,\evec{k}}^{-1} - \Sigma^r_\evec{k}) \convol S^s_\evec{k}
        \bigr](t_1, t_2) &&= (\Sigma^s_\evec{k} \convol S^a_\evec{k})(t_1, t_2)
        \text{,} \label{eq:KB-two-time-B}
    \end{alignat}
\end{subequations}
where $p = r, a$ and $s = \smallless, \smallgreater$. The convolution appearing here is defined by
\begin{equation}
    (F_\evec{k} \convol G_\evec{k})(t_1, t_2) \equiv
    \int_\tin^\infty \dd t \, F_\evec{k}(t_1, t) G_\evec{k}(t, t_2)
    \text{,} \label{eq:convolution-two-time-def}
\end{equation}
where $\tin$ is the initial time of the CTP, and we have taken the limit $\tf \to \infty$ for the final time. We will eventually also take the limit $\tin\to-\infty$ when deriving the kinetic transport equations. Finally, the inverse free propagator in equations~\cref{eq:KB-two-time} is given by
\begin{equation}
    S_{0,\evec{k}}^{-1}(t_1, t_2) \equiv \bigl[
        \im \gamma^0 \partial_{t_1} - \evec{\gamma} \cdot \evec{k} - m(t_1)
    \bigr] \delta(t_1 - t_2)
    \text{.} \label{eq:inverse-free-propagator-two-time}
\end{equation}
The time dependent, real mass matrix $m(t)$, can be viewed as a singular contribution to the Hermitian part of the self-energy, but with our immediate application to leptogenesis in mind, it is useful to write it explicitly.

Later on we make frequent use of the Hermiticity relations of the propagators and the self-energies. In the two-time representation we can write them as
\begin{subequations}
    \label{eq:hermiticity-two-time}
    \begin{align}
        \bar S^{\lt,\gt}_{\evec{k}}(t_1,t_2)^\dagger &= \bar S^{\lt,\gt}_{\evec{k}}(t_2,t_1)
        \text{,} \label{eq:wightman-hermiticity}
        \\
        \bar S^{\mathcal{A},{\rm H}}_{\evec{k}}(t_1,t_2)^\dagger &= \bar S^{\mathcal{A},{\rm H}}_{\evec{k}}(t_2,t_1)
        \text{,}
        \\
        \bar S^{r,a}_{\evec{k}}(t_1,t_2)^\dagger &= \bar S^{a,r}_{\evec{k}}(t_2,t_1)
        \text{,} \label{eq:pole-propagator-hermiticity}
    \end{align}
\end{subequations}
where we defined $\bar S^{\lt,\gt} \equiv \im S^{\lt,\gt} \gamma^0$, $\bar S^{r,a} \equiv S^{r,a} \gamma^0$ and $\bar S^{\mathcal{A},{\rm H}} \equiv S^{\mathcal{A},{\rm H}} \gamma^0$. The relations~\cref{eq:hermiticity-two-time} follow straightforwardly from the definitions~\cref{eq:wightman-functions-def,eq:ret-adv-propagators-def,eq:two-time-def}. Note that Hermitian conjugation exchanges the pole propagators in addition to their time arguments. Similar Hermiticity properties hold among the self-energy functions (with $S$ replaced by $\Sigma$ in equations~\cref{eq:hermiticity-two-time}) with the definitions $\bar \Sigma^{\lt,\gt} \equiv \gamma^0 \im \Sigma^{\lt,\gt}$, $\bar \Sigma^{r,a} \equiv \gamma^0 \Sigma^{r,a} $ and $\bar \Sigma^{\mathcal{A},{\rm H}} \equiv \gamma^0 \Sigma^{\mathcal{A},{\rm H}}$. We will also need the spectral sum rule, which in the two-time representation reads
\begin{equation}
    2\bar{\mathcal{A}}_\evec{k}(t,t) = \idmat \quad \text{for any $t$}
    \text{.} \label{eq:sum-rule-two-time}
\end{equation}
Equation~\cref{eq:sum-rule-two-time} is a direct consequence of the canonical equal-time anti-commutation relations, and it can be also derived from the definitions~\cref{eq:ret-adv-propagators-def}, the analogous relations for the self-energy functions and
equations~\cref{eq:inverse-free-propagator-two-time,eq:KB-two-time-A}.

The coupled integro-differential equations~\cref{eq:KB-two-time} would be difficult to solve even if the self-energies were some externally given functions. The fact that self-energies are in general functionals of the correlation functions $S^{r,a,\lt,\gt}$, makes~\cref{eq:KB-two-time} also non-linear and consequently much more complicated. Our main goal is to find a tractable and efficient approximation scheme for these equations, which still captures the relevant physics.

%%%%%%%%%%%%%%%%%%%%%%%%%%%%%%%%%%%%%%%%%%%%%%%%%%%%%%%%%%%%%%%%%%%%%%%%%%%%%%%
%
\subsection{Formal solutions}
\label{sec:formal-solutions}
%
%%%%%%%%%%%%%%%%%%%%%%%%%%%%%%%%%%%%%%%%%%%%%%%%%%%%%%%%%%%%%%%%%%%%%%%%%%%%%%%

Before introducing our approximations, it is useful to study some general properties of the solutions. We start by rewriting~\cref{eq:KB-two-time} in an alternative, but equivalent form of Schwinger--Dyson integral equations:
\begin{subequations}
    \label{eq:Schwinger-Dyson-two-time}
    \begin{align}
        S^p_\evec{k} &= S_{0,\evec{k}}^p
        + S_{0,\evec{k}}^p \nconvol \Sigma^p_\evec{k} \nconvol S^p_\evec{k}
        \text{,} \label{eq:Schwinger-Dyson-two-time-A}
        \\*
        S^s_\evec{k} &= S_{0,\evec{k}}^s
        + S_{0,\evec{k}}^s \nconvol \Sigma^a_\evec{k} \nconvol S^a_\evec{k}
        + S_{0,\evec{k}}^r \nconvol \Sigma^s_\evec{k} \nconvol S^a_\evec{k}
        + S_{0,\evec{k}}^r \nconvol \Sigma^r_\evec{k} \nconvol S^s_\evec{k}
        \text{,} \label{eq:Schwinger-Dyson-two-time-B}
    \end{align}
\end{subequations}
where the free propagators $S_0^p$ and $S_0^s$ satisfy
\begin{subequations}
    \label{eq:free-propagators}
    \begin{alignat}{2}
        & S_{0,\evec{k}}^{-1} \nconvol S_{0,\evec{k}}^p &&= \idmat
        \text{,} \label{eq:pole-free}
        \\*
        & S_{0,\evec{k}}^{-1} \nconvol S_{0,\evec{k}}^s &&= 0
        \text{,} \label{eq:Wightman-free}
    \end{alignat}
\end{subequations}
and again $p = r, a$ and $s = \smallless, \smallgreater$.%
\footnote{It is easy to show, using~\cref{eq:pole-free} and \cref{eq:Wightman-free} that any solution $S^s$ to~\cref{eq:Schwinger-Dyson-two-time-B} is also a solution to~\cref{eq:KB-two-time-B}.}
We suppressed the now obvious coordinate dependencies and used $\idmat$ to denote the functional identity matrix (\ie~the delta function in time variables).

We can turn~\cref{eq:Schwinger-Dyson-two-time} into formal solutions by iterating and rearranging the infinite series solutions suitably (intermediate steps of the procedure are given in appendix~\cref{sec:sd-resummation}). The formal pole propagator solutions are given by
\begin{equation}
    S^{p}_\evec{k} = \bigl(S_{0,\evec{k}}^{-1} - \Sigma^p_\evec{k}\bigr)^{-1}
    \text{.} \label{eq:formal-pole-solution}
\end{equation}
The Wightman functions $S^s_\evec{k}$ can be broken into \emph{homogeneous} and \emph{inhomogeneous} parts: $S^s_\evec{k} = S^s_{{\rm hom},\evec{k}} + S^s_{{\rm inh},\evec{k}}$, where
\begin{align}
    S^s_{{\rm hom},\evec{k}} &\equiv
    \bigl(\idmat + S^r_\evec{k} \nconvol \Sigma^r_\evec{k}\bigr)
    % \bigl(S^r_\evec{k} \nconvol S_{0,\evec{k}}^{-1}\bigr) % alternative form
    \nconvol S_{0,\evec{k}}^s \nconvol
    \bigl(\idmat + \Sigma^a_\evec{k} \nconvol S^a_\evec{k}\bigr)
    % \bigl(S_{0,\evec{k}}^{-1} \nconvol S^a_\evec{k}\bigr) % alternative form
    \text{,} \label{eq:formal-homog-solution}
    \\*
    S^s_{{\rm inh},\evec{k}} &\equiv
    S^r_\evec{k} \nconvol \Sigma^s_\evec{k} \nconvol S^a_\evec{k}
    \text{.} \label{eq:formal-inhomog-solution}
\end{align}
The inhomogeneous part $S^s_{{\rm inh},\evec{k}}$ is a particular solution to the full KB equation~\cref{eq:KB-two-time-B}, while the homogeneous part $S^s_{{\rm hom},\evec{k}}$ satisfies the same equation with the right-hand side put to zero. Solutions~\cref{eq:formal-pole-solution,eq:formal-homog-solution,eq:formal-inhomog-solution} are still completely general, but indeed purely formal because the self-energies in general depend on $S_\evec{k}$.%
\footnote{Of course, when one makes approximations, such as the 2PI expansion for the self-energy, the generality of these equations is restricted. In particular, when using the 2PIEA-method, one also makes an implicit assumption of Gaussianity of the initial state~\cite{Berges:2004yj,Garny:2011hg}. This is not always warranted, but the nontrivial correlations induced tend to be short lived, lasting only over $t \lesssim {\mathcal{O}}(1-10)/m$~\cite{Garny:2015oza}.}
However, they provide important insight as to how to define and set up approximative solutions and equations. For example, if the self-energy functions are dominated by some known part, such as the equilibrium contribution, they suggest how to split the pole-functions and the inhomogeneous part of the Wightman function to a leading part and a perturbation, following schematically the division $\Sigma_\evec{k} = \Sigma_{{\rm eq},\evec{k}} + \delta \Sigma_{\evec{k}}$. We shall put this observation into good use in section~\cref{sec:quantum-kinetic-equation} below.

%%%%%%%%%%%%%%%%%%%%%%%%%%%%%%%%%%%%%%%%%%%%%%%%%%%%%%%%%%%%%%%%%%%%%%%%%%%%%%%
%
\subsection{Homogeneous solutions}
\label{sec:homogeneous-solutions}
%
%%%%%%%%%%%%%%%%%%%%%%%%%%%%%%%%%%%%%%%%%%%%%%%%%%%%%%%%%%%%%%%%%%%%%%%%%%%%%%%

We can gain further insight to the homogeneous solution~\cref{eq:formal-homog-solution}, simplifying it further (still in full generality) by using the KB equations. To this end we first substitute the inverse free propagator~\cref{eq:inverse-free-propagator-two-time} to equations~\cref{eq:KB-two-time} and write them explicitly as
\begin{subequations}
    \label{eq:KB-barred}
    \begin{alignat}{2}
        & \bigl[\im\partial_{t_1} - H_\evec{k}(t_1)\bigr]
        \bar S^p_\evec{k}(t_1,t_2)
        - (\bar\Sigma^p_\evec{k} \convol \bar S^p_\evec{k})(t_1, t_2)
        &&= \delta(t_1 - t_2)
        \text{,} \label{eq:KB-barred-A}
        \\*
        & \bigl[\im\partial_{t_1} - H_\evec{k}(t_1)\bigr]
        \bar S^s_\evec{k}(t_1,t_2)
        - (\bar\Sigma^r_\evec{k} \convol \bar S^s_\evec{k})(t_1, t_2)
        &&= (\bar\Sigma^s_\evec{k} \convol \bar S^a_\evec{k})(t_1, t_2)
        \text{.} \label{eq:KB-barred-B}
    \end{alignat}
\end{subequations}
Here $H_\evec{k}(t) \equiv \evec{\alpha} \cdot \evec{k} + \gamma^0 m(t)$ is the free Dirac Hamiltonian with $\evec{\alpha} \equiv \gamma^0 \evec{\gamma}$, and we used the barred propagators $\bar S$ and self-energies $\bar \Sigma$ defined below equations~\cref{eq:hermiticity-two-time}. Using the pole equation~\cref{eq:KB-barred-A} and its Hermitian conjugate in equation~\cref{eq:formal-homog-solution} then leads to
\begin{align}
    &\bar S^s_{{\rm hom},\evec{k}}(t_1, t_2) \notag
    \\*
    &= \int_\tin^\infty \!\! \int_\tin^\infty \dd t \dd t' \,
    \bar S^r_\evec{k}(t_1, t)
    \bigl[-\im \overset\leftarrow{\partial_t} - H_\evec{k}(t)\bigr]
    \bar S^s_{0,\evec{k}}(t, t')
    \bigl[\im \partial_{t'} - H_\evec{k}(t')\bigr]
    \bar S^a_\evec{k}(t', t_2) \notag
    \\*
    &= 2\bar{\mathcal{A}}_\evec{k}(t_1, \tin)
    \bar S^s_{0, \evec{k}}(\tin, \tin)
    2\bar{\mathcal{A}}_\evec{k}(\tin, t_2) \notag
    \\*
    &= 2\bar{\mathcal{A}}_\evec{k}(t_1, \tin)
    \bar S^s_{{\rm hom}, \evec{k}}(\tin, \tin)
    2\bar{\mathcal{A}}_\evec{k}(\tin, t_2)
    \text{.} \label{eq:local-wightman-homog-tin}
\end{align}
To get the second equality we integrated the derivative terms by parts, used equation~\cref{eq:Wightman-free} and its Hermitian conjugate and the definitions~\cref{eq:ret-adv-propagators-def}. The only remaining term in the expression is then the boundary term involving the initial time $\tin$. Note that it is essential in this derivation that $\tin$ is smaller than both $t_1$ and $t_2$. Finally, we used the sum rule~\cref{eq:sum-rule-two-time} at $t = \tin$ to identify $\bar S^s_{0,\evec{k}}(\tin,\tin) = \bar S^s_{{\rm hom},\evec{k}}(\tin,\tin)$.

%------------------------------------------------------------------------------
%
\begin{figure}[t!]
    \centering
    \includegraphics[width=13.5cm]{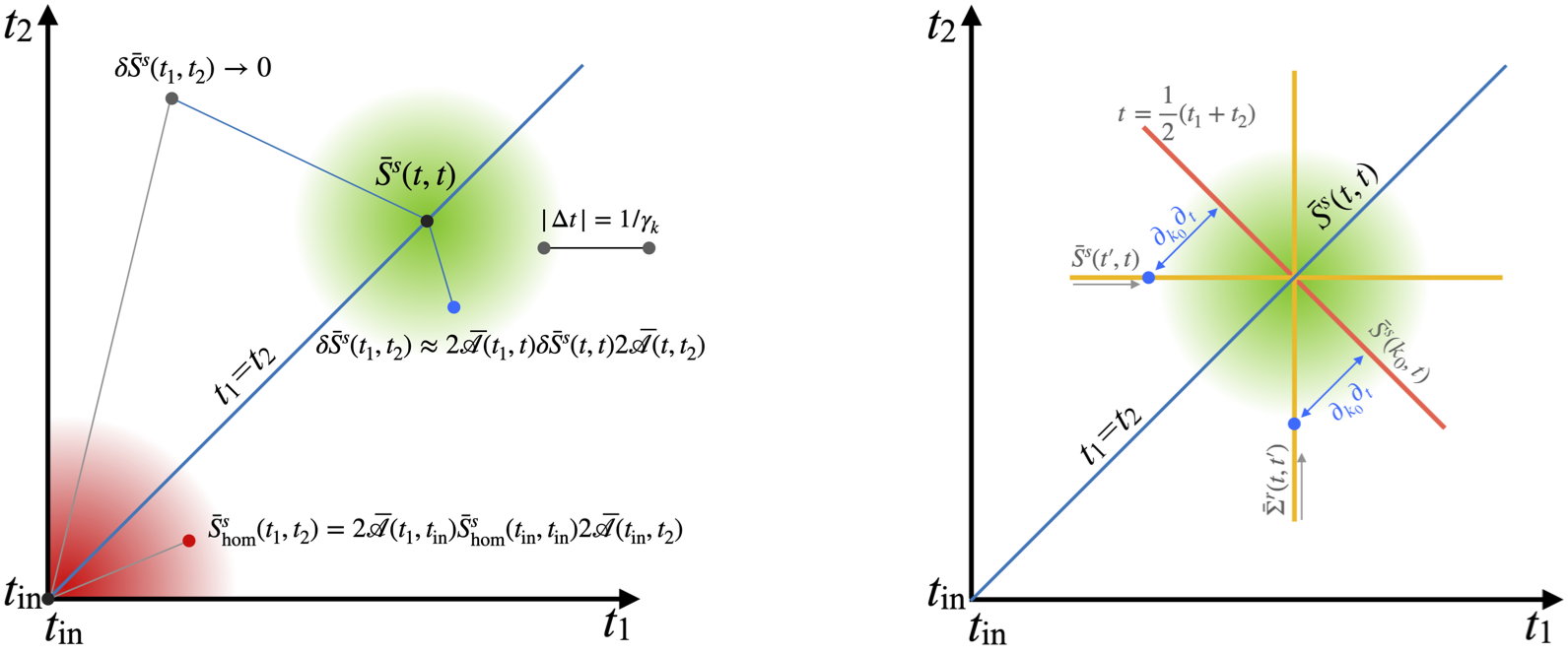}
    \caption{Left panel: schematics of the propagation of the homogeneous correlation from the initial condition set at $(t_1,t_2) = (\tin,\tin)$ (red) and the extent of the non-local correlation around a local solution defined at $(t_1,t_2) = (t,t)$ (green, see section~\cref{sec:local-approximation}). The strength of the colour illustrates the strength of the correlation. Outside the spheres, with $\abs{\Delta t} \gg 1/\gamma_{\evec{k}}$, points are no longer correlated (with $(\tin,\tin)$ and $(t,t)$ respectively) because of the dissipation. Right panel: shown are the particular time-contours of relevance for the local equation as discussed in section~\cref{sec:local-kinetic-equation}.}
    \label{fig:schematics-of-dissipation}
\end{figure}
%
%------------------------------------------------------------------------------

The result~\cref{eq:local-wightman-homog-tin} is exact and it shows that homogeneous solutions describe transients evolving from initial conditions set at a finite initial time $\tin$. Indeed, it is easy to show that the spectral function is the unitary time-evolution operator for two-point functions in the free theory limit: $2\bar{\mathcal{A}}_0(t_1,t_2) = U(t_1,t_2)$, for which the spectral sum rule~\cref{eq:sum-rule-two-time} imposes the correct normalisation $U(t,t) = 1$. However, in dissipative systems this time-evolution is no longer unitary. Dissipation shows up as a finite width of the pole functions in the Wigner representation. Going to the Wigner representation, finding quasiparticle poles and computing the corresponding quasiparticle widths $\gamma_{\evec{k}}$, and transforming back to the two-time representation, one could then obtain a more general $\mathcal{A}_\evec{k}$ effecting an explicit non-unitary decay of correlations in the relative time~\cite{Jukkala:2019slc}: $2\bar{\mathcal{A}}_\evec{k}(t_1,t_2) \sim \e^{-\gamma_\evec{k} \abs{t_1 - t_2}}$.%
\footnote{In this qualitative explanation we do not explicitly account for flavour dependence. It is clear however, that flavour dependent $\gamma_\evec{k}$'s correspond to the widths of quasi-states in the locally diagonalised matter basis.}
As a result, if the initial time is pushed to the past infinity, $\tin \to -\infty$, all memory of the initial conditions will vanish, leaving only the inhomogeneous solution. We illustrate this phenomenon schematically in figure~\cref{fig:schematics-of-dissipation}.

We find the homogeneous solutions of the form~\cref{eq:local-wightman-homog-tin} to be very useful still in another way, which is crucial to our scheme. We will return to this issue in section~\cref{sec:local-approximation} below.

%%%%%%%%%%%%%%%%%%%%%%%%%%%%%%%%%%%%%%%%%%%%%%%%%%%%%%%%%%%%%%%%%%%%%%%%%%%%%%%
%
\section{Local quantum kinetic equation}
\label{sec:quantum-kinetic-equation}
%
%%%%%%%%%%%%%%%%%%%%%%%%%%%%%%%%%%%%%%%%%%%%%%%%%%%%%%%%%%%%%%%%%%%%%%%%%%%%%%%

The main source of complexity in equations~\cref{eq:KB-barred} comes from their non-locality and the associated need for a complete accounting of the memory effects. Yet, all physical observables are expressible in terms of the local correlation function $S_\evec{k}(t,t)$, and moreover, we have seen that dissipative processes in general wash out memory effects over time intervals $\abs{\Delta t} \ge 1/\gamma_\evec{k}$. This suggests to try to find an approximative equation that involves \emph{only} the local correlation function. Such an equation is easy to set up formally: we first use the chain rule to write%
\footnote{In order to get an equation for the local correlation function, it is essential that the derivative acts on both time-variables. Otherwise the limiting procedure introduces other independent functions (first moment of the propagator in the mixed representation) to the left hand side of the equation~\cref{eq:general-local-eom}.}
\begin{equation}
    \partial_t S^\lt_\evec{k}(t,t) =
    \Bigl[
        (\partial_t + \partial_{t'})S^\lt_\evec{k}(t,t')
    \Bigr]_{t' = t} \,\text{.}
\end{equation}
Using this with the KB equation~\cref{eq:KB-barred-B} and its Hermitian conjugate together with the Hermiticity relations~\cref{eq:hermiticity-two-time} yields the equal-time equation
\begin{align}
    \im\partial_t \bar S^\lt_\evec{k}(t,t) =
    \comm[\big]{H_\evec{k}(t), \bar S^\lt_\evec{k}(t,t)}
    &+ (\bar\Sigma^r_\evec{k} \convol \bar S^\lt_\evec{k})(t,t)
    - (\bar S^\lt_\evec{k} \convol \bar\Sigma^a_\evec{k})(t,t) \notag
    \\*
    &+ (\bar\Sigma^\lt_\evec{k} \convol \bar S^a_\evec{k})(t,t)
    - (\bar S^r_\evec{k} \convol \bar\Sigma^\lt_\evec{k})(t,t)
    \text{.} \label{eq:general-local-eom}
\end{align}
This equation is still exact, but of course not closed because it still involves the non-local function $S^\lt_\evec{k}(t,t')$ with $t\neq t'$ explicitly in the interaction convolution terms $\bar\Sigma^r_\evec{k}\convol\bar S^\lt_\evec{k}$ and $\bar S^\lt_\evec{k}\convol\bar\Sigma^a_\evec{k}$ and implicitly within the self-energies. It also depends explicitly on the pole functions $S^{r,a}_\evec{k}(t_1,t_2)$. To make~\cref{eq:general-local-eom} self-contained, we need to supply it with enough information of these particular correlations, without going back to the full KB equations.

%%%%%%%%%%%%%%%%%%%%%%%%%%%%%%%%%%%%%%%%%%%%%%%%%%%%%%%%%%%%%%%%%%%%%%%%%%%%%%%
%
\subsection{Perturbation around the adiabatic solution}
\label{sec:perturbation-and-adiabatic-solution}
%
%%%%%%%%%%%%%%%%%%%%%%%%%%%%%%%%%%%%%%%%%%%%%%%%%%%%%%%%%%%%%%%%%%%%%%%%%%%%%%%

Let us first address the coupling between the Wightman functions and the pole functions. This issue is intimately connected to finding a good approximation for the inhomogeneous solution for the Wightman functions as is suggested by equations~\cref{eq:formal-inhomog-solution,eq:formal-pole-solution}. We start by formally dividing the correlation functions into some known adiabatic background solution and a perturbation:
\begin{equation}
    S_\evec{k}^\alpha = S^\alpha_{{\rm ad},\evec{k}} + \delta S_\evec{k}^\alpha
    \text{.} \label{eq:adiabatic-expansion}
\end{equation}
where $\alpha = r, a, \smallless, \smallgreater$. There is some freedom as to how to choose the adiabatic solutions. For example, in a system near thermal equilibrium, the instantaneous thermal equilibrium solutions would be an obvious choice. More formally, the adiabatic Wightman function can be defined as a solution that reduces to the stationary solution of~\cref{eq:general-local-eom}, when ignoring all local time dependence.%
\footnote{This corresponds to working to lowest order in gradients in the Wigner representation.}
When $S^\alpha_{{\rm ad},\evec{k}}$ fulfil this requirement, inserting the division~\cref{eq:adiabatic-expansion} into equation~\cref{eq:general-local-eom} leads to
\begin{align}
    \im\partial_t \delta\bar S^\lt_\evec{k}(t,t) \simeq
    \comm{H_\evec{k}(t), \delta\bar S^\lt_\evec{k}(t,t)}
    &+ (\bar\Sigma^r_\evec{k} \convol \delta\bar S^\lt_\evec{k})(t,t)
    - (\delta\bar S^\lt_\evec{k} \convol \bar\Sigma^a_\evec{k})(t,t) \notag
    \\*
    &+ (\bar\Sigma^\lt_\evec{k} \convol \delta\bar S^a_\evec{k})(t,t)
    - (\delta\bar S^r_\evec{k} \convol \bar\Sigma^\lt_\evec{k})(t,t) \notag
    \\*
    &- \im\partial_t \bar S^\lt_{{\rm ad},\evec{k}}(t,t)
    \text{.} \label{eq:general-local-eom-for-perturbation}
\end{align}
The source $-\im\partial_t \bar S^\lt_{{\rm ad},\evec{k}}(t,t)$ is the leading correction left from the terms involving the adiabatic solution in the full dynamical equation. Note that as the Wightman function can contain also a homogeneous transient, $\delta S^\lt_\evec{k}$ is not necessarily small even when the adiabatic solution is an equilibrium solution.

The equal-time pole functions, on the other hand, can be taken to be purely adiabatic with no dynamical perturbations: $\delta S_\evec{k}^p = 0$. In the equal-time limit this condition is actually strictly imposed by the spectral sum rule~\cref{eq:sum-rule-two-time} and the relations~\cref{eq:ret-adv-propagators-def} between the pole functions and the spectral function. One can also show that $\partial_t S^{p}_{{\rm ad},\evec{k}}(t,t) = 0$ exactly at least in a non-interacting theory. This suggests that no perturbations in the pole functions can be included in a truncation to the local limit. It is then remarkable, that in the truncation scheme for the collision terms developed in the next section, the non-local pole function perturbations vanish consistently with their vanishing equal-time counterparts. This means that pole functions become entirely non-dynamical quantities that account for the structure of the phase space only. With this information, equation~\cref{eq:general-local-eom-for-perturbation} further reduces to
\begin{equation}
    \im\partial_t \delta\bar S^\lt_\evec{k}(t,t) \simeq
    \comm[\big]{H_\evec{k}(t), \delta\bar S^\lt_\evec{k}(t,t)}
    - \im\partial_t \bar S^\lt_{{\rm ad},\evec{k}}(t,t)
    + \bigl[
        (\bar\Sigma^r_\evec{k} \convol \delta\bar S^\lt_\evec{k})(t,t) - \text{H.c.}
    \bigr]
    \text{.} \label{eq:local-eom-for-perturbation-1st}
\end{equation}
This equation no longer depends explicitly on the pole functions. They only have limited influence on~\cref{eq:local-eom-for-perturbation-1st} through the source function and the self-energy functions, as we shall discuss below.

%%%%%%%%%%%%%%%%%%%%%%%%%%%%%%%%%%%%%%%%%%%%%%%%%%%%%%%%%%%%%%%%%%%%%%%%%%%%%%%
\paragraph{On the choice of the adiabatic solution.}
%%%%%%%%%%%%%%%%%%%%%%%%%%%%%%%%%%%%%%%%%%%%%%%%%%%%%%%%%%%%%%%%%%%%%%%%%%%%%%%

Explicit forms for the adiabatic solutions are most conveniently given in the Wigner representation, which is defined as the Fourier transform of the two-time representation~\cref{eq:two-time-def} with respect to the relative time-coordinate:
\begin{equation}
    F(k,t) \equiv
    \int_{-\infty}^\infty \dd r^0 \e^{\im k^0 r^0}
    F_\evec{k}\bigl(t + \sfrac{r^0}{2}, t - \sfrac{r^0}{2}\bigr)
    \text{.} \label{eq:wigner-transform}
\end{equation}
Here $t = (t_1 + t_2)/2$ is the average time coordinate and $k^0$ is the internal energy conjugate to the relative time coordinate $r^0 = t_1 - t_2$. Note that we always take the limit $\tin \to -\infty$ before calculating any Wigner transforms. We now define the adiabatic solutions with instantaneous mass $m(t)$ and self-energies as follows%
\footnote{\label{footnote:out-transform-in-adiabatic}More precisely, we should replace the self-energies $\Sigma(k,t)$ in equations~\cref{eq:adiabatic-solutions-def} by their ``out-versions'' defined below equation~\cref{eq:HKR-ansatz-gen-wspace}. This is important when the self-energy depends on the perturbation $\delta S$ itself and may thus contain rapidly oscillating coherence functions~\cite{Herranen:2010mh,Fidler:2011yq,Jukkala:2019slc}. However, this phenomenon is not relevant for us in this paper where we eventually will average over such fast fluctuations and both definitions correspond to the same local correlator in the two-time representation.}:
\begin{subequations}
    \label{eq:adiabatic-solutions-def}
    \begin{align}
        S^p_{\rm ad}(k,t) &= \bigl[\slashed k - m(t) - \Sigma^p_{\rm ad}(k,t)\bigr]^{-1}
        \text{,} \\*
        S^s_{\rm ad}(k,t) &= S^r_{\rm ad}(k,t) \Sigma^s_{\rm ad}(k,t) S^a_{\rm ad}(k,t)
        \text{.} \label{eq:adiabatic-wightman}
    \end{align}
\end{subequations}

There still is significant freedom left in these solutions related to the choice of the self-energy functions $\Sigma^\alpha_{\rm ad}$. For example, one might choose to ignore or include the Hermitian part $\Sigma^{\rm H}_{\rm ad}$ of the pole self-energy functions, leading to solutions either with vacuum or quasiparticle dispersion relations. Moreover, if one neglects the finite width, setting $\Sigma^\mathcal{A}_{\rm ad} = 0$, the corresponding solutions become spectral (either vacuum or quasiparticle). This is the choice made in the derivation of cQPA-formalism~\cite{Herranen:2008hi,Herranen:2008hu,Herranen:2008yg,Herranen:2008di,Herranen:2010mh,Fidler:2011yq,Herranen:2011zg,Jukkala:2019slc}, as well as in the usual Boltzmann theory~\cite{Mahan:1987251}. If one includes a finite width however, the adiabatic part of the solution spreads out in phase space, with a consequent change in the source term in the equation~\cref{eq:local-eom-for-perturbation-1st} for the perturbation $\delta S^\lt_\evec{k}$. One can even include corrections from the perturbations $\delta S^\lt_\evec{k}$ in the self-energies, without changing the basic structure of the equation for the perturbation $\delta S^\lt_\evec{k}$ itself. The point is that the validity of all these approximations is controlled by the coupling constant expansion.

%%%%%%%%%%%%%%%%%%%%%%%%%%%%%%%%%%%%%%%%%%%%%%%%%%%%%%%%%%%%%%%%%%%%%%%%%%%%%%%
%
\subsection{Local approximation}
\label{sec:local-approximation}
%
%%%%%%%%%%%%%%%%%%%%%%%%%%%%%%%%%%%%%%%%%%%%%%%%%%%%%%%%%%%%%%%%%%%%%%%%%%%%%%%

The division into an adiabatic background and the perturbation simplified the original equal-time equation considerably. However, the problem of closure still remains: our equation describes the evolution of the perturbation only along the diagonal $t_1 = t_2$ in figure~\cref{fig:schematics-of-dissipation}, but the collision integrals depend on $\delta S^\lt_\evec{k}(t_1,t_2)$ everywhere in the two-time plane. In a system with dissipation, these \emph{memory effects} are suppressed however, and there is hope that a strictly local description can be found. We use the evolution of the homogeneous perturbation~\cref{eq:local-wightman-homog-tin} as our guiding principle to reach this goal.

Indeed, imagine first that we somehow have found the correct solution $\delta S^s_\evec{k}(t,t)$ along the diagonal. For $t_{1,2}$ not too far from the time $t$, the true non-local solution $\delta S^s_\evec{k}(t_1,t_2)$ should be correlated with $\delta S^s_\evec{k}(t,t)$, and even reasonably well approximated by a homogeneous solution similar to equation~\cref{eq:local-wightman-homog-tin}, with the initial time $t_{\rm in}$ replaced by the local time $t$. Even when we do not know the local solution beforehand, we can \emph{parametrise} the non-local solution with the local one. Specifically, we make the \emph{local ansatz}
\begin{equation}
    \delta\bar S^s_\evec{k}(t_1,t_2)
    =
    2\bar{\mathcal{A}}_\evec{k}(t_1,t)
    \delta\bar S^s_\evec{k}(t,t)
    2\bar{\mathcal{A}}_\evec{k}(t,t_2) \quad \text{for any } t
    \text{.} \label{eq:local-approximation}
\end{equation}
Note in particular that the spectral function is not a dynamical quantity in~\cref{eq:local-approximation}; following the discussion of the previous section, the generalised time evolution operator $2\bar{\mathcal{A}}_\evec{k}(t_1,t_2)$ is a non-dynamical adiabatic solution that can be computed to the desired accuracy independently from the local correlation function $\delta S^s_\evec{k}(t,t)$. We stress that the ansatz~\cref{eq:local-approximation} will only be used in the convolution terms describing the interactions. Indeed, if taken to hold universally for all $t$, it would generally be too restrictive and in contradiction with the local equation of motion~\cref{eq:local-eom-for-perturbation-1st}.

%%%%%%%%%%%%%%%%%%%%%%%%%%%%%%%%%%%%%%%%%%%%%%%%%%%%%%%%%%%%%%%%%%%%%%%%%%%%%%%
\paragraph{cQPA and Boltzmann theory limit.}
%%%%%%%%%%%%%%%%%%%%%%%%%%%%%%%%%%%%%%%%%%%%%%%%%%%%%%%%%%%%%%%%%%%%%%%%%%%%%%%

It should be stressed that the ansatz~\cref{eq:local-approximation} is an exact relation in free theory and for spectral quasiparticles, where the full solution is homogeneous and the free spectral function is the unitary time evolution operator. Indeed, we can derive the cQPA-correlation function~\cite{Fidler:2011yq, Jukkala:2019slc}, and eventually the Boltzmann theory limit directly from~\cref{eq:local-approximation}. First choosing $t = (t_1 + t_2)/2$ and then Wigner-transforming~\cref{eq:local-approximation}, one finds (here we write the results explicitly for $S^\lt$ but analogous results hold for $S^\gt$)
\begin{equation}
    \delta \bar S^\lt(k,t) = \int \frac{\dd p^0}{\pi}
    2\bar{\mathcal{A}}_{{\rm in},\evec{k}}(p^0,t)
    \delta \bar S^\lt_\evec{k}(t,t)
    2\bar{\mathcal{A}}_{{\rm out},\evec{k}}(2k^0 - p^0,t)
    \text{,} \label{eq:HKR-ansatz-gen-wspace}
\end{equation}
where $\mathcal{A}_{\rm out}(k,t)\equiv\e^{\frac{\im}{2}\partial_t \partial_{k^0}}\mathcal{A}(k,t)$ and $\mathcal{A}_{\rm in}(k,t)\equiv\e^{-\frac{\im}{2}\partial_t \partial_{k^0}}\mathcal{A}(k,t)$. This form is still valid for any adiabatic solution for the spectral function. Working to lowest order in gradients and using the free adiabatic spectral function,
\begin{equation}
    \bar{\mathcal{A}}^{(0)}_{\evec{k}ij}(k^0,t)
    = \pi \sgn(k^0) \bigl(\slashed k + m_i(t)\bigr) \gamma^0 \,
    \delta(k^2 - m_i^2) \delta_{ij}
    \text{,} \label{eq:vacuumA}
\end{equation}
equation~\cref{eq:HKR-ansatz-gen-wspace} reduces to the spectral form
\begin{equation}
    \delta \bar S^\lt_{ij}(k,t) = 2\pi \sum_{h,\pm,\pm'}
    P_{\evec{k}h}^{} P_{\evec{k}i}^\pm \, \delta \bar S^\lt_{\evec{k}hij}(t,t) \, P_{\evec{k}j}^{\pm'}
    \delta\bigl(
        k^0 \mp \sfrac{1}{2}\omega_{\evec{k}i} \mp' \sfrac{1}{2}\omega_{\evec{k}j}
    \bigr) \text{,} \label{eq:firstform}
\end{equation}
where the helicity and energy projection matrices are defined by
\begin{equation}
    P_{\evec{k}h}^{} = \frac{1}{2}\Bigl(\idmat + h \hat{h}_{\evec{k}}\Bigr)
    \text{,} \qquad
    P_{\evec{k}i}^s = \frac{1}{2}\biggl(\idmat + s\frac{H_{\evec{k}i}}{\omega_{\evec{k}i}}\biggr)
    \text{,} \label{eq:projector-def}
\end{equation}
with $\hat{h}_{\evec{k}} \equiv \evec{\alpha} \cdot \Uevec{k} \,\gamma^5$, $H_{\evec{k}i} = \evec{\alpha} \cdot \evec{k} + \gamma^0 m_i$ and $\omega_{\evec{k}i} = \sqrt{\abs{\evec{k}}^2 + {m_i}^2}$. The helicity and energy indices both take values $h, s = {\pm}$. It is easy to show that the projectors $P_{\evec{k}h}^{} P_{\evec{k}i}^\pm \gamma^0 P_{\evec{k}j}^{\pm'}$ form a complete basis of matrices consistent with homogeneity and isotropy (this will be elaborated further in section~\cref{sec:projector-parametrisation}). In the spectral limit the adiabatic solutions and the perturbations can be combined on the common shell functions. Expanding the corresponding full $\bar S^\lt_\evec{k}(t,t)$ in this basis (for the precise definition of $\mathcal{P}^{ss'}_{\evec{k}hij}$ see equation~\cref{eq:local-correlator-parametrisation} below), we can rewrite equation~\cref{eq:firstform} as
\begin{equation}
    \bar S^\lt_{ij}(k,t) = 2\pi \sum_{h,\pm} \Bigl(
        \mathcal{P}^{\pm\pm}_{\evec{k}hij} f^{m,\pm}_{\evec{k}hij}
        \delta(k^0 \mp \bar\omega_{\evec{k}ij})
        + \mathcal{P}^{\pm\mp}_{\evec{k}hij} f^{c,\pm}_{\evec{k}hij}
        \delta(k^0 \mp \sfrac{1}{2}\Delta\omega_{\evec{k}ij})
    \Bigr)
    \text{,} \label{eq:full_wightman_diag}
\end{equation}
where $\bar\omega_{\evec{k}ij} \equiv (\omega_{\evec{k}i} + \omega_{\evec{k}j})/2$ and $\Delta\omega_{\evec{k}ij} \equiv \omega_{\evec{k}i} - \omega_{\evec{k}j}$. This is the flavoured cQPA-propagator, up to normalisation, derived in~\cite{Fidler:2011yq} and it carries information of all coherence structures consistent with homogeneity and isotropy in the spectral limit.

If one ignores all coherence information, equation~\cref{eq:full_wightman_diag} reduces to a generalised KB-ansatz
\begin{equation}
    \bar S^\lt_{ij}(k,t) =
    \sum_{h} P_{\evec{k}h}^{} f^{m,\sgn(k^0)}_{\evec{k}hii} \,
    2 \bar{\mathcal{A}}^{(0)}_{\evec{k}ij}(k^0,t) \sgn(k^0)
    \text{,}
\end{equation}
which corresponds to the Boltzmann theory limit with distribution functions that are diagonal in flavour and helicity. The extra $\sgn(k^0)$ factor could be absorbed to normalisation, but the present normalisation will be more convenient later. Moreover, if one imposes the thermal equilibrium Kubo--Martin--Schwinger (KMS) condition for the cQPA Wightman functions, $S^\gt(k,t) = \e^{\beta k^0}S^\lt(k,t)$, which now is equivalent to $\bar S^\lt_{ij}(k,t) = 2 \smash{\bar{\mathcal{A}}^{(0)}_{\evec{k}ij}}(k^0,t) f_{\rm FD}(k^0)$, the distribution function further reduces to the thermal Fermi--Dirac distribution: $f^{m,\pm}_{\evec{k}hii} = \pm f_{\rm FD}(\pm\omega_{\evec{k}i})$ (\cf~\cite{Jukkala:2019slc}).

%%%%%%%%%%%%%%%%%%%%%%%%%%%%%%%%%%%%%%%%%%%%%%%%%%%%%%%%%%%%%%%%%%%%%%%%%%%%%%%
%
\subsection{Local transport equation}
\label{sec:local-kinetic-equation}
%
%%%%%%%%%%%%%%%%%%%%%%%%%%%%%%%%%%%%%%%%%%%%%%%%%%%%%%%%%%%%%%%%%%%%%%%%%%%%%%%

For us, the most important utility of the local approximation~\cref{eq:local-approximation} is that it allows closure in equation~\cref{eq:local-eom-for-perturbation-1st}, reducing all interaction convolutions containing the non-local function $\delta\bar S^\lt_\evec{k}(t,t')$ to simple matrix products involving only the local function $\delta\bar S^\lt_\evec{k}(t,t)$. For example,
\begin{align}
    (\bar\Sigma^r_\evec{k} \convol \delta\bar S^\lt_\evec{k})(t,t) &=
    \int_\tin^\infty \dd t'\, \bar\Sigma^r_\evec{k}(t,t') \delta\bar S^\lt_\evec{k}(t',t) \notag
    \\*
    &\simeq \biggl[
        \int_\tin^\infty \dd t' \, \bar\Sigma^r_\evec{k}(t,t') 2\bar{\mathcal{A}}_\evec{k}(t',t)
    \biggr] \delta\bar S^\lt_\evec{k}(t,t) \notag
    \\*
    &\equiv \bar\Sigma^r_{{\rm eff},\evec{k}}(t,t) \delta\bar S^\lt_\evec{k}(t,t)
    \text{,} \label{eq:local-approx-for-self-energy}
\end{align}
where we introduced the effective self-energy $\bar\Sigma^r_{{\rm eff},\evec{k}}$. We remind that the spectral function is adiabatic $\mathcal{A} \equiv \mathcal{A}_{\rm ad}$ in these expressions, consistent with our approximation scheme. While the effective self-energy
\begin{equation}
    \bar\Sigma^r_{{\rm eff},\evec{k}}(t,t)
    = \bigl(\bar\Sigma^r_\evec{k} \convol 2\bar{\mathcal{A}}_{\evec{k}}\bigr)(t,t)
    \label{eq:effective-sigma-ad}
\end{equation}
is still a convolution, it can be computed at any time during the solution based only on the local solution itself, or independently of it, depending on the approximation one uses for the adiabatic functions, as discussed in section~\cref{sec:perturbation-and-adiabatic-solution}.

We now use the local approximation~\cref{eq:local-approximation} to obtain closure in equation~\cref{eq:local-eom-for-perturbation-1st}. This amounts to using~\cref{eq:local-approx-for-self-energy} and its Hermitian conjugate with~\cref{eq:effective-sigma-ad} in equation~\cref{eq:local-eom-for-perturbation-1st}, resulting in the local equation of motion
\begin{equation}
    \partial_t \delta\bar S^\lt_\evec{k}(t,t)
    + \im \comm[\big]{H_\evec{k}(t), \delta\bar S^\lt_\evec{k}(t,t)}
    = - \partial_t \bar S^\lt_{{\rm ad},\evec{k}}(t,t) - \bigl(
        \im \bar\Sigma^r_{{\rm eff},\evec{k}}(t,t) \delta\bar S^\lt_\evec{k}(t,t) + \text{H.c.}
    \bigr) \text{.} \label{eq:closed-local-eom}
\end{equation}
Equation~\cref{eq:closed-local-eom} is our final quantum kinetic equation (QKE) for non-equilibrium evolution of mixing fermions. The non-local memory integrals have been truncated by the local approximation, so it is an ordinary (matrix) differential equation for the local non-equilibrium correlation function $\delta\bar S^\lt_\evec{k}(t,t)$. Equation~\cref{eq:closed-local-eom} still describes both flavour and particle-antiparticle coherence effects of the mixing fermions. It also takes into account quantum statistical effects of the thermal medium (within the weak coupling expansion), and it can accommodate thermal corrections to the dispersion relations via the effective self-energy and the adiabatic source term. We have shown that~\cref{eq:closed-local-eom} encompasses the coherent cQPA-formalism and consequently the usual Boltzmann theory including also semiclassical corrections~\cite{Jukkala:2019slc}, but it is a more general formulation in that it is not restricted to the spectral limit. We will apply this equation in the leptogenesis setting to describe the evolution of the right-handed Majorana neutrinos in the next section.

%%%%%%%%%%%%%%%%%%%%%%%%%%%%%%%%%%%%%%%%%%%%%%%%%%%%%%%%%%%%%%%%%%%%%%%%%%%%%%%
\paragraph{On the accuracy of the local ansatz.}
%%%%%%%%%%%%%%%%%%%%%%%%%%%%%%%%%%%%%%%%%%%%%%%%%%%%%%%%%%%%%%%%%%%%%%%%%%%%%%%

Despite its wide range of applicability, the ansatz~\cref{eq:local-approximation} should eventually break down if the system develops significant temporal correlations (a memory) over large time intervals. When would this happen and how large would the corrections be? Ultimately one would like to compare the results obtained using the local equation~\cref{eq:closed-local-eom} with a numerical solution of the full non-local two-time equations~\cref{eq:KB-barred}, but we can get a good idea of the size of the memory effects by studying their origin in the Wigner representation.

We first note that the Wigner transform~\cref{eq:wigner-transform} encodes all dependence on the relative time $r^0 = t_1 - t_2$ at constant average time slices $t = \sfrac{1}{2}(t_1 + t_2)$ in the frequency components $S_{\evec{k}}(k^0,t)$ (see the right panel of figure~\cref{fig:schematics-of-dissipation} for illustration). In the weak coupling limit this information gets concentrated on narrow shells in frequency space and eventually to spectral solutions when widths are neglected. The non-local information relevant for equations~\cref{eq:local-eom-for-perturbation-1st,eq:closed-local-eom} is contained in the convolution integrals that can be expressed as follows:
\begin{equation}
    (\bar \Sigma^r_\evec{k} \convol \bar G_\evec{k})(t,t)
      = \int \dd t' \,\bar \Sigma^r_\evec{k}(t,t') \bar G_\evec{k}(t',t)
      = \int \frac{\dd k^0}{2\pi} \,
      \bar \Sigma^r_{\rm out}\bigl(k + \sfrac{\im}{2}\partial_t, t\bigr) \bar G(k,t)
    \text{,} \label{eq:local-convolutions}
\end{equation}
where $\bar G=\delta \bar S^\lt$ in~\cref{eq:local-eom-for-perturbation-1st} and $\bar G = \bar{\mathcal{A}}$ in~\cref{eq:closed-local-eom}.%
\footnote{There is one subtlety if we take $\bar G = \delta \bar S^\lt$ as in equation~\cref{eq:local-eom-for-perturbation-1st}. In this case one has to account for the gradient operator in the argument of $\bar\Sigma^r_{\rm out}$, when it acts on the rapidly oscillating coherence solutions in $\delta \bar S^\lt$. This ensures that the coherence shell contributions get computed on correct frequency shells in the cQPA-formulation (see \eg~\cite{Jukkala:2019slc}). One of the nice features of the ansatz~\cref{eq:local-approximation} is that it fully automatises this resummation, also when evaluating higher loop self-energy functions~\cite{Fidler:2011yq}. Indeed, the issue clearly does not arise when $\bar G = \bar{\mathcal{A}}$, because $\mathcal{A}$ is an adiabatic function.}
Clearly, for a fixed $t$ the non-local information of $G_{\evec{k}}(t',t)$ contained in the two-time convolution (along the contour $t_2 = t$) is fully encoded in the gradients in the Wigner representation, since $G(k,t)$ only contains information along the contour $\sfrac{1}{2}(t_1 + t_2) = t$. This correspondence is schematically illustrated by the blue arrows in figure~\cref{fig:schematics-of-dissipation}.

Finally then, the validity of the local approximation~\cref{eq:local-approximation} boils down to the smallness of the gradient corrections and assuming that $\delta S(k,t)$ has a similar phase space structure as the adiabatic solution. In the leptogenesis application gradient corrections are controlled by the Hubble expansion and hence they are small since $H/T \ll 1$. The phase space structures of the adiabatic solution and the perturbation $\delta S(k,t)$ should be similar because the latter is created by the former. Also, both solutions become spectral when the width is zero, so this approximation becomes good also in the weak coupling limit.

%%%%%%%%%%%%%%%%%%%%%%%%%%%%%%%%%%%%%%%%%%%%%%%%%%%%%%%%%%%%%%%%%%%%%%%%%%%%%%%
%
\section{Leptogenesis}
\label{sec:leptogenesis}
%
%%%%%%%%%%%%%%%%%%%%%%%%%%%%%%%%%%%%%%%%%%%%%%%%%%%%%%%%%%%%%%%%%%%%%%%%%%%%%%%

We now apply our methods to study lepton asymmetry production in the early universe. Leptogenesis has different variants, including the original thermal leptogenesis~\cite{Fukugita:1986hr}, resonant leptogenesis~\cite{Pilaftsis:1997jf,Pilaftsis:2003gt} and the freeze-in, or Akhmedov--Rubakov--Smirnov (ARS) leptogenesis~\cite{Akhmedov:1998qx}. For more discussion see \eg~\cite{Davidson:2008bu,Blanchet:2012bk}. Our methods apply, with minor modifications, to all these variants, but we will focus to the resonant scenario in the minimal model with two heavy Majorana neutrinos coupled to a single light SM lepton doublet and a Higgs doublet. Generalisation to more neutrino flavours, or more SM lepton flavours necessary \eg~for low scale leptogenesis~\cite{Klaric:2020lov,Granelli:2020ysj} and the ARS-mechanism, would be straightforward. We will also only include the decay and inverse decay interactions, neglecting the $2 \to 2$ scattering processes. This limits the range of validity of our predictions, but our goal is not the maximal phenomenological reach, but the accuracy of the quantum transport formulation and detailed comparisons between different approximations. Again, generalisation to more complex interactions would be straightforward.

In principle all particle species involved in the problem could be treated on the same footing in the CTP-context, resulting in a network of local transport equations. However, a number of simplifying approximations can be made for the SM fields. For example we can neglect the decay widths of the lepton and Higgs fields and assume that they are in kinetic equilibrium due to the SM gauge interactions. To first order, we can also neglect the chemical potential of the Higgs field, which then decouples from the dynamics. The lepton chemical potential $\mu_\ell$ is essential of course, but we can assume $\mu_\ell/T$ to be small, which allows us to neglect the backreaction of $\mu_\ell$ to the dynamics of the Majorana neutrinos. We can then solve the coupled neutrino and lepton equations consecutively instead of simultaneously. For neutrinos we will use the local transport equation~\cref{eq:closed-local-eom} with full phase space structure and flavour coherence information. Since we are only accounting for one-loop self-energies, we include only the indirect or $\epsilon$-type self-energy contribution to the CP-violation. Including the direct $\epsilon'$-type contribution would require a two-loop self-energy calculation, but we refrain from doing this, because the indirect contribution is dominant in the resonant regime.

In summary, the objective of this section is to derive an explicit local quantum transport equation for the mixing Majorana neutrinos, coupled with an equation for the particle-antiparticle asymmetry of the SM lepton doublet. We will compute explicitly all self-energy functions and the adiabatic background solutions for the neutrinos, as well as the source and washout terms for the lepton asymmetry equation. We will work consistently to the leading order in the coupling constant expansion and discuss the renormalisation procedure necessary for the loop calculations.

%%%%%%%%%%%%%%%%%%%%%%%%%%%%%%%%%%%%%%%%%%%%%%%%%%%%%%%%%%%%%%%%%%%%%%%%%%%%%%%
%
\subsection{The minimal model}
%
%%%%%%%%%%%%%%%%%%%%%%%%%%%%%%%%%%%%%%%%%%%%%%%%%%%%%%%%%%%%%%%%%%%%%%%%%%%%%%%

Our model contains two singlet Majorana neutrino fields $N_i$, coupled to an active lepton $\SU(2)$-doublet $\ell$ and the complex scalar Higgs doublet $\phi$ via chiral Yukawa interactions:
\begin{equation}
    \mathcal{L} =
    \sum_{i}\Bigl[
        \frac{1}{2} \bar N_i (\im\slashed\partial - m_i) N_i
        - y_i^* (\bar\ell \widetilde\phi) P_{\rm R} N_i
        - y_i \bar N_i P_{\rm L} (\widetilde\phi^\dagger \ell)
    \Bigr]
    + \bar\ell \,\im \slashed\partial \,\ell
    + (\partial_\mu\phi^\dagger)(\partial^\mu\phi)
    \text{.} \label{eq:lagrangian}
\end{equation}
We work in the mass basis for the Majorana neutrinos, where $m_i$ are the lepton number violating real Majorana masses. The lepton and Higgs fields are assumed to be massless as leptogenesis must take place in a high temperature in the electroweak symmetric phase. Finally, the $\SU(2)$-conjugate Higgs doublet is defined by $\widetilde{\phi} \equiv \epsilon \phi^*$ where $\epsilon$ is the anti-symmetric $2 \times 2$ matrix with $\epsilon_{12} = 1$. The CP-violating phases necessary for leptogenesis are contained in the complex coupling constants $y_i$.

The CTP propagators of the neutrino, lepton and Higgs fields are given by
\begin{subequations}
    \begin{align}
        \im S_{ij}(x,y) &\equiv \expectval[\big]{
            \mathcal{T}_\mathcal{C}\bigl[N_i(x) \bar N_j(y)\bigr]
        } \text{,}
        \\*
        \im S_{\ell,AB}(x,y) &\equiv \expectval[\big]{
            \mathcal{T}_\mathcal{C}\bigl[\ell_A(x) \bar {\ell_B}(y)\bigr]
        } \text{,} \label{eq:lepton_CTP_propagator}
        \\*
        \im \Delta_{AB}(x,y) &\equiv \expectval[\big]{
            \mathcal{T}_\mathcal{C}\bigl[\phi_A^{}(x) \phi_B^\dagger(y)\bigr]
        } \text{,}
    \end{align}
\end{subequations}
respectively, with the various real-time propagators defined as in section~\cref{sec:kb-equations}.%
\footnote{The real-time components of the complex scalar propagator are defined similarly to equations~\cref{eq:wightman-functions-def,eq:ret-adv-propagators-def}: for example the Wightman functions are $\Delta^\lt(u,v) \equiv \expectval[\big]{\phi^\dagger(v) \phi(u)}$ and $\Delta^\gt(u,v) \equiv \expectval[\big]{\phi(u) \phi^\dagger(v)}$. The only difference to the fermionic case is that the bosonic spectral function is a commutator of the fields, rather than an anti-commutator, and we use the standard sign convention $\Delta^\lt = \Delta^{+-}$ for bosons.}
Here we have explicitly marked the $\SU(2)$-doublet indices ($A,B = 1,2$) of the lepton and Higgs propagators and the Majorana flavour indices ($i,j$) of the neutrino propagator. Because of the $\SU(2)$ symmetry the lepton and Higgs propagators are diagonal in the $\SU(2)$-indices: $S_{\ell,AB} \equiv S_\ell \,\delta_{AB}$ and $\Delta_{AB} \equiv \Delta \,\delta_{AB}$. In the following we work directly with the diagonal elements $S_\ell$ and $\Delta$.
On the other hand, the Majorana fields $N_i$ satisfy the Majorana condition
\begin{equation}
    N_i = N_i^c \equiv C \bar N_i^\transp
    \text{,}\label{eq:majorana-condition}
\end{equation}
where $C$ is the unitary and antisymmetric charge conjugation matrix. An important consequence of the condition~\cref{eq:majorana-condition} is that the neutrino propagator is constrained by
\begin{equation}
    S(x,y) = C S(y,x)^\transp C^{-1}
    \text{.} \label{eq:majorana-condition-propagator}
\end{equation}
Note that here the transpose acts on all of the flavour, Dirac and CTP indices of the propagator.

%%%%%%%%%%%%%%%%%%%%%%%%%%%%%%%%%%%%%%%%%%%%%%%%%%%%%%%%%%%%%%%%%%%%%%%%%%%%%%%
%
\subsection{Self-energy functions}
%
%%%%%%%%%%%%%%%%%%%%%%%%%%%%%%%%%%%%%%%%%%%%%%%%%%%%%%%%%%%%%%%%%%%%%%%%%%%%%%%

\begin{figure}[t!]
    \centering
    \begin{subfigure}[c]{0.35\linewidth}
        \centering
        \includegraphics{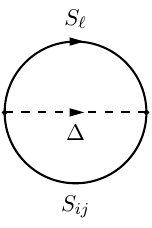}
        \caption{}\label{fig:2PI-Gamma2}
    \end{subfigure}%
    ~%
    \begin{subfigure}[c]{0.55\linewidth}
        \centering
        \includegraphics{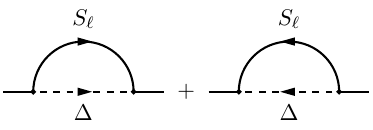}
        \caption{Neutrino self-energy}\label{fig:neutrino-self-energy}
        \vspace{1.2em}
        \includegraphics{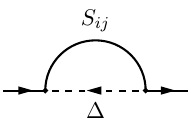}
        \caption{Lepton self-energy}\label{fig:lepton-self-energy}
    \end{subfigure}
    \vspace{3pt}
    \caption{Shown on the left (\subref{fig:2PI-Gamma2}) is the 2PI two-loop vacuum diagram in theory~\cref{eq:lagrangian}. The corresponding 1PI diagrams contributing to the Majorana neutrino (\subref{fig:neutrino-self-energy}) and lepton (\subref{fig:lepton-self-energy}) self-energies are shown on the right.}
    \label{fig:2PI}
\end{figure}

We calculate the self-energies in the 2PIEA formalism, where the self-energies are defined as functional derivatives of the non-trivial part $\Gamma_2$ of the effective action with respect to the propagators. The non-trivial part $\Gamma_2$ contains contributions of vacuum diagrams with two or more loops. The two-loop contribution (see figure~\cref{fig:2PI-Gamma2}) arising from the Yukawa interaction in the Lagrangian~\cref{eq:lagrangian} is given by
\begin{equation}
    \im \Gamma_2^{(2)} =
    \cw \sum_{i,j} y_i^* y_j \iint_{\mathcal{C}} \dd^4 x\dd^4 y
    \tr\bigl[
        P_{\rm R} \im S_{ij}(x,y) P_{\rm L} \im S_\ell(y,x)
    \bigr] \im \Delta(y,x)
    \text{.} \label{eq:two-loop-2piea}
\end{equation}
Here $\cw = 2$ is a multiplicity factor from the trace over the $\SU(2)$-doublet indices, and $\mathcal{C}$ denotes integration over the CTP.

%------------------------------------------------------------------------------
%
%
%------------------------------------------------------------------------------

To calculate the Majorana neutrino self-energy we need to take the functional derivative of $\Gamma_2$ with respect to the neutrino propagator $S$. As the propagator is constrained by the Majorana condition~\cref{eq:majorana-condition-propagator}, its functional derivative is defined by
\begin{alignat}{4}
    \frac{\delta S_{ij,\alpha\beta}(x,y)}{\delta S_{kl,\gamma\delta}(w,z)}
    = {}& \delta_{ik} \delta_{jl} && \delta_{\alpha\gamma} && \delta_{\beta\delta}
        && \delta^{(4)}(x-w) \delta^{(4)}(y-z) \notag
    \\*
    {}+{}& \delta_{il} \delta_{jk} && C_{\alpha\delta} && C^*_{\beta\gamma}
        && \delta^{(4)}(x-z) \delta^{(4)}(y-w) \text{,}
    \label{eq:majorana-functional-derivative}
\end{alignat}
where the flavour and Dirac indices have been indicated explicitly. For an unconstrained Dirac particle the second term on the right-hand side of equation~\cref{eq:majorana-functional-derivative} would not be present.

The one-loop Majorana neutrino self-energy $\Sigma_{ij}$ (figure~\cref{fig:neutrino-self-energy}), calculated using equations~\cref{eq:two-loop-2piea,eq:majorana-functional-derivative}, is then given in direct space by
\begin{align}
    \im \Sigma_{ij}(x,y) \equiv \frac{\delta\Gamma_2}{\delta S_{ji}(y,x)}
    = \cw \Bigl[
            & y_i y_j^* P_{\rm L} \im S_\ell(x,y) P_{\rm R} \im \Delta(x,y) \notag
            \\*
            {} + {} & y_i^* y_j P_{\rm R} C \im S_\ell(y,x)^\transp C^{-1} P_{\rm L} \im \Delta(y,x)
        \Bigr] \text{.}
    \label{eq:neutrino-self-energy}
\end{align}
The lepton self-energy (figure~\cref{fig:lepton-self-energy}) is calculated similarly, and the result for the $\SU(2)$-diagonal element is
\begin{equation}
    \im \Sigma_\ell(x,y)
    = \sum_{i,j} y_i^*y_j P_{\rm R} \im S_{ij}(x,y)P_{\rm L} \im \Delta(y,x)
    \text{.} \label{eq:lepton-self-energy}
\end{equation}
From these it is straightforward to calculate the real-time self-energies by inserting the CTP indices (which follow the spacetime arguments) and using the relations of the different propagators. Later we need the following neutrino self-energy functions (given now in the Wigner representation):
\begin{align}
    \im \Sigma_{ij}^{\lt,\gt}(k,t)
    = \cw \int \frac{\dd^4 p}{(2\pi)^4} \Bigl[
        & y_i y_j^* P_{\rm L} \im S_\ell^{\lt,\gt}(p,t) P_{\rm R}
            \im\Delta^{\lt,\gt}(k - p, t) \notag
            \\*
        {}-{} & y_i^* y_j P_{\rm R} C \im S_\ell^{\gt,\lt}(-p,t)^\transp C^{-1} P_{\rm L}
            \im\Delta^{\gt,\lt}(p - k, t)
    \Bigr]
    \text{,} \label{eq:neutrino-self-energy-wightman} \displaybreak[0]
    \\
    \Sigma_{ij}^{\rm H}(k,t)
    = \cw \int \frac{\dd^4 p}{(2\pi)^4} \Bigl[
        & y_i y_j^* P_{\rm L} S_\ell^{\rm H}(p,t) P_{\rm R}
            \im \Delta^{\rm F}(k-p,t) \notag
            \\*
        {}+{} & y_i y_j^* P_{\rm L} \im S_\ell^{\rm F}(p,t) P_{\rm R}
            \Delta^{\rm H}(k-p,t) \phantom{\Bigr]} \notag
            \\*
        {}+{} & y_i^* y_j P_{\rm R} C S_\ell^{\rm H}(-p,t)^\transp C^{-1} P_{\rm L}
            \im \Delta^{\rm F}(p-k,t) \phantom{\Bigr]} \notag
            \\*
        {}+{} & y_i^* y_j P_{\rm R} C \im S_\ell^{\rm F}(-p,t)^\transp C^{-1} P_{\rm L}
            \Delta^{\rm H}(p-k,t)
    \Bigr]
    \text{,} \label{eq:neutrino-self-energy-Hermitian}
\end{align}
where we defined the statistical propagators $S^{\rm F} \equiv \shalf(S^\gt - S^\lt)$ and $\Delta^{\rm F} \equiv \shalf(\Delta^\gt + \Delta^\lt)$.

%%%%%%%%%%%%%%%%%%%%%%%%%%%%%%%%%%%%%%%%%%%%%%%%%%%%%%%%%%%%%%%%%%%%%%%%%%%%%%%
%
\subsection{Tree level propagators}
\label{sec:tree_level_propagators}
%
%%%%%%%%%%%%%%%%%%%%%%%%%%%%%%%%%%%%%%%%%%%%%%%%%%%%%%%%%%%%%%%%%%%%%%%%%%%%%%%

In order to calculate the self-energies further we now introduce the tree level equilibrium approximations for the lepton and Higgs propagators $S_\ell$ and $\Delta$. We assume that the diagonal elements of the $\SU(2)$-symmetric lepton and Higgs correlators are given, in the Wigner representation, by
\begin{subequations}
    \label{eq:lepton-Higgs-tree-level-propagators}
    \begin{alignat}{2}
        S_{\ell,{\rm eq}}^\mathcal{A}(k) &=
        \pi \sgn(k^0) P_{\rm L} \slashed k \,\delta(k^2) \text{,} \hspace{2em} &
        \Delta_{\rm eq}^\mathcal{A}(k) &=
        \pi \sgn(k^0) \,\delta(k^2) \text{,}
        \\*
        S_{\ell,{\rm eq}}^{\rm H}(k) &=
        P_{\rm L} \slashed k \,\PV\bigl({\sfrac{1}{k^2}}\bigr) \text{,} \hspace{2em} &
        \Delta_{\rm eq}^{\rm H}(k) &=
        \PV\bigl({\sfrac{1}{k^2}}\bigr) \text{,}
        \\*
        \im S_\ell^{\lessgtr}(k,t) &=
        2 S_{\ell,{\rm eq}}^\mathcal{A}(k) f_{\rm FD}\bigl(\pm(k^0 - \mu_\ell)\bigr) \text{,} \hspace{2em} &
        \im \Delta_{\rm eq}^{\lessgtr}(k) &=
        \pm 2 \Delta_{\rm eq}^\mathcal{A}(k) f_{\rm BE}(\pm k^0) \text{,}
        \label{eq:lepton-Higgs-tree-level-Wightman}
    \end{alignat}
\end{subequations}
where $\PV$ denotes the Cauchy principal value and $\mu_\ell$ is the lepton chemical potential. The $\pm$ sign in equations~\cref{eq:lepton-Higgs-tree-level-Wightman} corresponds to the $\smash{\lessgtr}$ sign ($+$ for $\smallless$ and $-$ for $\smallgreater$). The Fermi--Dirac and Bose--Einstein phase space distribution functions are
\begin{equation}
    \label{eq:FD-and-BE-distributions}
    f_{\rm FD}(k^0) = \frac{1}{\e^{\beta k^0} + 1}
    \text{,} \hspace{3em}
    f_{\rm BE}(k^0) = \frac{1}{\e^{\beta k^0} - 1}
    \text{,}
\end{equation}
where $\beta = 1/T$ and we assume a common temperature $T$ for both $\ell$ and $\phi$. Note that the time-dependence of the lepton Wightman functions $S_{\ell}^{\lt,\gt}(k,t)$ in~\cref{eq:lepton-Higgs-tree-level-Wightman} comes solely from the chemical potential $\mu_\ell = \mu_\ell(t)$.

Similarly to the Majorana neutrino correlation function, it is convenient to split the lepton Wightman functions into the equilibrium and non-equilibrium parts,
\begin{equation}
    S^{\lt,\gt}_{\ell}(k,t) \equiv S^{\lt,\gt}_{\ell,{\rm eq}}(k) + \delta S^{\lt,\gt}_{\ell}(k,t)
    \text{,} \label{eq:lepton-Wightman-split}
\end{equation}
where $S^{\lt,\gt}_{\ell,{\rm eq}}(k)$ is given by equation~\cref{eq:lepton-Higgs-tree-level-Wightman} with $\mu_{\ell} = 0$ and the non-equilibrium parts satisfy $\delta S^\gt_{\ell} = -\delta S^\lt_{\ell}$. It then suffices to consider \eg~$\delta S^\lt_{\ell}$ only, for which we define the non-equilibrium lepton distribution
\begin{equation}
    \delta f^\lt_\ell(k^0) \, \equiv \,f_{\rm FD}(k^0 - \mu_\ell) - f_{\rm FD}(k^0)
    \, \simeq \, -f_{\rm FD}'(k^0) \mu_\ell
    \text{.}\label{eq:lepton-distribution-linearisation}
\end{equation}
Here we also assumed that the lepton chemical potential $\mu_\ell$ remains small during leptogenesis. We can now split the Majorana neutrino self-energies~\cref{eq:neutrino-self-energy-wightman,eq:neutrino-self-energy-Hermitian} using equation~\cref{eq:lepton-Wightman-split}:
\begin{equation}
    \Sigma(k,t) = \Sigma_{\rm eq}(k) + \delta\Sigma(k,t)
    \text{,} \label{eq:neutrino-self-energy-split}
\end{equation}
where $\Sigma_{\rm eq}$ is the thermal equilibrium part with vanishing $\mu_\ell$, and $\delta\Sigma$ is proportional to $\delta S_{\ell}$ and hence linear in $\mu_\ell$ in the approximation~\cref{eq:lepton-distribution-linearisation}.

Using equations~\cref{eq:neutrino-self-energy-wightman,eq:neutrino-self-energy-Hermitian,eq:lepton-Higgs-tree-level-propagators,eq:neutrino-self-energy-split,eq:lepton-distribution-linearisation,eq:FD-and-BE-distributions}, together with $\Sigma^\mathcal{A} \equiv \frac{\im}{2}(\Sigma^\gt + \Sigma^\lt)$, we can calculate all the needed Majorana neutrino self-energies. We parametrise them as
\begin{subequations}
\label{eq:neutrino-self-energy-parametrisations}
\begin{align}
    \Sigma^\mathcal{A}_{{\rm eq},ij}(k) &=
    \cw \bigl(y_i y_j^* P_{\rm L} + y_i^* y_j P_{\rm R}\bigr)
    \slashed{\mathfrak{S}}^{\mathcal{A}}_{{\rm eq},\mu}(k)
    \text{,} \label{eq:neutrino-self-energy-absorptive-equilibrium}
    \\*
    \Sigma^{{\rm H}(T)}_{{\rm eq},ij}(k) &=
    \cw \bigl(y_i y_j^* P_{\rm L} + y_i^* y_j P_{\rm R}\bigr)
    \slashed{\mathfrak{S}}^{{\rm H}(T)}_{{\rm eq},\mu}(k)
    \text{,} \label{eq:neutrino-self-energy-dispersive-equilibrium}
    \\*
    \im \delta\Sigma^{\lt,\gt}_{ij}(k,t) &=
    \cw \bigl(y_i y_j^* P_{\rm L} - y_i^* y_j P_{\rm R}\bigr)
    \delta\slashed{\mathfrak{S}}^{\lt,\gt}_{\mu}(k) \,\beta \mu_\ell
    \text{,} \label{eq:neutrino-self-energy-wightman-delta}
\end{align}
\end{subequations}
where the various $\mathfrak{S}_\mu$ functions are calculated explicitly in appendix~\cref{sec:neutrino-self-energies}. Due to homogeneity and isotropy there are only two independent functions $\mathfrak{S}_0(k) \equiv a(k^0,\abs{\evec{k}})$ and $\mathfrak{S}_i(k) \equiv b(k^0,\abs{\evec{k}}) \hat{k}^i$ (for all $i = 1,2,3$) for each $\Sigma$. Note that we give only the temperature dependent part $\Sigma^{{\rm H}(T)}_{\rm eq}$ of the dispersive self-energy $\Sigma^{\rm H}_{{\rm eq}}(k)$ in equation~\cref{eq:neutrino-self-energy-dispersive-equilibrium}. The vacuum part $\Sigma^{{\rm H}({\rm vac})}_{{\rm eq}}$ and its on-shell renormalisation are presented in section~\cref{sec:neutrino-renormalisation}. The renormalisation of the Majorana neutrinos is not trivial because they mix and are unstable. Nevertheless, the relevant outcome for this paper is simple: the renormalised (vacuum part) $\widehat\Sigma^{{\rm H(vac)}}_{{\rm eq}}$ does not contribute to our results in the leading order approximation considered in this work.

%%%%%%%%%%%%%%%%%%%%%%%%%%%%%%%%%%%%%%%%%%%%%%%%%%%%%%%%%%%%%%%%%%%%%%%%%%%%%%%
%
\subsection{Adiabatic neutrino solutions}
%
%%%%%%%%%%%%%%%%%%%%%%%%%%%%%%%%%%%%%%%%%%%%%%%%%%%%%%%%%%%%%%%%%%%%%%%%%%%%%%%

Now that we have specified the self-energy, we can calculate the adiabatic solutions~\cref{eq:adiabatic-solutions-def} for the neutrino which are needed in the kinetic equation~\cref{eq:closed-local-eom}. Working to the lowest order approximation we only take the equilibrium part of the neutrino self-energy~\cref{eq:neutrino-self-energy-split} into account:
\begin{subequations}
    \begin{align}
        S^p_{\rm ad}(k,t) &= \bigl[\slashed k - m(t) - \Sigma^p_{\rm eq}(k)\bigr]^{-1}
        \text{,} \label{eq:adiabatic-pole-eq} \\*
        S^s_{\rm ad}(k,t) &= S^r_{\rm ad}(k,t) \Sigma^s_{\rm eq}(k) S^a_{\rm ad}(k,t)
        \text{.} \label{eq:adiabatic-Wightman-eq}
    \end{align}
\end{subequations}
We are implicitly assuming that all quantities have been renormalised, so the pole self-energy $\Sigma^p_{\rm eq}$ appearing here is actually the on-shell renormalised pole self-energy $\widehat\Sigma^p_{\rm eq}$ given by equations~\cref{eq:renormalised-general-Sigma-p,eq:renormalised-general-Sigma-p-split}. Furthermore, using the KMS relation $\Sigma^\gt_{\rm eq}(k) = \e^{\beta k^0} \Sigma^\lt_{\rm eq}(k)$ and the exact identity $\mathcal{A} = S^r \Sigma^\mathcal{A }S^a$, which holds in direct space as a convolution and as a simple product to zeroth order in gradients in the Wigner representation, we can write the Wightman functions~\cref{eq:adiabatic-Wightman-eq} as
\begin{subequations}
    \label{eq:adiabatic-wightman-resummed}
    \begin{align}
        \im S^\lt_{\rm ad}(k,t) &= 2\mathcal{A}_{\rm ad}(k,t) f_{\rm FD}(k^0)
        \text{,} \label{eq:adiabatic-wightman-1} \\*
        \im S^\gt_{\rm ad}(k,t) &= 2\mathcal{A}_{\rm ad}(k,t) \bigl(1 - f_{\rm FD}(k^0)\bigr)
        \text{.} \label{eq:adiabatic-wightman-2}
    \end{align}
\end{subequations}
In perturbative expansions it is more convenient to use the form~\cref{eq:adiabatic-wightman-resummed} than equation~\cref{eq:adiabatic-Wightman-eq}. For example, a naive coupling constant expansion of the pole propagators in equation~\cref{eq:adiabatic-Wightman-eq} would appear to give a $\mathcal{O}(y_i^2)$ result, whereas the right-hand side of equation~\cref{eq:adiabatic-wightman-resummed} obviously gives the correct $\mathcal{O}(1)$ result. In other words, a consistent expansion of the Wightman function~\cref{eq:adiabatic-Wightman-eq} requires resumming the pole propagators, which the identity $\mathcal{A} = S^r \Sigma^\mathcal{A }S^a$ implements automatically in equations~\cref{eq:adiabatic-wightman-resummed}.

Given the adiabatic solution, we can now compute the source term $-\partial_t \bar S^\lt_{{\rm ad},\evec{k}}(t,t)$ in equation~\cref{eq:closed-local-eom}. To this end, we employ the inverse Wigner transform
\begin{equation}
    F_\evec{k}(t_1,t_2) = \int \frac{\dd k^0}{2\pi}
    \e^{-\im k^0(t_1 - t_2)}F\Bigl(k,\frac{t_1 + t_2}{2}\Bigr)
    \label{eq:inverse-wigner-transform}
\end{equation}
to calculate the two-time representation function $S^\lt_{{\rm ad},\evec{k}}(t_1,t_2)$ from the Wigner representation~\cref{eq:adiabatic-wightman-1}. Utilising Cauchy's integral theorem to perform the $k^0$-integral, we can write the two-time representation as
\begin{align}
    \bar S^\lt_{{\rm ad},\evec{k}}(t_1,t_2)
    &= \int \frac{\dd k^0}{2\pi} \e^{-\im k^0(t_1-t_2)}
        \im (\bar S^r_{\rm ad} - \bar S^a_{\rm ad})(k,t) f_{\rm FD}(k^0) \notag
    \\*
    &\simeq \sum_{p=r,a} \theta(-s_p r^0) \sum_{k^0}^{\rm poles}
        \e^{-\im k^0 r^0} f_{\rm FD}(k^0) \Res_{k^0}\bigl[\bar S^p_{\rm ad}(k,t)\bigr] \text{,}
    \label{eq:two-time-Sad}
\end{align}
where the $k^0$-sum is taken over the poles of $S^p_{\rm ad}(k,t)$, and $s_p = -1,1$ for $p = r,a$, respectively. We also used the short-hands $r^0 \equiv t_1 - t_2$ and $t \equiv (t_1 + t_2)/2$. The corresponding result for $\bar S^\gt_{{\rm ad},\evec{k}}$ follows by replacing $f_{\rm FD}(k^0)$ above with $f_{\rm FD}(-k^0) = 1 - f_{\rm FD}(k^0)$. Note that we neglected any branch cuts in equation~\cref{eq:two-time-Sad} and kept only the residue contributions of the single particle poles, but otherwise the result is general. Also, the poles of $f_{\rm FD}(k^0)$ at the imaginary axis do not contribute due to the KMS relation.

%%%%%%%%%%%%%%%%%%%%%%%%%%%%%%%%%%%%%%%%%%%%%%%%%%%%%%%%%%%%%%%%%%%%%%%%%%%%%%%
\paragraph{Leading order approximation.}
%%%%%%%%%%%%%%%%%%%%%%%%%%%%%%%%%%%%%%%%%%%%%%%%%%%%%%%%%%%%%%%%%%%%%%%%%%%%%%%

Equation~\cref{eq:two-time-Sad} is written with a general adiabatic pole function, but it will be eventually sufficient to compute it to lowest order in gradients and in the coupling constants. In this case we can use the free tree-level solution with vacuum dispersion relations for the adiabatic pole propagators $S^p_{\rm ad}(k,t)$ in equation~\cref{eq:adiabatic-pole-eq}, given by
\begin{equation}
    S^{p(0)}_{{\rm ad},ij}(k,t) =
    \frac{\slashed k + m_i(t)}{k^2 - m_i^2(t) - \im s_p \sgn(k^0) \epsilon} \delta_{ij}
    \text{.}
\end{equation}
The small imaginary part with $\epsilon > 0$ ensures the correct boundary conditions for $p = r,a$. Inserting this solution into equation~\cref{eq:two-time-Sad}, calculating the residues and taking the limit $\epsilon \to 0^{+}$ yields the leading order result
\begin{equation}
    \bar S^{\lt(0)}_{{\rm ad},\evec{k}ij}(t_1,t_2) =
    \sum_{s = \pm} \frac{\e^{-\im s (t_1 - t_2) \omega_{\evec{k}i}}}{2s\omega_{\evec{k}i}}
    f_{\rm FD}(s\omega_{\evec{k}i}) (\slashed k_{si} + m_i) \gamma^0 \delta_{ij}
    \text{.} \label{eq:adiabatic-wightman-approx}
\end{equation}
Here $\slashed k_{si} \equiv s\omega_{\evec{k}i} \gamma^0 - \evec{\gamma} \cdot \evec{k}$ and $\omega_i \equiv \sqrt{\abs{\evec{k}}^2 + {m_i}^2}$, and the time-dependent masses $m_i(t)$ are evaluated at the average time $t = (t_1 + t_2)/2$.%
\footnote{The \emph{leading order} result~\cref{eq:adiabatic-wightman-approx} can be obtained also by substituting the tree-level spectral function~\cref{eq:vacuumA} directly into the first line of equation~\cref{eq:two-time-Sad} and performing the integration with the delta function, without resorting to the residue formula. Including a finite width requires using the more general formula however.}
Note that the lowest order adiabatic pole and Wightman functions are diagonal in the mass basis in agreement with~\cite{Iso:2013lba}. We emphasize that in our approach the off-diagonal corrections to the adiabatic propagators are not required to leading order, as we do not need to calculate the homogeneous solution non-dynamically from equation~\cref{eq:local-wightman-homog-tin}, as was done in \eg~\cite{Garny:2011hg}. Instead, corrections from the full adiabatic pole propagators are already taken into account in the source term of our dynamical equation~\cref{eq:closed-local-eom} (see also equation~\cref{eq:general-local-eom-for-perturbation}). More discussion about these corrections is given at the end of section~\cref{sec:effective_neutrino_self_energy}.

Setting $t_1 = t_2 \equiv t$ in equation~\cref{eq:adiabatic-wightman-approx} we then get the equal-time adiabatic function $\bar S^\lt_{{\rm ad},\evec{k}}(t,t)$ needed in the kinetic equation~\cref{eq:closed-local-eom}. The result may also be cast into a compact form
\begin{equation}
    \bar S^{\lt(0)}_{{\rm ad},\evec{k}ij}(t,t) =
    \sum_{s = \pm} f_{\rm FD}(s\omega_{\evec{k}i}) P^s_{\evec{k}i} \delta_{ij}
    \text{,} \label{eq:adiabatic-local-wightman}
\end{equation}
where the energy projection matrix $P_{\evec{k}i}^s$ was defined in equation~\cref{eq:projector-def}. To calculate the time-derivative in order to get the source term $-\partial_t \bar S^\lt_{{\rm ad},\evec{k}ij}(t,t)$ for equation~\cref{eq:closed-local-eom} is now a simple matter, once the time-dependent masses $m_i(t)$ are specified. Equation~\cref{eq:adiabatic-local-wightman} is also needed to calculate the initial values for the (local) non-equilibrium Wightman function $\delta S^\lt \equiv S^\lt - S^\lt_{\rm ad}$ once the initial value for the full function $S^\lt$ has been specified.

%%%%%%%%%%%%%%%%%%%%%%%%%%%%%%%%%%%%%%%%%%%%%%%%%%%%%%%%%%%%%%%%%%%%%%%%%%%%%%%
%
\subsection{Effective neutrino self-energy}
\label{sec:effective_neutrino_self_energy}
%
%%%%%%%%%%%%%%%%%%%%%%%%%%%%%%%%%%%%%%%%%%%%%%%%%%%%%%%%%%%%%%%%%%%%%%%%%%%%%%%

We now calculate the effective self-energy~\cref{eq:effective-sigma-ad}, which appears in the local kinetic equation~\cref{eq:closed-local-eom} for Majorana neutrinos. We use the general result for a convolution in terms of the Wigner representation correlation functions:
\begin{align}
    (F_\evec{k} \convol G_\evec{k})(t_1, t_2)
    &= \int \frac{\dd k^0}{2\pi} \e^{-\im k^0(t_1 - t_2)} \e^{
        -\ihalf(\partial_t^F \partial_{k^0}^G - \partial_{k^0}^F \partial_t^G)
    } F(k,t) G(k,t) \Bigr|_{t = (t_1 + t_2)/2} \notag
    \\*
    % &= \int \frac{\dd k^0}{2\pi} \e^{-\im k^0(t_1 - t_2)} \e^{
    %     -\ihalf(\partial_t^F - \partial_t^G) \partial_{k^0}
    % } F_{\rm out}(k,t) G_{\rm in}(k,t) \Bigr|_{t = (t_1 + t_2)/2} \notag
    % \\*
    &= \int \frac{\dd k^0}{2\pi} \e^{-\im k^0(t_1 - t_2)}
        F_{\rm out}(k,t) G_{\rm in}(k,t) \Bigr|_{t = (t_1 + t_2)/2}
    \text{,} \label{eq:local-convolution-in-wigner-rep}
\end{align}
where $F_{\rm out}(k,t)$ and $G_{\rm in}(k,t)$ were defined below equation~\cref{eq:HKR-ansatz-gen-wspace}. The first line of equation~\cref{eq:local-convolution-in-wigner-rep} is the inverse Wigner transform of the Moyal product (often denoted by $\Diamond$). When the gradients are small we may further approximate $F_{\rm out}(k,t) \simeq F(k,t)$ and $G_{\rm in}(k,t) \simeq G(k,t)$. We will again assume an adiabatic, equilibrium self-energy function and expand to zeroth order in gradients. The general result is then similar to equation~\cref{eq:two-time-Sad}:
\begin{align}
    \bar\Sigma^r_{{\rm eq,eff},\evec{k}}(t_1,t_2)
    &= \int \frac{\dd k^0}{2\pi} \e^{-\im k^0(t_1-t_2)}
        \bar\Sigma^r_{\rm eq}(k) \im (\bar S^r_{\rm ad} - \bar S^a_{\rm ad})(k,t)
    \\*
    &\simeq \sum_{p=r,a}\theta(-s_p r^0) \sum_{k^0}^{\rm poles}
        \e^{-\im k^0 r^0}\bar\Sigma^r_{\rm eq}(k)\Res_{k^0}\bigl[\bar S^p_{\rm ad}(k,t)\bigr] \text{.}
    \label{eq:two-time-Sigma-eff}
\end{align}
On the last line we again kept only the residue contribution from the poles of the propagator and neglected contributions from possible branch cuts. To calculate the leading tree level approximation we will again use the free pole propagators like in equations~\cref{eq:adiabatic-wightman-approx,eq:adiabatic-local-wightman}. The calculation proceeds as before and one finds the lowest order coupling constant result
\begin{equation}
    \bar\Sigma^{r(0)}_{{\rm eq,eff},\evec{k}ij}(t,t) =
    \sum_{s = \pm} \bar\Sigma^r_{{\rm eq},ij}(k_{sj}) P_{\evec{k}j}^s
    \text{.} \label{eq:local-Sigma-eff-LO}
\end{equation}
Corrections resulting from resummed pole propagators could be included by using the residue formula~\cref{eq:two-time-Sigma-eff} with complex poles, but as will be argued below, their effect would be formally of higher order ($\sim \mathcal{O}(y_i^4)$) in the coupling constants.

%%%%%%%%%%%%%%%%%%%%%%%%%%%%%%%%%%%%%%%%%%%%%%%%%%%%%%%%%%%%%%%%%%%%%%%%%%%%%%%
\paragraph{Beyond the leading approximation.}
%%%%%%%%%%%%%%%%%%%%%%%%%%%%%%%%%%%%%%%%%%%%%%%%%%%%%%%%%%%%%%%%%%%%%%%%%%%%%%%

To improve on the leading approximation in coupling constants used above, one should solve the complex $k^0$-poles from the determinant of $S^p_{\rm ad}(k,t)$, keeping the self-energy corrections and use them to calculate the residues in equations~\cref{eq:two-time-Sad,eq:two-time-Sigma-eff}. This is in principle straightforward, but laborious because the block-wise inversion of equation~\cref{eq:adiabatic-pole-eq} results in complicated formulae due to the flavour mixing. We will not implement these corrections in this paper, because we use the weak coupling approximation where these corrections should be small. We give details of the inversion procedure in appendix~\cref{sec:neutrino-pole-propagator-inversion} for completeness however, and note that a similar analysis was presented in~\cite{Garny:2011hg}, with a non-relativistic approximation for the self-energy. It should also be noted that these higher order corrections to~\cref{eq:adiabatic-local-wightman,eq:local-Sigma-eff-LO} remain parametrically small also in the maximally resonant regime where $\Delta m_{21}/m \sim \abs{y}^2$. We have verified this explicitly in a generic power counting expansion of the formula~\cref{eq:two-time-Sad} where $\Delta m_{21}/m$ and $\abs{y}^2$ are expanded simultaneously, under the assumption that the difference of the full $k^0$-poles to the corresponding free theory poles is $\mathcal{O}(y^2)$. To get the correct \emph{perturbative} result it is also crucial to start from the resummed form of equation~\cref{eq:adiabatic-wightman-1}.

The self-energy corrections to the pole propagators can be divided into dispersive corrections due to the Hermitian self-energy and dissipative ones due to the antihermitian self-energy part. The main effect of the latter was already discussed qualitatively in section~\cref{sec:homogeneous-solutions}. While the actual formulae are very complex, it is easy to see that the main qualitative effect of evaluating~\cref{eq:two-time-Sad} at a complex pole is to introduce complex damping terms $\sim \e^{-\gamma_{ii} \abs{t_1-t_2}}$ into equation~\cref{eq:adiabatic-wightman-approx}. Such factors represent the dissipation in the relative time coordinate, but eventually do not affect the source term in the local equation~\cref{eq:closed-local-eom}. Indeed, the only effect on the source, and likewise on the self-energy convolutions~\cref{eq:two-time-Sigma-eff} appearing in~\cref{eq:closed-local-eom}, from a finite width then amounts to shifts $\sim y_i^2$ to the energy shells where these terms are evaluated. In the weak coupling limit such corrections should be small, which we have verified numerically.

The dispersive corrections could be more interesting. Including Hermitian self-energy corrections would lead to new real-frequency poles for the adiabatic functions. Combining the vacuum Hamiltonian in equation~\cref{eq:closed-local-eom} with the effective Hermitian self-energy evaluated at these poles, would give rise to an effective matter Hamiltonian for the quasistates. The effective Hamiltonian would in general be no longer diagonal in the mass basis and the energy difference between the matter eigenstates would be a function of time, similar to the case of mixing light neutrinos in the early universe~\cite{Enqvist:1990ad,Kainulainen:1990ds,Enqvist:1991qj}. Such a dynamical modification of the energy level splitting could be relevant for resonant leptogenesis, although the analysis of ref.~\cite{Hohenegger:2014cpa} (performed in a simplified setup) does not suggest that the effect is quantitatively significant.

%%%%%%%%%%%%%%%%%%%%%%%%%%%%%%%%%%%%%%%%%%%%%%%%%%%%%%%%%%%%%%%%%%%%%%%%%%%%%%%
%
\subsection{Lepton transport equation}
%
%%%%%%%%%%%%%%%%%%%%%%%%%%%%%%%%%%%%%%%%%%%%%%%%%%%%%%%%%%%%%%%%%%%%%%%%%%%%%%%

The equation for the lepton asymmetry can be derived from the KB equations~\cref{eq:KB-two-time} for the lepton propagator. However, as the lepton is massless and we use the tree-level spectral approximation for its propagator, the lowest order adiabatic source term in equation~\cref{eq:general-local-eom-for-perturbation} vanishes. To derive the leading source for the lepton asymmetry, it is convenient to use a different but equivalent formulation of the KB equations (see \eg~equations~17 and~18 in~\cite{Herranen:2008hu}). The appropriate form of the equation where the finite width and dispersive contributions have been neglected is
\begin{equation}
    \bigl(
        S_{\ell,0,\evec{k}}^{-1} \convol S^{\lessgtr}_{\ell,\evec{k}}
    \bigr)(t_1,t_2)
    = \pm\frac{1}{2} \bigl(
        \Sigma^\gt_{\ell,\evec{k}} \convol S^\lt_{\ell,\evec{k}}
        -\Sigma^\lt_{\ell,\evec{k}} \convol S^\gt_{\ell,\evec{k}}
    \bigr)(t_1,t_2)
    \text{.} \label{eq:kb-two-time-lepton}
\end{equation}
The corresponding local equation for the local correlation function of the lepton can be derived in the same way that we derived equation~\cref{eq:general-local-eom}. The result is
\begin{align}
    \im\partial_t \bar S^\lt_{\ell,\evec{k}}(t,t) =
    \comm[\big]{\evec{\alpha}\cdot\evec{k}, \bar S^\lt_{\ell,\evec{k}}(t,t)}
    & -\frac{\im}{2} \bigl(
        \bar\Sigma^\gt_{\ell,\evec{k}} \convol \bar S^\lt_{\ell,\evec{k}}
        - \bar\Sigma^\lt_{\ell,\evec{k}} \convol \bar S^\gt_{\ell,\evec{k}}
    \bigr)(t,t) \notag
    \\*
    & -\frac{\im}{2} \bigl(
        \bar S^\lt_{\ell,\evec{k}} \convol \bar\Sigma^\gt_{\ell,\evec{k}}
        - \bar S^\gt_{\ell,\evec{k}} \convol \bar\Sigma^\lt_{\ell,\evec{k}}
    \bigr)(t,t)
    \text{,} \label{eq:local-eom-lepton}
\end{align}
where $\evec{\alpha}\cdot\evec{k}$ is the free Hamiltonian for the massless lepton. We also remind here that equations~\cref{eq:kb-two-time-lepton,eq:local-eom-lepton} are formulated for the diagonal element $S_{\ell}$ of the $\SU(2)$-symmetric lepton correlator~\cref{eq:lepton_CTP_propagator}. This equation could be solved as such, coupled with the local equation for the mixing Majorana states. However, in practice we can make several further approximations, eventually converging to a simple scalar equation for the lepton asymmetry.

%%%%%%%%%%%%%%%%%%%%%%%%%%%%%%%%%%%%%%%%%%%%%%%%%%%%%%%%%%%%%%%%%%%%%%%%%%%%%%%
\paragraph{Lepton asymmetry.}
%%%%%%%%%%%%%%%%%%%%%%%%%%%%%%%%%%%%%%%%%%%%%%%%%%%%%%%%%%%%%%%%%%%%%%%%%%%%%%%

The lepton asymmetry we are interested in this work can be related to the chiral current density of the left-handed leptons, which is defined by
\begin{equation}
    j^\mu_{\ell}(x) \equiv \expectval[\big]{\bar \ell_A(x) \gamma^\mu P_{\rm L} \ell_A(x)}
    \text{,} \label{eq:lepton-current-1}
\end{equation}
where an implicit summation over the $\SU(2)$-index is assumed. Since we consider a spatially homogeneous and isotropic system, the current depends only on the time $x^0 = t$ and it can be further related to the local two-time Wightman function. Using definitions~\cref{eq:lepton-current-1,eq:wightman-functions-def,eq:two-time-def}, we then get
\begin{equation}
    j^\mu_{\ell}(t) = \cw \int \frac{\dd^3\evec{k}}{(2\pi)^3}\tr\Bigl[
        \gamma^\mu P_{\rm L} \im S^\lt_{\ell,\evec{k}}(t,t)
    \Bigr] \text{.} \label{eq:lepton-current-2}
\end{equation}
Because no asymmetry can exist in thermal equilibrium~\cite{Kolb:1979qa}, we can define the lepton asymmetry density $n_\ell - {\bar n}_\ell$ as the zeroth component of the non-equilibrium part of the current:
\begin{equation}
    n_\ell - {\bar n}_\ell \equiv \delta j^0_{\ell}(t)
    = \cw \int \frac{\dd^3\evec{k}}{(2\pi)^3}\tr\Bigl[
        P_{\rm L} \delta \bar S^\lt_{\ell,\evec{k}}(t,t)
    \Bigr] \text{.} \label{eq:def-lepton-asymmetry}
\end{equation}
We can further relate the asymmetry to the chemical potential of the tree-level lepton propagator. Using equations~\cref{eq:lepton-Wightman-split,eq:lepton-Higgs-tree-level-Wightman,eq:FD-and-BE-distributions,eq:inverse-wigner-transform} we calculate the trace on the right-hand side of equation~\cref{eq:def-lepton-asymmetry}. The result, written in terms of lepton and anti-lepton phase space distributions, is
\begin{equation}
    n_\ell - {\bar n}_\ell =
    \cw \int \frac{\dd^3\evec{k}}{(2\pi)^3}
    \Bigl(f_{\rm FD}(\abs{\evec{k}} - \mu_\ell) - f_{\rm FD}(\abs{\evec{k}} + \mu_\ell)\Bigr)
    \text{,}
\end{equation}
where $\cw = 2$ is the weak isospin multiplicity factor. There is no additional spin multiplicity factor because of the chiral projection, \ie~only one spin state couples to the Majorana neutrino and develops an asymmetry in the massless limit. A standard calculation of the integral gives the relation between the asymmetry and the chemical potential:
\begin{equation}
    n_\ell - {\bar n}_\ell
    = \frac{\cw T^3}{6 \pi^2} \biggl[\pi^2 \Bigl(\frac{\mu_\ell}{T}\Bigr) + \Bigl(\frac{\mu_\ell}{T}\Bigr)^3 \biggr]
    \simeq \frac{\cw T^3}{6} \Bigl(\frac{\mu_\ell}{T}\Bigr)
    \text{,} \label{eq:lepton-asymmetry-vs-chemical-potential}
\end{equation}
where in the last step we used the linear approximation for $\mu_\ell$ like in equation~\cref{eq:lepton-distribution-linearisation}.

%%%%%%%%%%%%%%%%%%%%%%%%%%%%%%%%%%%%%%%%%%%%%%%%%%%%%%%%%%%%%%%%%%%%%%%%%%%%%%%
\paragraph{Lepton asymmetry equation.}
%%%%%%%%%%%%%%%%%%%%%%%%%%%%%%%%%%%%%%%%%%%%%%%%%%%%%%%%%%%%%%%%%%%%%%%%%%%%%%%

We can get the equation of motion for the lepton asymmetry~\cref{eq:def-lepton-asymmetry}
by using the split~\cref{eq:lepton-Wightman-split} in equation~\cref{eq:local-eom-lepton}, taking the trace over spinor indices and integrating over momentum. The trace of the commutator term vanishes due to the Dirac structure of the tree level propagator, so that we are left with
\begin{equation}
    \partial_t(n_\ell - {\bar n}_\ell) =
    -\frac{\cw}{2} \int \frac{\dd^3\evec{k}}{(2\pi)^3}
    \tr\Bigl[
        (\bar\Sigma^\gt_{\ell,\evec{k}} \convol \bar S^\lt_{\ell,\evec{k}})(t,t)
        - (\bar\Sigma^\lt_{\ell,\evec{k}} \convol \bar S^\gt_{\ell,\evec{k}})(t,t)
    \Bigr]
    + \text{H.c.} \label{eq:lepton-local-equation-traced}
\end{equation}
Substituting the two-time representation of the lepton self-energy~\cref{eq:lepton-self-energy} to the right-hand side then yields
\begin{align}
    \partial_t(n_\ell - {\bar n}_\ell) ={}&
    -\frac{\cw}{2} \sum_{i,j} y_i^* y_j
    \int \frac{\dd^3\evec{k}}{(2\pi)^3} \frac{\dd^3\evec{p}}{(2\pi)^3} \int \dd t' \notag
    \\*
    &{}\times
    \tr\Bigl[
        P_{\rm R} \im S^\gt_{\evec{p}ij}(t,t') P_{\rm L}
        \im \Delta^\lt_{{\rm eq},\evec{p}-\evec{k}}(t',t)
        \im S^\lt_{\ell,\evec{k}}(t',t)
        - \bigl({>} \leftrightarrow {<}\bigr)
    \Bigr]
    + \text{H.c.}
    \label{eq:lepton-asymmetry-1}
\end{align}
This equation can actually be expressed using the already calculated Majorana neutrino self-energy~\cref{eq:neutrino-self-energy-wightman}. Indeed, because of the trace and the equal time arguments it is just a matter of combining the terms differently in the right-hand side of equation~\cref{eq:lepton-asymmetry-1} to rewrite the integral in terms of the Majorana neutrino correlation function and the chirally projected Majorana self-energy $\Sigma_{\rm L}$, which results in:
\begin{equation}
    \partial_t(n_\ell - {\bar n}_\ell) =
    \frac{1}{2} \int \frac{\dd^3\evec{k}}{(2\pi)^3}
    \Tr\Bigl[
        (\bar\Sigma^\gt_{{\rm L},\evec{k}} \convol \bar S^\lt_\evec{k})(t,t)
        - (\bar\Sigma^\lt_{{\rm L},\evec{k}} \convol \bar S^\gt_\evec{k})(t,t)
    \Bigr]
    + \text{H.c.} \label{eq:lepton-asymmetry-2}
\end{equation}
One should note that the trace is now taken over both the Majorana neutrino flavours and the Dirac indices and we defined the barred chiral Majorana neutrino self-energy as $\bar\Sigma^{\lt,\gt}_{\rm L} \equiv \gamma^0 P_{\rm L} \im\Sigma^{\lt,\gt}$. Also note that the lepton doublet multiplicity $\cw$ is now included in the neutrino self-energy.

Results similar to~\cref{eq:lepton-asymmetry-2} are already known in the literature~\cite{Garny:2011hg}, but the novelty of our approach is in the use of the local approximation of section~\cref{sec:quantum-kinetic-equation} to evaluate the convolution integrals. This is now straightforward because equation~\cref{eq:lepton-asymmetry-2} is written explicitly in terms of the Majorana neutrino propagator. We first use equations~\cref{eq:adiabatic-expansion,eq:neutrino-self-energy-split} to split the neutrino Wightman functions and self-energies into the equilibrium and non-equilibrium parts. Then we invoke the local approximation~\cref{eq:local-approximation} to compute convolution integrals with the perturbations $\delta S^{\lt,\gt}$, along with the constraint $\delta S^\gt_\evec{k}(t,t) = -\delta S^\lt_\evec{k}(t,t)$, which is imposed by the sum rule. Finally, we write the convolution integrals involving $S^{\lt,\gt}_{\rm ad}$ in the Wigner representation using equation~\cref{eq:local-convolution-in-wigner-rep} and expand consistently to the leading order in coupling constants and gradients to obtain:
\begin{align}
    \partial_t(n_\ell - {\bar n}_\ell) \simeq
    \int \frac{\dd^3 \evec{k}}{(2\pi)^3}&
    \Tr\Bigl[
        \bar\Sigma^\mathcal{A}_{{\rm L,eq,eff},\evec{k}}(t,t)
        \delta\bar S^\lt_\evec{k}(t,t)
        + \delta\bar\Sigma^\mathcal{A}_{{\rm L,eff},\evec{k}}(t,t)
        \delta\bar S^\lt_\evec{k}(t,t) \notag
        \\*
        {}+{} &\frac{1}{2} \int \frac{\dd k^0}{2\pi}
        \Bigl(
            \delta\bar\Sigma^\gt_{\rm L}(k,t)\bar S^\lt_{\rm ad}(k,t)
            - \delta\bar\Sigma^\lt_{\rm L}(k,t)\bar S^\gt_{\rm ad}(k,t)
        \Bigr)
    \Bigr] + \text{H.c.} \label{eq:lepton-asymmetry-3}
\end{align}
The first term on the right-hand side of equation~\cref{eq:lepton-asymmetry-3}, proportional to $\Sigma^{\mathcal{A}}_{\rm eq}$, does not contain the lepton chemical potential $\mu_\ell$. It is therefore the source term for the lepton asymmetry. The remaining terms, proportional to $\delta\Sigma^\alpha$, are linear in $\mu_\ell \propto n_\ell - {\bar n}_\ell$ (in the approximation~\cref{eq:lepton-distribution-linearisation}) and so they contribute to the washout.

%%%%%%%%%%%%%%%%%%%%%%%%%%%%%%%%%%%%%%%%%%%%%%%%%%%%%%%%%%%%%%%%%%%%%%%%%%%%%%%
%
\subsection{General leptogenesis equations}
\label{sec:leptogenesis-final-equations}
%
%%%%%%%%%%%%%%%%%%%%%%%%%%%%%%%%%%%%%%%%%%%%%%%%%%%%%%%%%%%%%%%%%%%%%%%%%%%%%%%

To summarise, we use the local equation~\cref{eq:closed-local-eom} with equilibrium self-energies to solve the evolution of the Majorana neutrinos and equation~\cref{eq:lepton-asymmetry-3} to subsequently calculate the lepton asymmetry. Our coupled equations for leptogenesis therefore read
\begin{align}
    \partial_t \delta\bar S^\lt_\evec{k}(t,t)
    + \im \comm[\big]{H_\evec{k}(t), \delta\bar S^\lt_\evec{k}(t,t)}
    &= - \partial_t \bar S^\lt_{{\rm ad},\evec{k}}(t,t) - \bigl(
        \im \bar\Sigma^r_{{\rm eq,eff},\evec{k}}(t,t) \delta \bar S^\lt_\evec{k}(t,t) + \text{H.c.}
    \bigr)
    \text{,} \label{eq:neutrino-kinetic-final}
    \\*
    \partial_t(n_\ell - {\bar n}_\ell) &= S_{CP} + \delta W + W_{\rm ad}
    \text{,} \label{eq:lepton-asymmetry-final}
\end{align}
where the CP-violating source term $S_{CP}$ and the washout terms $\delta W$ and $W_{\rm ad}$ of the lepton equation are given by
\begin{align}
    S_{CP} &= \int \frac{\dd^3\evec{k}}{(2\pi)^3}
    \Tr\Bigl[
        \bar\Sigma^\mathcal{A}_{{\rm L,eq,eff},\evec{k}}(t,t)
        \delta\bar S^\lt_\evec{k}(t,t) + \text{H.c.}
    \Bigr]
    \text{,} \label{eq:cp-source}
    \\
    \delta W &= \int \frac{\dd^3\evec{k}}{(2\pi)^3}
    \Tr\Bigl[
        \delta\bar\Sigma^\mathcal{A}_{{\rm L,eff},\evec{k}}(t,t)
        \delta\bar S^\lt_\evec{k}(t,t) + \text{H.c.}
    \Bigr] \text{,} \label{eq:washout-term-A}
    \\
     W_{\rm ad} &= \int \frac{\dd^4 k}{(2\pi)^4}\frac{1}{2}
    \Tr\Bigl[\delta\bar\Sigma^\gt_{\rm L}(k,t)\bar S^\lt_{\rm ad}(k,t)
        - \delta\bar\Sigma^\lt_{\rm L}(k,t)\bar S^\gt_{\rm ad}(k,t) + \text{H.c.}
    \Bigr] \text{.} \label{eq:washout-term-B}
\end{align}
The washout terms are proportional to the lepton asymmetry $n_\ell - {\bar n}_\ell$ via equation~\cref{eq:lepton-asymmetry-vs-chemical-potential}. The adiabatic source term $-\partial_t \bar S^\lt_{{\rm ad},\evec{k}}(t,t)$ of the neutrino equation~\cref{eq:neutrino-kinetic-final} is calculated from equation~\cref{eq:adiabatic-local-wightman}, and the effective self-energy $\bar\Sigma^r_{{\rm eq},{\rm eff},\evec{k}}(t,t)$ is given by equation~\cref{eq:local-Sigma-eff-LO}. We expect $W_{\rm ad}$ to be the dominant washout term because it is proportional to the adiabatic functions, as opposed to $\delta W$ which depends on the non-equilibrium perturbation $\delta S^<$ only.

Equations~\cref{eq:neutrino-kinetic-final,eq:lepton-asymmetry-final,eq:cp-source,eq:washout-term-A,eq:washout-term-B} are fully general apart from our using the local ansatz~\cref{eq:local-approximation} to compute the collision terms and the simplifications made in the reduction of the SM sector. The effective self-energies in the lepton source and washout terms~\cref{eq:cp-source,eq:washout-term-A,eq:washout-term-B} are all calculated expanding consistently to the leading order in gradients and in the coupling constant expansion (more precisely: they are first order in $\abs{y_i}^2$, $\partial_t m$ and $\partial_t \mu_\ell$ combined). This is the most compact form of the equations relevant for the leptogenesis problem. They correspond to an initial value problem for a set of coupled first order equations which is straightforward to solve numerically by discretising the momentum variable. We shall recast these equations into a set of coupled Boltzmann-like equations for the generalised phase space functions in section~\cref{sec:noneq_distribution_functions}, after a short discussion of the issue of renormalisation.

%%%%%%%%%%%%%%%%%%%%%%%%%%%%%%%%%%%%%%%%%%%%%%%%%%%%%%%%%%%%%%%%%%%%%%%%%%%%%%%
%
\subsection{Vacuum on-shell renormalisation}
\label{sec:neutrino-renormalisation}
%
%%%%%%%%%%%%%%%%%%%%%%%%%%%%%%%%%%%%%%%%%%%%%%%%%%%%%%%%%%%%%%%%%%%%%%%%%%%%%%%

So far we have implicitly assumed that we are working with finite, renormalised quantities. The renormalisation procedure is slightly intricate due to the flavour mixing. For completeness, we perform the one-loop vacuum renormalisation in our model, following the on-shell method of ref.~\cite{Kniehl:1996bd}. This is sufficient for our purposes since we do not consider gauge interactions~\cite{Kniehl:2014dra,Fuchs:2016swt}. In this section we denote renormalised quantities by a hat. The renormalised pole self-energies $\widehat\Sigma^p_{{\rm eq},ij}(k,t)$ can be written in terms of the unrenormalised functions and the counterterms as follows:
\begin{align}
    \widehat\Sigma^p_{{\rm eq},ij}(k,t) = \Sigma^p_{{\rm eq},ij}(k)
    &{}- (\slashed k - m_i)(P_{\rm L}\textstyle{\frac{1}{2}}\delta Z^{\rm L}_{ij}
    + P_{\rm R}\textstyle{\frac{1}{2}}\delta Z^{\rm R}_{ij}) \notag
    \\*
    &{}- (P_{\rm L}\textstyle{\frac{1}{2}}{\delta Z^{\rm R}_{ji}}^*
    + P_{\rm R}\textstyle{\frac{1}{2}}{\delta Z^{\rm L}_{ji}}^*)(\slashed k - m_j)
    + \delta m_i \delta_{ij}
   \text{.} \label{eq:renormalised-general-Sigma-p}
\end{align}
The complex conjugation is here understood element-wise and not in the matrix sense. The mass counterterms $\delta m_i$ are diagonal in the vacuum mass basis, but the wave function renormalisation factors $\delta Z^{\rm L,R}_{ij}$ are in general flavour matrices~\cite{Kniehl:1996bd}. Because our neutrinos are Majorana fields, the counterterms for different chiralities are related by
\begin{equation}
    \delta Z^{L}_{ij} = {\delta Z^{\rm R}_{ij}}^* \text{.}
\end{equation}
Because of the Hermiticity of the counterterm Lagrangian, only the dispersive part of the self-energy $\Sigma^p_{\rm eq} = \Sigma^{\rm H}_{\rm eq} + \im s_p \Sigma^{\mathcal{A}}_{\rm eq}$ contributes to renormalisation. The absorptive parts are finite as such and can be understood as being computed in terms of renormalised parameters throughout. Also thermal corrections are purely finite and may be split from the vacuum parts according to equations~\cref{eq:Sigma-H-integral-vacuum-split,eq:Sigma-H-appendix} given in appendix~\cref{sec:neutrino-self-energies}. Renormalisation is then associated only with the vacuum part of the Hermitian self-energy function $\Sigma_{\rm eq}^{\rm H(vac)}$.

The on-shell renormalisation conditions which guarantee that there is no mixing in the external legs, that $m_i$ are the renormalised masses and that the residue of the diagonal propagator is unity, are given by~\cite{Aoki:1982ed,Kniehl:1996bd,Espriu:2000fq}
\begin{subequations}
\begin{align}
    \widehat\Sigma^{\rm H(vac)}_{{\rm eq},ij}(k)u_j^s(k) &\to 0\text{,}\quad\text{when } k^2 \to m_j^2
    \text{,} \label{eq:renormalisation-condition-1}
    \\*
    \bar{u}_i^s(k)\widehat\Sigma^{\rm H(vac)}_{{\rm eq},ij}(k) &\to 0\text{,}\quad\text{when } k^2 \to m_i^2
    \text{,} \label{eq:renormalisation-condition-2}
    \\*
    \frac{1}{\slashed k - m_i}\widehat\Sigma^{\rm H(vac)}_{{\rm eq},ii}(k)u_i^s(k) &\to 0
    \text{,}\quad\text{when } k^2 \to m_i^2 \text{,} \label{eq:renormalisation-condition-3}
    \\*
    \bar{u}_i^s(k)\widehat\Sigma^{\rm H(vac)}_{{\rm eq},ii}(k)\frac{1}{\slashed k - m_i} &\to 0
    \text{,}\quad\text{when } k^2 \to m_i^2 \text{,} \label{eq:renormalisation-condition-4}
\end{align}
\end{subequations}
where $u^s_i$ satisfies $(\slashed k - m_i)u^s_i(k) = 0$ when $k^2 = m_i^2$. Note that there is no summation over repeated indices here. The dimensionally regularised vacuum self-energy is given by $\smash{\Sigma^{{\rm H(vac)}}_{{\rm eq},ij}}(k) = \cw (y_i y_j^* P_{\rm L} + y_i^* y_j P_{\rm R}) \smash{\slashed{\mathfrak{S}}^{{\rm H(vac)}}_{\rm eq}}(k)$ with
\begin{equation}
    \slashed{\mathfrak{S}}^{{\rm H(vac)}}_{\rm eq}(k)
    = -\frac{\slashed k}{32\pi^2}\biggl(
        \frac{1}{\bar\epsilon} + 2 - \log\abs[\bigg]{\frac{k^2}{\mu^2}}
    \biggr) + \mathcal{O}(\epsilon)
    \text{,} \label{eq:Sigma-H-integral-vacuum-part}
\end{equation}
where $1/\bar\epsilon \equiv 1/\epsilon - \gamma_{\rm E} + \log(4\pi)$, $D = 4 - 2\epsilon$ is the spacetime dimension, $\gamma_{\rm E}$ is the Euler--Mascheroni constant and $\mu$ is the renormalisation scale parameter. Using these results we find the following counterterms for the Majorana neutrinos:
\begin{subequations}
\begin{align}
    \delta Z^{\rm R}_{ij} &\overset{i\neq j}{=} \frac{\cw}{32\pi^2} \frac{2 m_j}{m_i^2 - m_j^2}
    \bigl(m_j \,y_i y_j^* + m_i \,y_i^* y_j\bigr)
    \biggl[\frac{1}{\bar\epsilon} + 2 - \log\biggl(\frac{m_j^2}{\mu^2}\biggr) \biggr]
    \text{,}
    \\*
    \delta Z^{\rm R}_{ii} &= -\frac{\cw}{32\pi^2} \abs{y_i}^2
    \biggl[\frac{1}{\bar\epsilon} - \log\biggl(\frac{m_i^2}{\mu^2}\biggr) \biggr]
    \text{,}
    \\*
    \delta m_i &= \frac{\cw}{32\pi^2} m_i \abs{y_i}^2
    \biggl[\frac{1}{\bar\epsilon} + 2 - \log\biggl(\frac{m_i^2}{\mu^2}\biggr) \biggr]
    \text{,}
\end{align}
\end{subequations}
where $i \neq j$ in the first equation. The corresponding renormalised vacuum self-energy is
\begin{gather}
    \begin{aligned}
        \widehat\Sigma^{{\rm H}({\rm vac})}_{{\rm eq},ij}(k,t) = \cw \Bigl[
            & P_{\rm L} \slashed k \bigl(y_i y_j^* c^{\rm H}_{ij} - y_i^ *y_j d^{\rm H}_{ij}\bigr)
            + P_{\rm L}\bigl(y_i y_j^* m_i - y_i^* y_j m_j\bigr)d^{\rm H}_{ij}
            \\*
            {}+{} & P_{\rm R} \slashed k\bigl(y_i^* y_j c^{\rm H}_{ij} - y_i y_j^* d^{\rm H}_{ij}\bigr)
            + P_{\rm R}\bigl(y_i^* y_j m_i - y_i y_j^* m_j\bigr)d^{\rm H}_{ij}
        \Bigr] \text{,}
    \end{aligned}
\shortintertext{with}
    c^{\rm H}_{ij} = \frac{1}{32\pi^2}\log\abs[\bigg]{\frac{k^2}{m_i^2}}
    - \frac{m_j}{m_i}d^{\rm H}_{ij} \text{,}
    \qquad
    d^{\rm H}_{ij} = \frac{1}{32\pi^2}\frac{m_i m_j}{m_i^2-m_j^2}
    \log\biggl(\frac{m_i^2}{m_j^2}\biggr)
    \text{.}
\end{gather}
The full renormalised pole self-energy~\cref{eq:renormalised-general-Sigma-p} can now be written as
\begin{equation}
    \widehat\Sigma^p_{{\rm eq},ij}(k,t) =
    \widehat\Sigma^{{\rm H}({\rm vac})}_{{\rm eq},ij}(k,t)
    + \Sigma^{{\rm H}(T)}_{{\rm eq},ij}(k)
    + \im s_p \Sigma^\mathcal{A}_{{\rm eq},ij}(k)
    \text{.} \label{eq:renormalised-general-Sigma-p-split}
\end{equation}

Note that the renormalisation procedure associated with the processes relevant for leptogenesis is not affected by the expansion of the universe. Indeed, as long as curvature effects are not relevant for the physical processes involved, renormalisation can be carried out in the local orthonormal coordinate system, which is locally a Minkowski space. We shall introduce the extension of our equations to the expanding Friedman--Robertson--Walker spacetime in section~\cref{sec:expansion-of-universe}.

%%%%%%%%%%%%%%%%%%%%%%%%%%%%%%%%%%%%%%%%%%%%%%%%%%%%%%%%%%%%%%%%%%%%%%%%%%%%%%%
%
\section{Non-equilibrium distribution functions}
\label{sec:noneq_distribution_functions}
%
%%%%%%%%%%%%%%%%%%%%%%%%%%%%%%%%%%%%%%%%%%%%%%%%%%%%%%%%%%%%%%%%%%%%%%%%%%%%%%%

The local correlation function $\delta\bar S^\lt_\evec{k}(t,t)$ is a matrix in both Dirac and flavour indices and its components have a complicated time dependence involving many different scales. These scales reflect the complicated phase space structure of the underlying Wigner function $\delta\bar S^\lt_\evec{k}(k^0,t)$, and they ultimately arise from the different physical phenomena the correlation function describes. In particular, the vast difference between the particle-antiparticle oscillation time $\sim 1/\omega_\evec{k}$ and the flavour oscillation time $\sim 1/\Delta m$ makes equation~\cref{eq:neutrino-kinetic-final} challenging for studying resonant leptogenesis as such. To overcome this problem we will parametrise $\delta\bar S^\lt_\evec{k}(t,t)$ in terms of phase space distribution functions $\delta f^{ss'}_{\evec{k}hij}$, and derive their coupled equations of motion. The benefit of this parametrisation, first introduced in~\cite{Fidler:2011yq}, is that each phase space function $\delta f^{ss'}_{\evec{k}hij}$ describes separate, clearly defined physics with characteristic time-dependence. This allows us to isolate the physics that we are interested in and to write simplified and yet accurate versions of equation~\cref{eq:neutrino-kinetic-final}, that are amenable to efficient numerical solution.

%%%%%%%%%%%%%%%%%%%%%%%%%%%%%%%%%%%%%%%%%%%%%%%%%%%%%%%%%%%%%%%%%%%%%%%%%%%%%%%
%
\subsection{Projection matrix parametrisation}
\label{sec:projector-parametrisation}
%
%%%%%%%%%%%%%%%%%%%%%%%%%%%%%%%%%%%%%%%%%%%%%%%%%%%%%%%%%%%%%%%%%%%%%%%%%%%%%%%

Since we consider a spatially homogeneous and isotropic system, we can construct $\delta\bar S^\lt_\evec{k}(t,t)$ using only 8 of the 16 basis elements of the full Dirac algebra. The basis matrices of this subalgebra commute with the momentum representation of the Dirac helicity operator,
\begin{equation}
    \hat{h}_{\evec{k}} \equiv \evec{\alpha} \cdot \Uevec{k} \,\gamma^5
    \text{,}
\end{equation}
where $\Uevec{k} \equiv \evec{k}/\abs{\evec{k}}$ is the unit three-momentum vector. As mentioned above, we will use the parametrisation introduced in~\cite{Fidler:2011yq}, and which we already used in the spectral solution~\cref{eq:full_wightman_diag}:
\begin{subequations}
    \label{eq:local-correlator-parametrisation}
    \begin{align}
        \delta\bar S^\lt_{\evec{k}ij}(t,t) &=
        \sum_{\mathclap{h,s,s'=\pm}} \mathcal{P}_{\evec{k}hij}^{ss'}
        \,\delta f_{\evec{k}hij}^{ss'}(t) \text{,}
\shortintertext{with}
        \mathcal{P}_{\evec{k}hij}^{ss'} &\equiv N_{\evec{k}hij}^{ss'}
        \,P_{\evec{k}h}^{} P_{\evec{k}i}^s \gamma^0 P_{\evec{k}j}^{s'} \text{.}
    \end{align}
\end{subequations}
Here $P_{\evec{k}h}$ and $P_{\evec{k}i}^s$ are the helicity and energy projection matrices defined in equation~\cref{eq:projector-def} and are labelled by helicity $h = \pm1$, neutrino flavour $i,j$ and the energy sign indices $s,s' = \pm1$. These matrices obviously satisfy the completeness relations $P_{\evec{k}{+}} + P_{\evec{k}{-}} = P_{\evec{k}i}^{+} + P_{\evec{k}i}^{-} = \idmat$ and it is easy to show that they also satisfy the idempotence and orthogonality relations $P_{\evec{k}h} P_{\evec{k}h'} = \delta_{hh'} P_{\evec{k}h}$ and $P_{\evec{k}i}^s P_{\evec{k}i}^{s'} = \delta_{ss'} P_{\evec{k}i}^s$. It can also be shown that for any given $\evec{k},i,j$, the four different matrices $P_{\evec{k}i}^s \gamma^0 P_{\evec{k}j}^{s'}$ (with $s,s' = \pm1$) span the same set as $\{\idmat, \gamma^0, \evec{\gamma} \cdot \Uevec{k}, \evec{\alpha} \cdot \Uevec{k}\}$ which commute with the helicity operator $\hat{h}_{\evec{k}}$. Thus the eight matrices $P_{\evec{k}h}^{} P_{\evec{k}i}^s \gamma^0 P_{\evec{k}j}^{s'}$ (with $h, s, s' = \pm1$) in equation~\cref{eq:local-correlator-parametrisation} can be used as a basis for the entire homogeneous and isotropic Dirac subalgebra.

The normalisation factors $N_{\evec{k}hij}^{ss'}$, which in part define the perturbations $\delta f_{\evec{k}hij}^{ss'}$, can be chosen freely in~\cref{eq:local-correlator-parametrisation}. The choice which gives the most symmetric relations between the phase space distribution functions and the correlation function as well as between the different distribution function components, and leads to simplest source terms in the evolution equations is%
\footnote{In the massless limit~\cref{eq:local-correlator-normalisation} becomes singular. This is a technical problem similar to the one encountered with massless spinors, and it can be avoided by using a different normalisation. Alternatively one can use~\cref{eq:local-correlator-normalisation} with finite masses, and take the limit $m_i \to 0$ at the end of the calculation when needed.}
\begin{equation}
    N_{\evec{k}hij}^{ss'} \equiv
    \tr\bigl(P_{\evec{k}h}^{} P_{\evec{k}i}^s \gamma^0 P_{\evec{k}j}^{s'} \gamma^0\bigr)^{-\frac{1}{2}}
    = \sqrt{
        \frac{2 \omega_{\evec{k}i} \omega_{\evec{k}j}}
        {\omega_{\evec{k}i} \omega_{\evec{k}j} + ss'(m_i m_j - \abs{\evec{k}}^2)}
    } \text{.} \label{eq:local-correlator-normalisation}
\end{equation}
With this choice the phase space distributions are also correctly normalised in the thermal limit. Note that despite the fact that the definition~\cref{eq:local-correlator-normalisation} involves the helicity projector, $N_{\evec{k}hij}^{ss'}$ does not depend on helicity. It is also symmetric in the energy and flavour indices. We can invert the parametrisation~\cref{eq:local-correlator-parametrisation} to express the phase space distribution functions in terms of the matrix form of the correlation function. Using~\cref{eq:local-correlator-normalisation} this relation reads simply
\begin{equation}
    \delta f_{\evec{k}hij}^{ss'} = \tr\bigl[
        \mathcal{P}_{\evec{k}hji}^{s's} \,\delta\bar S^\lt_{\evec{k}ij}(t,t)
    \bigr] \text{.} \label{eq:local-correlator-inverse-parametrisation}
\end{equation}
That is, with the normalisation~\cref{eq:local-correlator-normalisation}, $\mathcal{P}_{\evec{k}hji}^{s's}$ is a ``correctly normalised'' projection operator in our basis.

The basis spanned by $\mathcal{P}_{\evec{k}hji}^{s's}$ can be used to define generalised distribution functions for any local correlation function. Below we need the adiabatic distribution functions, which can now be defined analogously to~\cref{eq:local-correlator-inverse-parametrisation}:
\begin{equation}
    f_{{\rm ad},\evec{k}hij}^{ss'} \equiv
    \tr\bigl[\mathcal{P}_{\evec{k}hji}^{s's} \,\bar S^\lt_{{\rm ad},\evec{k}ij}(t,t)\bigr]
    \text{.} \label{eq:adiabatic-particle-distribution-functions}
\end{equation}
Substituting here the leading order $\bar S^\lt_{{\rm ad},\evec{k}}(t,t)$ given by equation~\cref{eq:adiabatic-local-wightman} (\ie~the free case) we get
\begin{equation}
    f_{{\rm ad},\evec{k}hij}^{ss'} \simeq
    s f_{\rm FD}(s\omega_{\evec{k}i}) \delta_{ss'} \delta_{ij}
    \text{.} \label{eq:thermal-limit-of-adiabatic-solution}
\end{equation}
This shows that the parametrisation~\cref{eq:local-correlator-parametrisation} with the normalisation~\cref{eq:local-correlator-normalisation} naturally matches to the Fermi--Dirac distribution in the free theory (up to a sign for negative frequency states).

%%%%%%%%%%%%%%%%%%%%%%%%%%%%%%%%%%%%%%%%%%%%%%%%%%%%%%%%%%%%%%%%%%%%%%%%%%%%%%%
%
\subsection{Generalised density matrix equation}
\label{sec:delta-f-equations}
%
%%%%%%%%%%%%%%%%%%%%%%%%%%%%%%%%%%%%%%%%%%%%%%%%%%%%%%%%%%%%%%%%%%%%%%%%%%%%%%%

Using the parametrisation~\cref{eq:local-correlator-parametrisation} with the normalisation~\cref{eq:local-correlator-normalisation} for $\delta\bar S^\lt_{\evec{k}}(t,t)$ in equation~\cref{eq:neutrino-kinetic-final} we can derive an equation for the non-equilibrium distribution functions $\delta f_{\evec{k},hij}^{ss'}(t)$. The calculation is complicated by the fact that also the normalisation factors and the energy projection operators depend on time due to the time dependent masses $m_i(t)$, but the final master equation is structurally very simple:
\begin{align}
    \partial_t\delta f_{\evec{k}hij}^{ss'} ={}&
        -\im (s\omega_{\evec{k}i} - s'\omega_{\evec{k}j})\delta f_{\evec{k}hij}^{ss'}
        - \partial_t f_{{\rm ad},\evec{k}hij}^{ss'} \notag
        \\*
        &{} + ss'\frac{\abs{\evec{k}}}{2} \Bigl(
            \frac{\dot{m}_i}{\omega_{\evec{k}i}^2} f_{\evec{k}hij}^{-s,s'}
            + \frac{\dot{m}_j}{\omega_{\evec{k}j}^2} f_{\evec{k}hij}^{s,-s'}
        \Bigr) \notag
        \\*
        &{} - \sum_{l,\eta} \Bigl[
            C^{s \eta s'}_{\evec{k}hilj} \delta f_{\evec{k}hlj}^{\eta s'}
            + \bigl(C^{s' \eta s}_{\evec{k}hjli}\bigr)^* \delta f_{\evec{k}hil}^{s \eta}
        \Bigr]
    \text{.} \label{eq:delta-f-equation}
\end{align}
Here $\dot{m}_i \equiv \partial_t m_i$ and we combined $f_{\evec{k}} = f_{{\rm ad},\evec{k}} + \delta f_{\evec{k}}$ in the second line, with the adiabatic distribution functions $f_{{\rm ad},\evec{k}}$ defined in equation~\cref{eq:adiabatic-particle-distribution-functions}, and the collision term is given by
\begin{equation}
    C^{s \eta s'}_{\evec{k}hilj} \equiv \im \tr\bigl[
        \mathcal{P}_{\evec{k}hji}^{s's} \,\bar\Sigma^r_{{\rm eq,eff},\evec{k}il}(t,t)
        \,\mathcal{P}_{\evec{k}hlj}^{\eta s'}
    \bigr] \text{.}
    \label{eq:collision_term_trace_def}
\end{equation}

Equation~\cref{eq:delta-f-equation} describes all particle-antiparticle and flavour-coherence effects in the local limit and includes helicity. This universality is reflected in the large number of indices, which may appear overwhelming at first. However, all terms in~\cref{eq:delta-f-equation} have simple interpretations. For example the first term on the right-hand side corresponds to the Hamiltonian commutator term in equation~\cref{eq:neutrino-kinetic-final}. It falls into this simple form because $H_{\evec{k}i}^{} P_{\evec{k}i}^s = P_{\evec{k}i}^s H_{\evec{k}i}^{} = s \omega_{\evec{k}i} P_{\evec{k}i}^s$. The Hamiltonian term causes oscillations in the off-diagonal components and its simple form is pivotal in distinguishing the relevant time scales. For $s = -s'$ the oscillations are very fast. These oscillations are essential \eg~for vacuum particle production, but they can be a problem if one is interested only in flavour oscillations, which correspond to $s = s'$ but $i \neq j$, and usually evolve much more slowly. In the next section we derive an effective equation for flavour oscillations by averaging over the fast oscillations.

The second line in equation~\cref{eq:delta-f-equation} arises from the time dependence of the basis matrices $\mathcal{P}_{\evec{k}hij}^{ss'}$ in the parametrisation~\cref{eq:local-correlator-parametrisation}. The precise form of these terms depends on the normalisation, and the choice~\cref{eq:local-correlator-normalisation} turns out to give the most compact form. These terms are again essential for vacuum particle production, but they can be neglected in the leptogenesis application. The source terms $\partial_t f_{{\rm ad},\evec{k}hij}^{ss'}$ result from
the projection of the adiabatic matrix source in~\cref{eq:neutrino-kinetic-final}. Lastly, the collision terms $\smash{C^{s \eta s'}_{\evec{k}hilj}}$ can in general be separated into dispersive and absorptive parts, just by using $\Sigma^r = \Sigma^{\rm H} - \im \Sigma^\mathcal{A}$, and consequently we define
\begin{equation}
    C^{s \eta s'}_{\evec{k}hilj} \equiv C^{{\rm H},s \eta s'}_{\evec{k}hilj} + C^{\mathcal{A},s \eta s'}_{\evec{k}hilj}.
\end{equation}
The dispersive term $C^{{\rm H},s \eta s'}_{\evec{k}hilj}$ can be broadly characterised as a generalised matter Hamiltonian. It is of course a very general structure, but \eg~in the case of ordinary light neutrino mixing it can be shown~\cite{JKP_in_progress} to exactly reproduce the neutrino effective potential in matter. It could be interesting for resonant leptogenesis as well, because it would make the energy splitting $\Delta \omega_{\evec{k}}$ between the neutrinos a dynamical quantity. We will not consider the dispersive corrections numerically in this paper, but also this topic will be pursued elsewhere. Finally, the absorptive part $C^{\mathcal{A},s\eta s'}_{\evec{k}hilj}$ of the collision term contains both flavour-diagonal and off-diagonal damping rates, and importantly for leptogenesis, cross couplings between the diagonal and off-diagonal distribution functions. These cross coupling functions together with the CP-violating flavour oscillation are the mechanism which generates the lepton asymmetry.

%%%%%%%%%%%%%%%%%%%%%%%%%%%%%%%%%%%%%%%%%%%%%%%%%%%%%%%%%%%%%%%%%%%%%%%%%%%%%%%
%
\subsection{Lepton source and washout terms}
%
%%%%%%%%%%%%%%%%%%%%%%%%%%%%%%%%%%%%%%%%%%%%%%%%%%%%%%%%%%%%%%%%%%%%%%%%%%%%%%%

It is easy to write also the CP-violating source term and the washout terms of the lepton equation using the parametrisation~\cref{eq:local-correlator-parametrisation} and normalisation~\cref{eq:local-correlator-normalisation} for the Majorana neutrino correlator. For the source term~\cref{eq:cp-source} and the washout term~\cref{eq:washout-term-A} we get, respectively,
\begin{align}
    S_{CP} &= \sum_{h,s,s',i,j} \int \frac{\dd^3\evec{k}}{(2\pi)^3}
    \tr\Bigl[
        \Bigl(
            P_{\rm R} \bar\Sigma^\mathcal{A}_{{\rm eq,eff},\evec{k}ji}
            + (\bar\Sigma^\mathcal{A}_{{\rm eq,eff},\evec{k}ij})^\dagger P_{\rm R}
        \Bigr)
        \mathcal{P}_{\evec{k}hij}^{ss'}
    \Bigr]
    \delta f_{\evec{k}hij}^{ss'}
    \text{,} \label{eq:cp-source-with-distributions} \displaybreak[0]
    \\*
    \delta W &= \sum_{h,s,s',i,j} \int \frac{\dd^3\evec{k}}{(2\pi)^3}
    \tr\Bigl[
        \Bigl(
            P_{\rm R} \delta\bar\Sigma^\mathcal{A}_{{\rm eff},\evec{k}ji}
            + (\delta\bar\Sigma^\mathcal{A}_{{\rm eff},\evec{k}ij})^\dagger P_{\rm R}
        \Bigr)
        \mathcal{P}_{\evec{k}hij}^{ss'}
    \Bigr]
    \delta f_{\evec{k}hij}^{ss'}
    \text{.} \label{eq:washout-1-with-distributions}
\end{align}
For the washout term~\cref{eq:washout-term-B}, we again expand the adiabatic Wightman functions to leading order to get the $\mathcal{O}(y_i^2)$ result
\begin{equation}
    W_{\rm ad} = \sum_{s,i} \int \frac{\dd^3\evec{k}}{(2\pi)^3} \frac{1}{2} \tr\Bigl[
        P_{\rm R}\Bigl(
            \anticomm[\big]{\delta\bar\Sigma^\gt_{ii}(k_{si}), f_{\rm FD}(s\omega_{\evec{k}i}) P_{\evec{k}i}^s}
            - \anticomm[\big]{\delta\bar\Sigma^\lt_{ii}(k_{si}), f_{\rm FD}(-s\omega_{\evec{k}i}) P_{\evec{k}i}^s}
        \Bigr)
    \Bigr] \text{.} \label{eq:washout-2-with-distributions}
\end{equation}
Expanded forms of equations~\cref{eq:cp-source-with-distributions,eq:washout-1-with-distributions,eq:washout-2-with-distributions} are given in appendix~\cref{sec:cp-source-and-washout-appendix}, where we use the leading order approximation~\cref{eq:local-Sigma-eff-LO} for the effective self-energies and perform the traces.

%%%%%%%%%%%%%%%%%%%%%%%%%%%%%%%%%%%%%%%%%%%%%%%%%%%%%%%%%%%%%%%%%%%%%%%%%%%%%%%
%
\subsection{Mass shell equations}
\label{sec:mass_shell_equation}
%
%%%%%%%%%%%%%%%%%%%%%%%%%%%%%%%%%%%%%%%%%%%%%%%%%%%%%%%%%%%%%%%%%%%%%%%%%%%%%%%

As stated above, the most important benefit of the parametrisation~\cref{eq:local-correlator-parametrisation} is that it allows to separate the physics corresponding to different time scales. In particular we can distinguish the \emph{mass shell} functions corresponding to $s = s'$ (but including the flavour coherences $i \neq j$) from the fast oscillating \emph{coherence shell} functions for which $s \neq s'$.%
\footnote{This naming scheme follows the earlier cQPA notation~\cite{Fidler:2011yq} and the one we already used in~\cref{eq:full_wightman_diag}, although in our current treatment the phase space structures are not a priori restricted to a spectral form.}
For a graphical illustration see figure~2 in~\cite{Fidler:2011yq}. We denote these functions by
\begin{align}
    \delta f^{m,\pm}_{\evec{k}} &\equiv \delta f^{\pm\pm}_{\evec{k}}
    \text{,} \label{eq:mass-shell-functions} \\*
    \delta f^{c,\pm}_{\evec{k}} &\equiv \delta f^{\pm\mp}_{\evec{k}}
    \text{.} \label{eq:coherence-shell-functions}
\end{align}
Indeed, from equation~\cref{eq:delta-f-equation} we see that for $\delta f^{c,\pm}_{\evec{k}hij}$ the Hamiltonian term is proportional to $\mp \im(\omega_{\evec{k}i} + \omega_{\evec{k}j}) = \mp 2\im\bar\omega_{\evec{k}ij}$, corresponding to very fast particle-antiparticle oscillations (zitterbewegung). For the mass shell solutions $\delta f^{m,\pm}_{\evec{k}hij}$ the Hamiltonian term is proportional to $\mp \im(\omega_{\evec{k}i} - \omega_{\evec{k}j}) = \mp \im\Delta\omega_{\evec{k}ij}$, corresponding to flavour oscillations for $i \neq j$, at a frequency which is suppressed for large $\abs{\evec{k}}$ or a small mass difference $\abs{m_i - m_j}$. This is the case of interest for resonant leptogenesis.

If particle-antiparticle oscillations are much faster than the flavour oscillations, we can drop the coherence functions $\delta f^c_\evec{k}$ in equations~\cref{eq:delta-f-equation}, since their effect on the flavour oscillations averages out.%
\footnote{\label{footnote:fc-elimation}We can formally justify this as follows. First, generically $\delta f^{c\pm}_{\evec{k}}(t) = A^\pm_{\evec{k}}(t) \exp(\mp 2\im\bar\omega_{\evec{k}}t)$, where $A^\pm_\evec{k}(t)$ are some functions that vary only in the flavour scale. Next, take the convolution of~\cref{eq:delta-f-equation} with some appropriate normalised weight function $W$, \eg~the Weierstrass transform with $W(t,t') \sim \exp(-(t - t')^2/2 \sigma^2)$, where we can choose $1/\Delta\omega_\evec{k} \gg \sigma \gg 1/2 \bar\omega_\evec{k}$. This has no effect on the mass-shell contributions, because they do not vary significantly over the time $\sigma$. However, the terms involving the coherence solutions behave as
\[
    \int \dd t' \, W(t,t') D^\pm_\evec{k}(t')\delta f^{c\pm}_{\evec{k}}(t')
    \sim D^\pm_\evec{k}(t)\delta f^{c\pm}_{\evec{k}}(t) \exp\bigl(-2(\bar\omega_\evec{k}\sigma)^2\bigr)
    \text{,}
\]
where $D^\pm_\evec{k}$ represents whatever term that is multiplying the coherence solution. Because $\bar\omega \sigma \gg 1$, these terms are extremely suppressed and completely drop from the averaged mass-shell equations.}
This results in a much simpler coarse-grained master equation for the mass shell functions:
\begin{align}
    \partial_t\delta f_{\evec{k}hij}^{m,s} ={}&
    -\im s(\omega_{\evec{k}i} - \omega_{\evec{k}j}) \delta f_{\evec{k}hij}^{m,s}
    - \partial_t f_{{\rm ad},\evec{k}hij}^{m,s} \notag
    \\*
    &{} - \sum_l \Bigl[
        C^{sss}_{\evec{k}hilj} \delta f_{\evec{k}hlj}^{m,s}
        + \bigl(C^{sss}_{\evec{k}hjli}\bigr)^* \delta f_{\evec{k}hil}^{m,s}
    \Bigr] \text{.} \label{eq:delta-f-m-equation}
\end{align}
Equation~\cref{eq:delta-f-m-equation} still holds complete information of flavour mixing and the helicity degree of freedom in the local limit. In particular it contains as limiting cases the familiar Boltzmann theory, light neutrino density matrix formalism and the cQPA-formulation of the generic flavour and spin dependent problem. It also agrees with or encompasses a number of other quantum transport approaches in the literature, of which we give a more detailed comparisons in section~\cref{sec:comparison_and_discussion}.

Removing the coherence shell solutions greatly facilitates the numerical solution, in particular for quasi-degenerate Majorana neutrinos. In addition, a number of relations hold between different components of $\delta f_\evec{k}$, which further simplify the numerical task; we will give these relations in appendix~\cref{sec:delta_f_components}. In the following sections we will solve equations~\cref{eq:delta-f-m-equation} numerically using the leading order expansion~\cref{eq:local-Sigma-eff-LO} for the effective self-energies in the collision terms functions. Detailed expressions for $C^{sss}_{\evec{k}hilj}$ are given in appendix~\cref{sec:collision-term-traces}.

%%%%%%%%%%%%%%%%%%%%%%%%%%%%%%%%%%%%%%%%%%%%%%%%%%%%%%%%%%%%%%%%%%%%%%%%%%%%%%%
\paragraph{Hierarchical limit.}
%%%%%%%%%%%%%%%%%%%%%%%%%%%%%%%%%%%%%%%%%%%%%%%%%%%%%%%%%%%%%%%%%%%%%%%%%%%%%%%

Let us briefly comment on the validity of our master equations in the hierarchical limit of leptogenesis~\cite{Davidson:2008bu,Beneke:2010wd}, where $\Delta m/m \sim \mathcal{O}(1)$. The general master equation~\cref{eq:delta-f-equation} is of course applicable also in this limit. However, the condition $\Delta \omega_{\evec{k}ij} \ll 2\bar \omega_{\evec{k}ij}$ might hold only to a limited degree (see footnote~\cref{footnote:fc-elimation}), so that neglecting the coherence shell functions $\delta f^c_\evec{k}$ might not be warranted, possibly reducing the accuracy of the mass-shell equations~\cref{eq:delta-f-m-equation}. However, using equation~\cref{eq:delta-f-equation} to model this case would be numerically very challenging, because due to the large hierarchy $H \ll m$, both oscillation time scales are very fast compared to the heavy neutrino decoupling time. Luckily, due to the very same reason, one can in this case work in the decoupling limit (see section~\cref{sec:helicity_symmetric_equations} below) and derive an effective Boltzmann equation for the lepton asymmetry. To our knowledge the contribution from the coherence shell functions $\delta f^c_\evec{k}$ has never been included, however. This could be done by starting from the master equation~\cref{eq:delta-f-equation} and it would indeed be interesting to study the size of these corrections.

%%%%%%%%%%%%%%%%%%%%%%%%%%%%%%%%%%%%%%%%%%%%%%%%%%%%%%%%%%%%%%%%%%%%%%%%%%%%%%%
%
\section{Expansion of the universe}
\label{sec:expansion-of-universe}
%
%%%%%%%%%%%%%%%%%%%%%%%%%%%%%%%%%%%%%%%%%%%%%%%%%%%%%%%%%%%%%%%%%%%%%%%%%%%%%%%

So far all our equations have been formulated in the flat Minkowski spacetime, but we eventually need to work in the expanding Friedmann--Lemaître--Robertson--Walker (FLRW) background. In this section we show how the generalisation to an expanding universe can be done by a simple reinterpretation of all variables in the comoving frame. We finish this section by rewriting our master equations explicitly in the expanding universe in terms of the scaled inverse temperature.

%%%%%%%%%%%%%%%%%%%%%%%%%%%%%%%%%%%%%%%%%%%%%%%%%%%%%%%%%%%%%%%%%%%%%%%%%%%%%%%
%
\subsection{Lagrangian in curved spacetime}
\label{sec:conformal}
%
%%%%%%%%%%%%%%%%%%%%%%%%%%%%%%%%%%%%%%%%%%%%%%%%%%%%%%%%%%%%%%%%%%%%%%%%%%%%%%%

First we need to generalise the Minkowski Lagrangian~\cref{eq:lagrangian} to curved spacetime:
\begin{align}
    \mathcal{L} = \sqrt{-g} \biggl\{
        &\sum_{i=1}^2 \Bigl[
            \frac{1}{2} \bar N_i(\im \gamma^\mu \nabla_\mu - m_i) N_i
            - y_i^* (\bar\ell \widetilde\phi) P_{\rm R} N_i
            - y_i \bar N_i P_{\rm L} (\widetilde\phi^\dagger \ell)
        \Bigr] \notag
        \\*
        &{} + \bar\ell \,\im \gamma^\mu \nabla_\mu \ell
        + (\partial_\mu\phi^\dagger)(\partial^\mu\phi) - \xi R\phi^\dagger\phi
    \biggr\} \text{.} \label{eq:curved-spacetime-lagrangian}
\end{align}
Here $g \equiv \det(g_{\mu\nu})$ is the determinant of the metric and $\nabla_\mu$ is the covariant derivative given by the spin connection. We also added the non-minimal coupling of the Higgs field $\phi$ to the scalar curvature $R$ and the factor $\sqrt{-g}$ originates from the volume form of the curved spacetime action integral. We then assume a spatially flat FLRW metric
\begin{equation}
    {\dd s}^2 = g_{\mu\nu} \dd x^\mu \dd x^\nu
    = a(\eta)^2({\dd \eta}^2 - {\dd \evec{x}}^2)
    \text{,} \label{eq:flat-FLRW-metric}
\end{equation}
where $a = a(\eta)$ is the dimensionless scale factor and the conformal time $\eta$ is defined by $\dd t = a(\eta) \dd \eta$. With the metric~\cref{eq:flat-FLRW-metric} the contracted covariant derivative in equation~\cref{eq:curved-spacetime-lagrangian} becomes
\begin{equation}
    \gamma^\mu\nabla_\mu = \frac{1}{a}\Bigl(
        \gamma^0 \partial_0 + \evec{\gamma} \cdot \evec{\nabla} + \frac{3}{2}\frac{a'}{a}\gamma^0
    \Bigr) \text{,} \label{eq:contracted-spin-connection}
\end{equation}
where $\evec{\nabla}$ is the usual flat spatial derivative vector, $\partial_0 = \partial/\partial \eta$ and $a' \equiv \partial a/\partial\eta$.

Using equations~\cref{eq:flat-FLRW-metric,eq:contracted-spin-connection} with $\sqrt{-g} = a^4$ and scaling all fermion fields $\psi$ according to $\psi \to a^{-3/2} \psi$ and the Higgs field as $\phi \to a^{-1}\phi$, the Lagrangian~\cref{eq:curved-spacetime-lagrangian} is transformed to
\begin{align}
    \mathcal{L} = {}& \sum_{i=1}^2 \Bigl[
        \frac{1}{2} \bar N_i \bigl(\im \gamma^0 \partial_0
        + \im \evec{\gamma} \cdot \evec{\nabla} - a m_i\bigr) N_i
        - y_i^* (\bar\ell \widetilde\phi) P_{\rm R} N_i
        - y_i \bar N_i P_{\rm L} (\widetilde\phi^\dagger \ell)
    \Bigr] \notag
    \\*
    &{}+ \bar\ell \bigl(\im \gamma^0 \partial_0
    + \im \evec{\gamma} \cdot \evec{\nabla}\bigr) \ell
    + \phi^\dagger \Bigl[
        -\partial_0^2 + \evec{\nabla}^2
        - a^2 \Bigl(\xi R - \frac{a''}{a^3}\Bigr)
    \Bigr] \phi
    \text{,} \label{eq:scaled-lagrangian}
\end{align}
where the scalar curvature is given by $R = 6a''/a^3$. From the Lagrangian~\cref{eq:scaled-lagrangian} we see that the only effects of the expanding universe compared to the Minkowski theory~\cref{eq:lagrangian} are that the time variable is replaced by the conformal time $\eta$, spatial coordinates become the comoving ones, neutrino masses are multiplied by the scale factor $a$ and the Higgs field gets a geometric mass term. This mass term vanishes for a conformal coupling $\xi = 1/6$ or when the universe is radiation-dominated: $a(\eta) \propto \eta$, which is the case to a high accuracy in leptogenesis. We shall thus continue to assume that the Higgs field is massless.

%%%%%%%%%%%%%%%%%%%%%%%%%%%%%%%%%%%%%%%%%%%%%%%%%%%%%%%%%%%%%%%%%%%%%%%%%%%%%%%
\paragraph{The comoving frame.}
%%%%%%%%%%%%%%%%%%%%%%%%%%%%%%%%%%%%%%%%%%%%%%%%%%%%%%%%%%%%%%%%%%%%%%%%%%%%%%%

Based on the above, all expressions and equations in the earlier sections given in the Minkowski background remain valid also in the expanding flat spacetime when we make the replacements
\begin{equation}
    t \to \eta \text{,} \qquad
    \evec{k} \to \evec{k}_{\rm com} \text{,} \qquad
    T \to T_{\rm com} \text{,} \qquad
    m_i(t) \to a(\eta) m_i \text{,}
    \label{eq:comoving-variables}
\end{equation}
where $\evec{k}_{\rm com} \equiv \evec{k} a$ is the comoving momentum, $T_{\rm com} \equiv T a$ is the comoving temperature of the relativistic SM heat bath and $m_i$ are the physical constant masses. Note that the phase space distribution functions $f$ are dimensionless scalars and have the same values in both physical and comoving variables.

We assume that the universe is dominated by relativistic SM particles, which are kept in equilibrium by fast gauge interactions, and that the universe expands adiabatically. In the absence of entropy production the physical temperature then scales as $T \propto a^{-1}$, so that $T_{\rm com}$ remains constant. The comoving momentum $\evec{k}_{\rm com}$ is also constant, as the physical momentum redshifts as $\evec{k} \propto a^{-1}$. We can also write the Hubble parameter as
\begin{equation}
   H(T) = \biggl(\frac{4\pi^3}{45} g_*\biggr)^{1/2} \frac{T^2}{m_{\rm Pl}}
   \text{,} \label{eq:Hubble_parameter_from_T}
\end{equation}
where $m_{\rm Pl}$ is the Planck mass and $g_*$ is the effective number of relativistic degrees of freedom, which is $g_* = 110.25$ when including all SM fields and two massless Majorana neutrinos. Note that the scale factor can now be written as $a(\eta) = H(T_{\rm com}) \eta$. The entropy density is given by $s = 2\pi^2 g_* T^3/45$. Overall, the leptogenesis equations retain the same form they had when formulated with a generic time dependent mass term in the Minkowski background, when using the replacements~\cref{eq:comoving-variables} and scaling the equations by appropriate powers of the scale factor $a$.

%%%%%%%%%%%%%%%%%%%%%%%%%%%%%%%%%%%%%%%%%%%%%%%%%%%%%%%%%%%%%%%%%%%%%%%%%%%%%%%
%
\subsection{Final master equation in expanding space time}
\label{sec:z-equations}
%
%%%%%%%%%%%%%%%%%%%%%%%%%%%%%%%%%%%%%%%%%%%%%%%%%%%%%%%%%%%%%%%%%%%%%%%%%%%%%%%

For the numerical implementation it is convenient to formulate the equations using dimensionless variables. The most relevant temperature scale for leptogenesis is around $T = m_1$, where $m_1$ is the mass of the lightest Majorana neutrino. We thereby use the variables
\begin{gather}
    z = \frac{m_1}{T} \text{,} \qquad
    x_i = \frac{m_i}{m_1} \text{,} \qquad
    \kappa = \frac{\abs{\evec{k}}}{T} \text{,}
    \label{eq:dimensionless_variables}
\shortintertext{with}
    \frac{\dd}{\dd t} = z H \frac{\dd}{\dd z}
    = \frac{H_1}{z} \frac{\dd}{\dd z} \text{,}
    \label{eq:cosmic_d-dt}
\end{gather}
where $H_1 \equiv H(m_1)$ is the Hubble parameter~\cref{eq:Hubble_parameter_from_T} evaluated at $T = m_1$. The $z$-parameter is directly proportional to the scale factor so it serves as the time evolution parameter. Due to the constraints~\cref{eq:delta-f-m-c-component-constraints} there are only four independent mass shell functions for the two mixing Majorana neutrinos. We choose them as follows:
\begin{equation}
    \delta f_{\evec{k}h} \equiv \bigl(
        \delta f^{m,+}_{\evec{k}h11}, \,
        \delta f^{m,+}_{\evec{k}h22}, \,
        \Re\delta f^{m,+}_{\evec{k}h12}, \,
        \Im\delta f^{m,+}_{\evec{k}h12}
    \bigr) \text{.}
    \label{eq:delta_f_independent_components}
\end{equation}
We can now formulate the $\evec{k}$-dependent neutrino equation~\cref{eq:delta-f-m-equation} as a vector equation for the components~\cref{eq:delta_f_independent_components}. Including also the lepton asymmetry equation~\cref{eq:lepton-asymmetry-final}, our final master equations written using the dimensionless variables are
\begin{align}
    \frac{\dd \delta f_{\evec{k}h}}{\dd z} &= \bigl(
        \Delta\tilde\omega_{\evec{k}} - \tilde{C}_{\evec{k}h}
    \bigr)\delta f_{\evec{k}h} - \frac{\dd f_{{\rm ad},\evec{k}h}}{\dd z} \text{,}
    \label{eq:final-numerical-neutrino-equation}
    \\*
    \frac{\dd Y_L}{\dd z} &= \tilde S_{CP} + (\delta \tilde W + \tilde W_{\rm ad}) Y_L \text{,}
    \label{eq:final-numerical-lepton-equation}
    \end{align}
where $Y_L \equiv (n_{\ell} - \bar n_{\ell})/s$, and we used~\cref{eq:lepton-asymmetry-vs-chemical-potential} to relate the lepton chemical potential to the asymmetry. The dimensionless tree level oscillation coefficient $\Delta\tilde\omega_{\evec{k}}$ and the collision term coefficient $\tilde{C}_{\evec{k}h}$ of the neutrino equation, as well as the CP-violating lepton source term $\tilde S_{CP}$ and the washout term coefficients $\tilde W$ (with $W = \delta W, W_{\rm ad}$) are given by
\begin{alignat}{2}
    \Delta\tilde\omega_{\evec{k}} &\equiv \frac{z \,\Delta\omega_{\evec{k}}}{H_1} \text{,} \qquad &
    \tilde{C}_{\evec{k}h} &\equiv \frac{z \,C_{\evec{k}h}}{H_1} \text{,}
    \label{eq:dimensionless_osc_and_coll_terms}
    \\
    \tilde S_{CP} &\equiv \frac{z \,S_{CP}}{s H_1} \text{,} \qquad &
    \tilde W &\equiv \frac{6z \,W}{c_w H_1 \mu_\ell T^2} \text{.}
    \label{eq:dimensionless_SCP_and_W}
\end{alignat}

Equations~\cref{eq:final-numerical-neutrino-equation,eq:final-numerical-lepton-equation} are formally analogous to the momentum dependent Boltzmann equations, which we present in equations~\cref{eq:Boltzmann_fi_dimensionless,eq:Boltzmann_nL_dimensionless} of appendix~\cref{sec:Boltzmann_equations}. The difference is that the quantities $\Delta\omega_{\evec{k}}$ and $C_{\evec{k}h}$ in equations~\cref{eq:final-numerical-neutrino-equation,eq:dimensionless_osc_and_coll_terms} are $4 \times 4$ matrices consisting of the coefficients for different components of the column vector $\delta f_{\evec{k}h}$. To avoid confusion with earlier notation, we denote these components (when needed) with bracketed indices: for example $\delta f_{\evec{k}h[3]} = \Re\delta f^{m,+}_{\evec{k}h12}$, and $C_{\evec{k}h[12]}$ is the coefficient of $\delta f_{\evec{k}h[2]}$ in the equation of $\delta f_{\evec{k}h[1]}$. The matrix $\Delta\omega_{\evec{k}}$ corresponds to the tree level flavour oscillation term and its only non-vanishing elements are $\Delta\omega_{\evec{k}[34]} = -\Delta\omega_{\evec{k}[43]} = \omega_{\evec{k}1} - \omega_{\evec{k}2}$. The collision term coefficient matrix is given by
\begin{gather}
    C_{\evec{k}h} = \begin{pmatrix*}[c]
        2\Re(C^+_{111}) & 0 & 2\Re(C^+_{121}) & 2\Im(C^+_{121})
        \\
        0 & 2\Re(C^+_{222}) & 2\Re(C^+_{212}) & -2\Im(C^+_{212})
        \\
        \Re(C^+_{211}) & \Re(C^+_{122}) & \Re(C^+_{112}+C^+_{221}) & -\Im(C^+_{112}-C^+_{221})
        \\
        -\Im(C^+_{211}) & \Im(C^+_{122}) & \Im(C^+_{112}-C^+_{221}) & \Re(C^+_{112}+C^+_{221})
    \end{pmatrix*} \text{,} \label{eq:collision-term-matrix}
\end{gather}
where $C^{+}_{ilj} \equiv C^{+++}_{\evec{k}hilj}$ were defined in equation~\cref{eq:collision_term_trace_def}. Explicit expressions for the collision terms $C^{+++}_{\evec{k}hilj}$ are given in appendix~\cref{sec:collision-term-traces}, computed using self-energies in the leading order expansion~\cref{eq:local-Sigma-eff-LO} and explicitly written in appendix~\cref{sec:neutrino-self-energies}. Finally, explicit expressions for the lepton source and washout terms can be found in appendix~\cref{sec:cp-source-and-washout-appendix}.

%%%%%%%%%%%%%%%%%%%%%%%%%%%%%%%%%%%%%%%%%%%%%%%%%%%%%%%%%%%%%%%%%%%%%%%%%%%%%%%
%
\section{Numerical results}
\label{sec:Numerical_results}
%
%%%%%%%%%%%%%%%%%%%%%%%%%%%%%%%%%%%%%%%%%%%%%%%%%%%%%%%%%%%%%%%%%%%%%%%%%%%%%%%

In this section we solve numerically the system of equations~\cref{eq:final-numerical-neutrino-equation,eq:final-numerical-lepton-equation} for the Majorana neutrino distribution functions $\delta f_{\evec{k}h}$ and the (normalised) lepton asymmetry density $Y_L$ in the case of two Majorana neutrinos and one lepton flavour. We start by considering possible initial conditions. Then, before going to the discussion of the final lepton asymmetry and its dependence on model parameters, we establish the time scales relevant for the problem and study the momentum dependent neutrino distribution functions.

%%%%%%%%%%%%%%%%%%%%%%%%%%%%%%%%%%%%%%%%%%%%%%%%%%%%%%%%%%%%%%%%%%%%%%%%%%%%%%%
%
\subsection{Initial conditions}
\label{sec:initial_conditions}
%
%%%%%%%%%%%%%%%%%%%%%%%%%%%%%%%%%%%%%%%%%%%%%%%%%%%%%%%%%%%%%%%%%%%%%%%%%%%%%%%

While we took $\tin \to -\infty$ for the CTP to calculate the interactions, we can of course choose any finite time $t = t_0$ (or $z = z_0$) as the starting point of our calculation with arbitrary initial conditions for the correlation functions. In particular, we will consider both vacuum and thermal initial conditions for the Majorana neutrinos, while we assume that the lepton and Higgs distributions stay in local equilibrium in the common SM plasma temperature $T$. In both cases we assume a vanishing initial lepton asymmetry: $Y_L(t_0) = 0$.

%%%%%%%%%%%%%%%%%%%%%%%%%%%%%%%%%%%%%%%%%%%%%%%%%%%%%%%%%%%%%%%%%%%%%%%%%%%%%%%
\paragraph{Vacuum initial conditions.}
%%%%%%%%%%%%%%%%%%%%%%%%%%%%%%%%%%%%%%%%%%%%%%%%%%%%%%%%%%%%%%%%%%%%%%%%%%%%%%%

Here we assume that at $t = t_0$ the Majorana neutrinos are decoupled and thus effectively in zero temperature (\eg~if only the SM particles get reheated after inflation). The full local neutrino correlator is then given by
\begin{equation}
    \bar S^\lt_{\evec{k}ij}(t_0,t_0) = \bar S^\lt_{{\rm ad,vac},\evec{k}ij}(t_0,t_0)
    \simeq P^{-}_{\evec{k}i}(t_0) \delta_{ij} \text{,}
    \label{eq:vacuum-initial-value-for-correlator}
\end{equation}
which is calculated from~\cref{eq:adiabatic-local-wightman} by taking the limit $T \to 0$. For the non-equilibrium part we then get, using equations~\cref{eq:adiabatic-expansion,eq:vacuum-initial-value-for-correlator},
\begin{equation}
    \delta\bar S^\lt_{\evec{k}}(t_0,t_0) =
    \bar S^\lt_{{\rm ad,vac},\evec{k}}(t_0,t_0) - \bar S^\lt_{{\rm ad},\evec{k}}(t_0,t_0)
    \text{.} \label{eq:vacuum-initial-value-for-noneq-correlator}
\end{equation}
Note that this perturbation is not necessarily small as it measures the deviation of the full correlator $S^\lt$ (now initially in vacuum with zero temperature for the Majorana species) from the adiabatic equilibrium correlator $S^\lt_{\rm ad}$ (with the high temperature $T$ of the SM particle species). For the mass shell functions the initial condition~\cref{eq:vacuum-initial-value-for-noneq-correlator} with~\cref{eq:vacuum-initial-value-for-correlator} becomes
\begin{align}
    \delta f_{\evec{k}hij}^{m,+}(t_0) &= - f_{{\rm ad},\evec{k}hij}^{m,+}(t_0)
    = -f_{\rm FD}(\omega_{\evec{k}i}) \delta_{ij}
    \text{,} \label{eq:vacuum_initial_value_delta_f}
    \\*
    \delta f_{\evec{k}hij}^{m,-}(t_0) &= -\delta_{ij} - f_{{\rm ad},\evec{k}hij}^{m,-}(t_0)
    = -f_{\rm FD}(\omega_{\evec{k}i}) \delta_{ij}
    \text{,}
\end{align}
where we used equations~\cref{eq:local-correlator-inverse-parametrisation,eq:thermal-limit-of-adiabatic-solution}.

%%%%%%%%%%%%%%%%%%%%%%%%%%%%%%%%%%%%%%%%%%%%%%%%%%%%%%%%%%%%%%%%%%%%%%%%%%%%%%%
\paragraph{Thermal initial conditions.}
%%%%%%%%%%%%%%%%%%%%%%%%%%%%%%%%%%%%%%%%%%%%%%%%%%%%%%%%%%%%%%%%%%%%%%%%%%%%%%%

Here we assume that also the Majorana neutrinos are in local thermal equilibrium with the SM particles at $t = t_0$, corresponding to some high initial temperature $T_0 \gg m_i$. The full local neutrino correlator is then given by
\begin{equation}
    \bar S^\lt_{\evec{k}}(t_0,t_0) = \bar S^\lt_{{\rm ad},\evec{k}}(t_0,t_0)
\end{equation}
which trivially implies that $\delta\bar S^\lt_{\evec{k}}(t_0,t_0) = 0$ and that all non-equilibrium distribution functions vanish initially: $\delta f_{\evec{k}hij}^{m,\pm}(t_0) = 0$. Note also that in this case the Majorana neutrinos deviate from equilibrium only due to the dynamical source term $-\partial_t f_{\rm ad}$.

%%%%%%%%%%%%%%%%%%%%%%%%%%%%%%%%%%%%%%%%%%%%%%%%%%%%%%%%%%%%%%%%%%%%%%%%%%%%%%%
%
\subsection{Physical scales and parameters}
\label{sec:parameters}
%
%%%%%%%%%%%%%%%%%%%%%%%%%%%%%%%%%%%%%%%%%%%%%%%%%%%%%%%%%%%%%%%%%%%%%%%%%%%%%%%

The minimal leptogenesis mechanism is mainly controlled by four time scales, corresponding to the expansion rate of the universe $H$, the Majorana neutrino decay rate $\Gamma$, and the oscillation frequencies $\Delta \omega_{\evec{k}ij}$ and $2\bar \omega_{\evec{k}ij}$. We already discussed the role of the flavour and zitterbewegung oscillations in section~\cref{sec:mass_shell_equation}, finding that the latter may be relevant in the hierarchical limit, but can be safely ignored when $\Delta \omega_{\evec{k}ij}\ll 2\bar \omega_{\evec{k}ij}$. We work in this limit here, but the question of the relative sizes of the three slow time scales $H^{-1}$, $\Gamma^{-1}$ and $\Delta \omega_{\evec{k}ij}^{-1}$ still remains. The resonant leptogenesis mechanism turns out to be most efficient when all these scales are roughly comparable.

Indeed, if $H \gg \Delta \omega_{\evec{k}ij}$, there is no time for flavour oscillations to develop, while for $H \ll \Delta \omega_{\evec{k}ij}$ the source terms $\sim H$ become suppressed. Also, in the very strong washout limit, $\Gamma \gg H$ the asymmetry is suppressed by thermalisation due to interactions, whereas for $\Gamma \ll \Delta\omega_{\evec{k}ij}$, the magnitude of CP-violation is suppressed, since the source in the asymmetry equation is proportional to the couplings $y_i$. Additionally, if $\Gamma \gg \Delta \omega_{\evec{k}ij}$ the flavour oscillations become over-damped~\cite{Drewes:2016gmt}, analogously to light mixing neutrino systems~\cite{Enqvist:1990ad,Kainulainen:1990ds,Enqvist:1991qj}. This leaves us to seek parameters in the range $\Gamma \sim \Delta\omega_{\evec{k}ij} \sim H$ for maximal resonant enhancement.

%%%%%%%%%%%%%%%%%%%%%%%%%%%%%%%%%%%%%%%%%%%%%%%%%%%%%%%%%%%%%%%%%%%%%%%%%%%%%%%
\paragraph{Benchmark parameters.}
%%%%%%%%%%%%%%%%%%%%%%%%%%%%%%%%%%%%%%%%%%%%%%%%%%%%%%%%%%%%%%%%%%%%%%%%%%%%%%%

Our simple leptogenesis model has 5 physical input parameters: the magnitudes of the two Yukawa couplings $y_1$ and $y_2$, their relative CP-violating phase $\theta_{12}$, the lighter Majorana neutrino mass $m_1$ and the relative mass-squared difference $\Delta m^2_{21}/m_1^2 \equiv (m_2^2 - m_1^2)/m_1^2$. We define the Yukawa phase as
\begin{equation}
    \sin(\theta_{ij}) \equiv \frac{\Im(y_i^* y_j)}{\abs{y_i}\abs{y_j}}
    \text{.} \label{eq:Yukawa_CP_phase}
\end{equation}
We use the following set of benchmark values as a baseline:
\begin{subequations}
    \label{eq:benchmark-parameters}
    \begin{alignat}{2}
        \begin{split}
            \abs{y_1} &= 0.06 \text{,}
            \\
            \abs{y_2} &= 0.1 \text{,}
        \end{split} \hspace{3em} &
        \theta_{12} &= \frac{\pi}{4} \text{,}
        \\*
        m_1 &= \SI{1e13}{GeV} \text{,} \hspace{3em} &
        \frac{m_2^2}{m_1^2} &= 1 + \frac{\abs{y_1}^2 + \abs{y_2}^2}{16\pi} \approx 1.00027 \text{.}
    \end{alignat}
\end{subequations}
In addition one has to define the number of effective relativistic degrees of freedom $g_*$ and the initial temperature $T_0$ where we start the calculation. For these we use $g_* = 110$ and $z_0 = 10^{-2}$. For the benchmark parameters this corresponds to the initial temperature $T_0 = m_1/z_0 = \SI{1e15}{GeV}$ and the Hubble expansion rate $H_1 \equiv H(m_1) \approx \SI{1.428e8}{GeV}$.

Some more comments are still in order. First, we have chosen a high mass scale $\SI{1e13}{GeV}$ for the lightest Majorana neutrino, typical for traditional thermal leptogenesis~\cite{Davidson:2008bu}. However, it can be seen that all terms in equations~\cref{eq:dimensionless_osc_and_coll_terms,eq:dimensionless_SCP_and_W} scale either as $\sim \Gamma/H_1$ or $\sim \Delta m/H_1$, so that the dynamics does not depend on the mass scale $m_1$ as long as $\Gamma/H_1$ is kept constant (and if $\Delta m \sim \Gamma$). More precisely, the dynamics can be effectively characterised by three parameters: the washout strength parameters
\begin{equation}
    K_i \equiv \frac{\Gamma^{(0)}_i}{H_1}
    \sim \frac{\abs{y_i}^2}{m_1} m_{\rm Pl}
    \label{eq:washout_strength_params}
\end{equation}
and the number of flavour oscillations in a Hubble time: $N_{12} \equiv \abs{\Delta m_{12}}/(2\pi H_1)$. Using $\Gamma^{(0)}_i = |y_i|^2m_1/(8\pi )$ (see appendix~\cref{sec:mom-dep-equations-appendix}), we find that in the benchmark case $K_i \sim \mathcal{O}(10)$ (corresponding to strong washout) along with $N_{12} \approx 1.5$.

These estimates agree qualitatively with what is previously known from the semiclassical Boltzmann approach, where the CP-asymmetry parameter $\epsilon_i^{CP}\propto \sin(2\theta_{12})$ is resonantly enhanced for $\Delta m \sim \Gamma$. The CP-violating angle $\theta_{12}$ was chosen to be maximal in this sense in the benchmark case, but it can be used to adjust the value of the final asymmetry downwards, as it affects the results mainly as an overall scaling factor with only a small impact on the dynamics.

%%%%%%%%%%%%%%%%%%%%%%%%%%%%%%%%%%%%%%%%%%%%%%%%%%%%%%%%%%%%%%%%%%%%%%%%%%%%%%%
%
\subsection{Neutrino distribution functions}
\label{sec:neutrino-results}
%
%%%%%%%%%%%%%%%%%%%%%%%%%%%%%%%%%%%%%%%%%%%%%%%%%%%%%%%%%%%%%%%%%%%%%%%%%%%%%%%

Solving the master equations~\cref{eq:final-numerical-neutrino-equation} accurately requires of the order of hundred discrete momentum variables. Thousands of collision integrals $C^{+++}_{\evec{k}hilj}$ are then needed at each time-step, each of which contains a one-dimensional integral. It is clear that these cannot be feasibly computed during the evaluation. Fortunately, to the order we are working, they can be computed and fitted before solving~\cref{eq:final-numerical-neutrino-equation}. Moreover, the source terms in the neutrino equation~\cref{eq:final-numerical-neutrino-equation} are localised in momenta and temperature, which facilitates the fitting process.

%------------------------------------------------------------------------------
%
\begin{figure}[t!]
    \centering
    \includegraphics{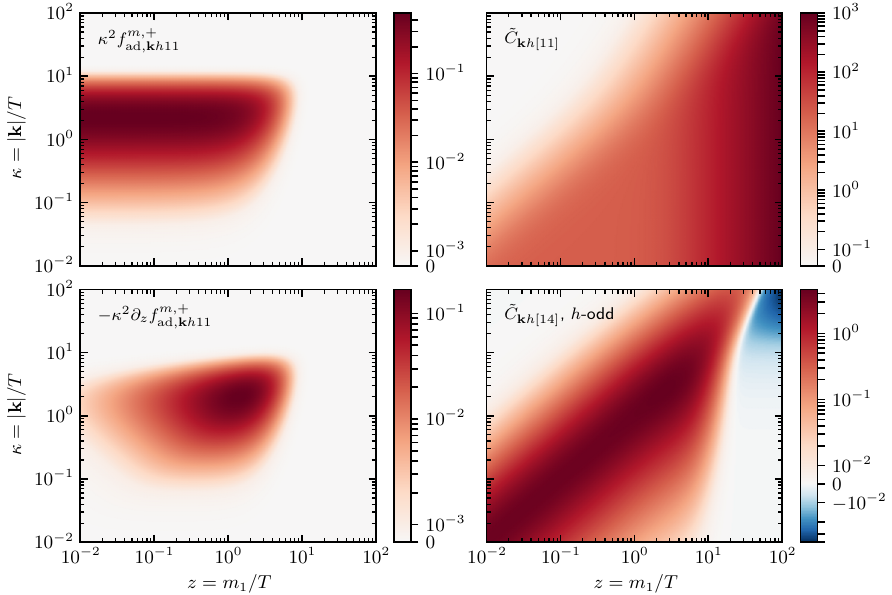}
    \caption{Shown are select intermediate functions of the Majorana neutrino equation~\cref{eq:final-numerical-neutrino-equation}. On the left: the adiabatic distribution function $\kappa^2 \smash{f_{{\rm ad},\evec{k}h11}^{m,{+}} \simeq \kappa^2 f_{\rm FD}(\omega_{\evec{k}1})}$ (top) and the source term $\smash{-\kappa^2 \partial_z f_{{\rm ad},\evec{k}h11}^{m,{+}}}$ (bottom). On the right: the collision term coefficient functions $\tilde C_{\evec{k}h[11]}$, which is the same for both helicities (top), and the helicity-odd combination $(\tilde C_{\evec{k},+1[14]} - \tilde C_{\evec{k},-1[14]})/2$ (bottom).}
    \label{fig:neutrino_fAd_source_Coll}
\end{figure}
%
%------------------------------------------------------------------------------

In the left panels of figure~\cref{fig:neutrino_fAd_source_Coll} we show heat maps of the flavour diagonal adiabatic distribution function~\cref{eq:thermal-limit-of-adiabatic-solution} (top) multiplied by a phase space factor, $\kappa^2 f_{{\rm ad},\evec{k}h11}^{m,+}$, and the corresponding source term (bottom) $-\kappa^2 \dd f_{{\rm ad},\evec{k}h11}^{m,+}/\dd z$ as a functions of $z=m_1/T$ and $\kappa = |\evec{k}|/T$. Both functions are indeed localised around $\abs{\evec{k}} \simeq T$ in momentum and the source term is also localised around $z \simeq 1$, while the distribution function $f_{\rm FD}(\omega_{\evec{k}1})$ exhibits exponential fall off to zero in the same region. Plots for $f_{{\rm ad},\evec{k}h22}^{m,+}$ are qualitatively similar, with the fall off region moving according to the value of $m_2$.

In the right panels of figure~\cref{fig:neutrino_fAd_source_Coll} we show heat maps of $\tilde C_{\evec{k}h[11]} = \,2\Re(C^{+++}_{\evec{k}h111}) \,z/H_1$ and $(\tilde C_{\evec{k},{+1}[14]} - \tilde C_{\evec{k},{-1}[14]})/2 = \Im(C^{+++}_{\evec{k},{+1},121} - C^{+++}_{\evec{k},{-1},121}) \,z/H_1$, given by equations~\cref{eq:collision-term-matrix,eq:dimensionless_osc_and_coll_terms}. The former is the damping rate for the flavour-diagonal distribution function $\delta f_{\evec{k}h11}$, and the latter is the helicity-odd part of the function which couples the flavour diagonal distribution to the off-diagonal distribution $\Im(\delta f_{\evec{k}h12})$. The diagonal damping rate is the same for both helicities in our leading order approximation. Note that the colour bar scales in the figure are logarithmic for both positive and negative directions separately, except for a small region around zero (up to one tick in both directions), where linear scaling is used.

Figure~\cref{fig:neutrino_fAd_source_Coll} was created using the benchmark resonant leptogenesis parameters given in subsection~\cref{sec:parameters}. Note that we have dropped the dispersive self-energy $\Sigma^{\rm H}$ everywhere, as was discussed earlier. All other components of $C^{+++}_{\evec{k}hilj}$ are qualitatively similar to the ones shown here. In particular, the helicity even parts of the off-diagonal coupling functions $2\Re(C^{+++}_{\evec{k}h121})$ and $2\Re(C^{+++}_{\evec{k}h211})$, which are important for leptogenesis, have similar forms to the diagonal damping rates, and in the quasi-degenerate case $m_1 \simeq m_2$ also their scales are similar. Also, when $\Sigma^{\rm H}$ is dropped, the diagonal damping rate $C_{\evec{k}h[ii]}$ is actually just the momentum and temperature dependent decay rate of the Majorana neutrinos. This is because the leading order Majorana neutrino self-energy $\Sigma^{\mathcal{A}}$ used in this work corresponds only to the decay and inverse decay processes.

%------------------------------------------------------------------------------
%
\begin{figure}[t!]
    \centering
    \includegraphics{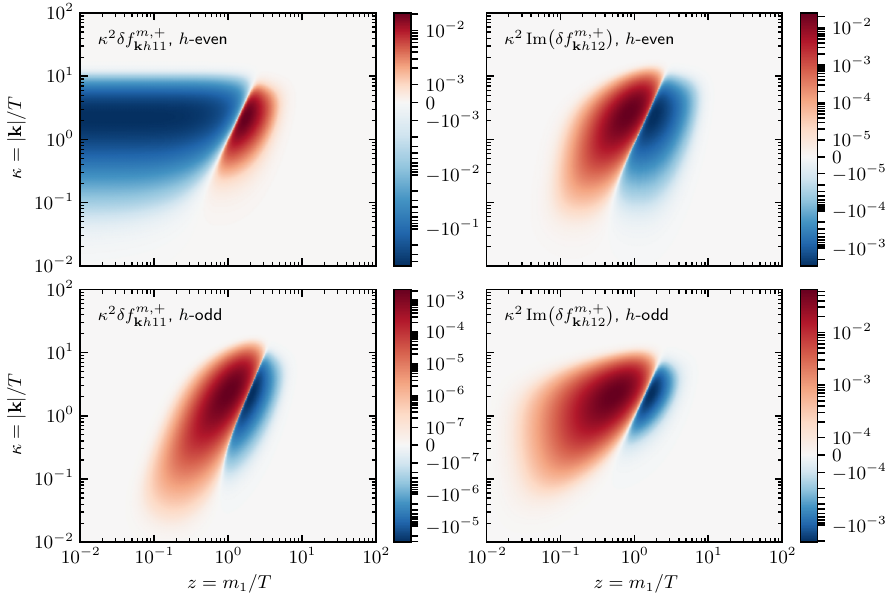}
    \caption{Shown are select components of the numerical solution $\delta f_{\evec{k},hij}^{m,+}$ of the Majorana neutrino equation~\cref{eq:final-numerical-neutrino-equation} with vacuum initial conditions. Left and right sides show the flavour-diagonal function $\delta f_{\evec{k},h11}$ and the off-diagonal function $\Im(\delta f_{\evec{k},h12})$, respectively. Top row features the helicity-even combination $(\delta f_{\evec{k},+1} + \delta f_{\evec{k},-1})/2$ and the bottom row the helicity-odd combination $(\delta f_{\evec{k},+1} - \delta f_{\evec{k},-1})/2$ of the corresponding functions. All functions have been scaled by $\kappa^2$.}
    \label{fig:neutrino_deltaf}
\end{figure}
%
%------------------------------------------------------------------------------

In figure~\cref{fig:neutrino_deltaf} we show similar heat maps of some non-equilibrium distribution functions $\delta f_{\evec{k}hij}^{m,+}$, obtained from a numerical solution of equation~\cref{eq:final-numerical-neutrino-equation}, using the same benchmark parameters as in figure~\cref{fig:neutrino_fAd_source_Coll}. In the top row we show the helicity-even parts of the flavour-diagonal and off-diagonal functions: $\kappa^2 (\delta f_{\evec{k},{+1},11}^{m,+} + \delta f_{\evec{k},{-1},11}^{m,+})/2$ (left panel) and $\kappa^2 \Im(\delta f_{\evec{k},{+1},12}^{m,+} + \delta f_{\evec{k},{-1},12}^{m,+})/2$ (right panel). In the bottom row we show the corresponding helicity-odd parts, $\kappa^2 (\delta f_{\evec{k},{+1},11}^{m,+} - \delta f_{\evec{k},{-1},11}^{m,+})/2$ (left panel) and $\kappa^2 \Im(\delta f_{\evec{k},{+1},12}^{m,+} - \delta f_{\evec{k},{-1},12}^{m,+})/2$ (right panel). We will later see that the helicity-even part of $\Im(\delta f_{\evec{k}h12}^{m,+})$ gives the main contribution to the CP-violating lepton source term. The results show that the off-diagonal components are localised around $z \simeq 1$ and complete approximately one period of flavour oscillation before being exponentially suppressed (the blue colour indicates negative and the red colour positive values). The localisation around $\abs{\evec{k}} \simeq T$ is due to the phase space factor $\kappa^2$ together with the exponential suppression at high momenta. The negative values of the $h$-even diagonal distribution (top left panel), extending to very small $z$ result from the vacuum initial conditions used here. The other independent components of $\delta f_{\evec{k}hij}^{m,+}$ are again qualitatively similar to the ones shown here.

%%%%%%%%%%%%%%%%%%%%%%%%%%%%%%%%%%%%%%%%%%%%%%%%%%%%%%%%%%%%%%%%%%%%%%%%%%%%%%%
%
\subsection{Results on lepton asymmetry}
\label{sec:results}
%
%%%%%%%%%%%%%%%%%%%%%%%%%%%%%%%%%%%%%%%%%%%%%%%%%%%%%%%%%%%%%%%%%%%%%%%%%%%%%%%

Having now established the scales of the problem and that the numerical solution of the neutrino correlation function is under control, we turn to study the lepton asymmetry evolution. Because of the smallness of the lepton chemical potential $\mu_\ell$, we can solve the equations~\cref{eq:final-numerical-neutrino-equation,eq:final-numerical-lepton-equation} sequentially. The neutrino equation is solved first and its solutions $\delta f_{\evec{k}h}$ are used to calculate the lepton source term $\tilde S_{CP}$ and the washout term coefficients $\delta \tilde W$ and $\tilde W_{\rm ad}$ according to equations~\cref{eq:dimensionless_SCP_and_W,eq:cp-source-simplified,eq:washout-1-simplified,eq:washout-2-simplified}. These are then used to solve the lepton equation~\cref{eq:final-numerical-lepton-equation}. Note that equation~\cref{eq:final-numerical-neutrino-equation} does not couple helicities or momenta so we can solve each mode $\delta f_{\evec{k}h}$ separately. In practice we reformulate the neutrino equations in terms of the helicity-even and helicity-odd combinations $(\delta f_{\evec{k},+} \pm \delta f_{\evec{k},-})/2$, which are more convenient for the calculation of $S_{CP}$ and $\delta W$, as described in appendix~\cref{sec:cp-source-and-washout-appendix}.

%%%%%%%%%%%%%%%%%%%%%%%%%%%%%%%%%%%%%%%%%%%%%%%%%%%%%%%%%%%%%%%%%%%%%%%%%%%%%%%
\paragraph{Benchmark case.}
%%%%%%%%%%%%%%%%%%%%%%%%%%%%%%%%%%%%%%%%%%%%%%%%%%%%%%%%%%%%%%%%%%%%%%%%%%%%%%%

%------------------------------------------------------------------------------
%
\begin{figure}[t!]
    \centering
    \includegraphics{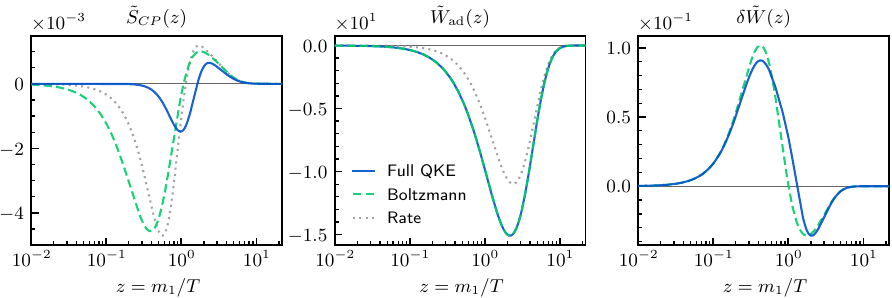}
    \caption{Numerically calculated CP-violating source term $\tilde{S}_{CP}(z)$ and the washout term coefficients $\tilde{W}_{\rm ad}(z)$ and $\delta \tilde{W}(z)$ of our main lepton equation~\cref{eq:final-numerical-lepton-equation} (solid line), Boltzmann equation~\cref{eq:Boltzmann_nL_dimensionless} (dashed line) and rate equation~\cref{eq:integrated_Boltzmann_nL_dimensionless} (dotted line). We used the benchmark parameter values~\cref{eq:benchmark-parameters} and vacuum initial conditions.}
    \label{fig:lepton_SCP_and_W_benchmark}
\end{figure}
%
%------------------------------------------------------------------------------

In section~\cref{sec:neutrino-results} we presented some intermediate results for the phase space functions $\delta f_{\evec{k}h}$ using the vacuum initial condition~\cref{eq:vacuum_initial_value_delta_f} and the benchmark parameter values~\cref{eq:benchmark-parameters}. In figure~\cref{fig:lepton_SCP_and_W_benchmark} we show the source term $\tilde S_{CP}$ and the washout term coefficients $\tilde W_{\rm ad}$,  $\delta \tilde W$ of the lepton equation~\cref{eq:final-numerical-lepton-equation} as functions of $z$ for the same parameters and initial conditions. We find that dividing the dimensionless momentum $\kappa = \abs{\evec{k}}/T$ to 100 bins in logarithmic scale between $10^{-2}$ and $10^4$ already ensures that the results are not sensitive to the cutoff or the discretisation. Indeed, one can see from figures~\cref{fig:neutrino_fAd_source_Coll,fig:neutrino_deltaf} that the largest contributions come from the range $\kappa \in [0.1, 10]$. In figure~\cref{fig:lepton_SCP_and_W_benchmark} we also compare the results from our full quantum kinetic equations (QKEs)~\cref{eq:final-numerical-neutrino-equation,eq:final-numerical-lepton-equation} to those following from the traditional semiclassical Boltzmann equations (BEs) given in~\cref{eq:Boltzmann_fi_dimensionless,eq:Boltzmann_nL_dimensionless} and the corresponding momentum integrated rate equations (REs)~\cref{eq:integrated_Boltzmann_fi_dimensionless,eq:integrated_Boltzmann_nL_dimensionless}. The latter two sets of equations are well known, but we provided them explicitly for completeness. Also, we wrote all equations using a similar notation, which greatly facilitates the comparisons. Both BEs and REs require an externally provided CP-violating parameter $\epsilon_i^{CP}$. At this point we are using $\epsilon_i^{CP}$ corresponding to the ``mixed'' regulator~\cref{eq:epsilon-mixed-regulator}~\cite{Pilaftsis:1997jf,Pilaftsis:2003gt} (see also \eg~\cite{Anisimov:2005hr}). Other regulators will be discussed in detail below.

It turns out that both the BEs and the REs significantly overestimate the source term $\tilde S_{CP}$ in the relativistic region $z \lesssim 1$. The QKE-source starts to grow and changes the sign later, but all sources start to converge for $z\gtrsim 2$. On the other hand, the washout terms $\tilde W_{\rm ad}$ and $\delta \tilde W$ have only minor differences. As expected, $\tilde W_{\rm ad}$ is by far the dominant of the two. It also appears to be identical in the full QKE and in the BE approach, and indeed it is: this term originates from the flavour-diagonal equilibrium part of the Majorana neutrino distributions, which we have calculated to the zeroth order in gradients in the QKEs. The function $\delta \tilde W$, which is proportional to the non-equilibrium perturbation $\delta f$, is approximately two orders of magnitude smaller and could be neglected with practically no effect on the final asymmetry. Also, there is no corresponding function $\delta \tilde W$ in the leading order rate equations~\cref{eq:integrated_Boltzmann_fi_dimensionless,eq:integrated_Boltzmann_nL_dimensionless} because in the Boltzmann approach this part results from the Pauli blocking and stimulated emission factors, which are dropped from the rate equations.

%------------------------------------------------------------------------------
%
\begin{figure}[t!]
    \centering
    \includegraphics{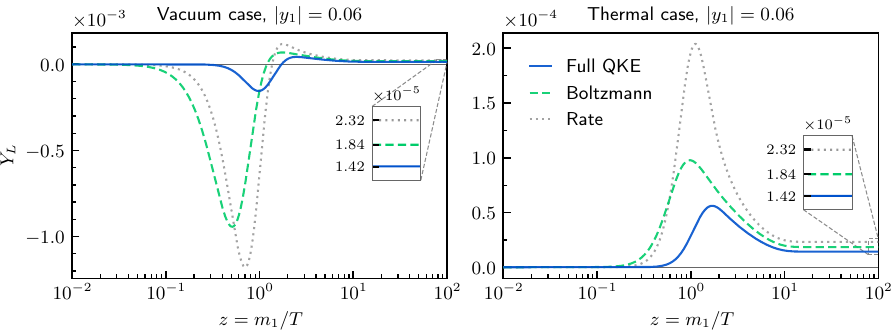}
    \caption{Numerical solutions for the lepton asymmetry $Y_L(z)$ from our main equation~\cref{eq:final-numerical-lepton-equation} (solid line), Boltzmann equation~\cref{eq:Boltzmann_nL_dimensionless} (dashed line) and rate equation~\cref{eq:integrated_Boltzmann_nL_dimensionless} (dotted line). We used the benchmark parameter values~\cref{eq:benchmark-parameters}. The left (right) panel has vacuum (thermal) initial conditions for the Majorana neutrinos. The final values of the asymmetries are shown as insets.}
    \label{fig:lepton_YL_benchmark}
\end{figure}
%
%------------------------------------------------------------------------------

In figure~\cref{fig:lepton_YL_benchmark} we show the lepton asymmetries $Y_L$ as a function of $z$ in the benchmark case. Left panel corresponds to the vacuum initial conditions used above. The asymmetry behaves as expected from the source term $\tilde S_{CP}$ shown in the left panel of figure~\cref{fig:lepton_SCP_and_W_benchmark}. While the $Y_L(z)$-evolution obtained using the BE or the RE deviate strongly from that found using the full QKE, the final asymmetries differ by less than a factor of 2. In the right panel we show the case with thermal initial conditions for the Majorana neutrinos. The early evolution of the asymmetry is of course very different from the vacuum case, but the final asymmetries are identical with both initial conditions. This behaviour is due to the strong washout assumed in the benchmark case ($K_1 \approx 10.0$ and $K_2 \approx 27.9$), which efficiently erases the early evolution of the lepton asymmetry. The final asymmetry then mostly depends on the source at the very end of the integration range, where the $\tilde S_{CP}$ computed using different methods were found to agree better.

%------------------------------------------------------------------------------
%
\begin{figure}[t!]
    \centering
    \includegraphics{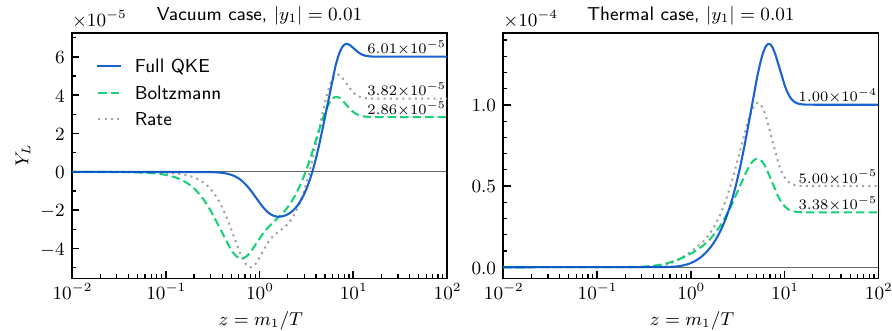}
    \caption{Numerical solutions for the lepton asymmetry $Y_L(z)$, with the same setup and mass difference as figure~\cref{fig:lepton_YL_benchmark} except $\abs{y_1} = 0.01$. The final values of the asymmetries are indicated on the corresponding graphs.}
    \label{fig:lepton_YL_with_Y1_001}
\end{figure}
%
%------------------------------------------------------------------------------

In figure~\cref{fig:lepton_YL_with_Y1_001} we show for comparison similar plots in the weak washout case, with $\abs{y_1} = 0.01$, corresponding to $K_1 \approx 0.279$. The left panel again corresponds to the vacuum and the right panel to thermal initial conditions. Now the early evolution deviates less in different approaches. Also, as expected, the final asymmetries are no longer the same for vacuum and thermal initial conditions. The differences between the BE and RE predictions and our QKE results for the final asymmetry also remain significant. These results show that the lepton asymmetry evolution and its asymptotic value depend in an essential way on the model parameters. Also, it is clear that to obtain accurate results, one should use the full QKEs instead of the less accurate BE- or RE-approaches.

%%%%%%%%%%%%%%%%%%%%%%%%%%%%%%%%%%%%%%%%%%%%%%%%%%%%%%%%%%%%%%%%%%%%%%%%%%%%%%%
\paragraph{Varying mass difference.}
%%%%%%%%%%%%%%%%%%%%%%%%%%%%%%%%%%%%%%%%%%%%%%%%%%%%%%%%%%%%%%%%%%%%%%%%%%%%%%%

\begin{figure}[t!]
    \centering
    \includegraphics{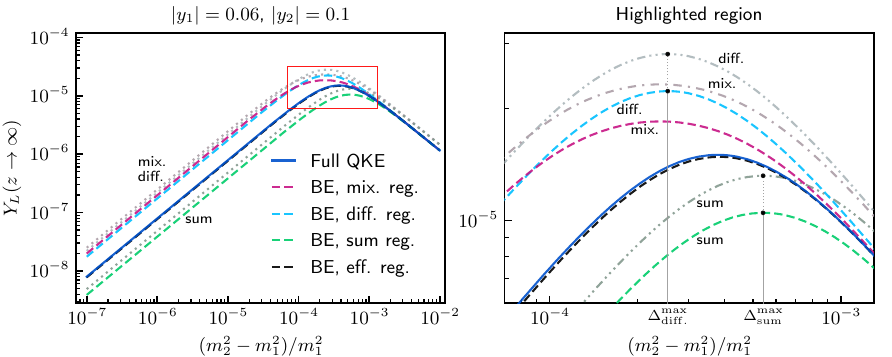}\par
    \vspace{0.7em}
    \includegraphics{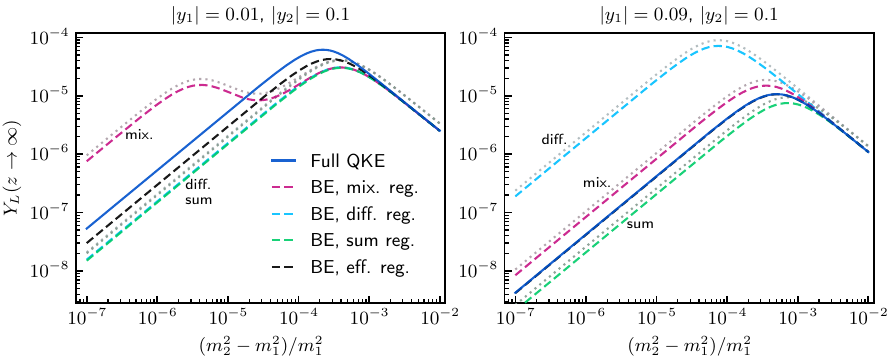}
	\caption{Shown is the asymptotic lepton asymmetry $Y_L$ as a function of the relative mass-squared difference $\Delta m^2_{21}/m_1^2$ for different Yukawa couplings. The Boltzmann equation (BE) results are shown with four different regulators for $\epsilon^{CP}_i$. The dotted and dash-dotted lines show the rate equation results for the corresponding BE results. The top-right panel shows the region highlighted in red in the left. The bottom panels have hierarchical (left) and almost degenerate (right) Yukawa couplings. Other parameters have the benchmark values and vacuum initial conditions were used.}
	\label{fig:final_YL_vs_massdiff}
\end{figure}

In the top panels of figure~\cref{fig:final_YL_vs_massdiff} we show the asymptotic lepton asymmetry as a function of $(m_2^2 - m_1^2)/m_1^2 \approx 2\Delta m_{21}/m_1$, keeping other benchmark parameters fixed. We again compare the full QKE results with the BE and RE predictions, now for four different CP-asymmetry parameters $\epsilon^{CP}_i$ that have been discussed in the literature~\cite{Anisimov:2005hr,Garny:2011hg,Garbrecht:2011aw,Dev:2014laa,Garbrecht:2014aga,Dev:2017wwc}. The different choices, which we denote as `mixed', `difference', `sum' and `effective' vary in how the resonance near $m_1 = m_2$ is regulated. Explicit forms for the $\epsilon^{CP}_i$-parameter and the regulators are given in appendix~\cref{sec:CP_parameter}. All approaches give qualitatively similar $\Delta m_{21}$-dependence with a single maximum at $\Delta m_{21} \sim \Gamma$. However, a more close look reveals significant quantitative differences between the full QKE results and the BE and RE approximations, as well as between using different regulators in the latter two approaches.

%------------------------------------------------------------------------------
%
%
%------------------------------------------------------------------------------

The location of the resonance peak is approximatively given by $\Delta m^2_{21} = \abs{m_1 \Gamma_1^{(0)} \mp m_2 \Gamma_2^{(0)}}$ for the difference and sum regulators in the Boltzmann approaches. We have shown these locations in the top-right panel of figure~\cref{fig:final_YL_vs_massdiff}, denoting them by $\Delta^{\rm max}_{\rm diff.}$ and $\Delta^{\rm max}_{\rm sum}$. The BE and RE results using different regulators fall either above or below the correct QKE-result shown by the solid blue line, varying by an almost order of magnitude for $\Delta m_{21} \lesssim \Gamma$. The effective sum regulator given in~\cite{Dev:2017wwc} (see equation~\cref{eq:effective_sum_regulator} below) is designed to work in the strong washout case; it is thus not surprising that it works best in our benchmark case. On the other hand, for $\Delta m_{21} > \Gamma$, where we enter the rapid flavour oscillation regime, all results converge. This is expected, since the regulators become irrelevant in the CP-parameter~\cref{eq:epsilon-type-CP-asymmetry} and the diagonal elements decouple from the off-diagonals in the QKEs in this limit (we will show this explicitly below). We also observe that the BEs always give a slightly lower final asymmetry than do the REs, as was also observed in~\cite{Basboll:2006yx}. The difference between the BE and RE results is smaller, however, than the difference arising from using different regulators and eventually the correct QKEs. That is, treating the quantum physics part of the problem correctly is more important than the momentum dependence of the phase space distributions.

The situation gets even more interesting when we begin to vary the couplings. In the bottom panels of figure~\cref{fig:final_YL_vs_massdiff} we show the results for hierarchical (left panel) and for almost degenerate (right panel) Yukawa couplings. In the left panel the washout is weaker, $K_1 \approx 0.279$ and $K_2 \approx 27.9$, whereas in the right panel the washout is strong, $K_1 \approx 22.6$. The different CP-asymmetry regulators now lead to even more dispersion in the BE and the RE results when $\Delta m_{21} \lesssim \Gamma$. In the hierarchical case using the mixed regulator leads to two peaks at $\Delta m_{21} \simeq \Gamma_1$ and $\Delta m_{21} \simeq \Gamma_2$, corresponding to the different regulators used for the two Majorana neutrinos in this case. The full QKE result, again shown by the solid blue line, shows no such structure. Also the effective regulator is somewhat less accurate here. In the right panel, with almost degenerate Yukawas, the mixed and sum regulators are in better agreement with the QKEs, but the difference regulator has an extra spurious enhancement because the regulator vanishes and the CP-asymmetry is unbounded in the double limit $m_2 \to m_1$ and $\Gamma_2 \to \Gamma_1$.

The main take-home message from this section is that the Boltzmann equation and the rate equation approaches are inaccurate and strongly sensitive to the choice of the regulator in the resonant and quasidegenerate region $\Delta m_{21} \lesssim \Gamma_i$. The Boltzmann approach reproduces the full QKE results accurately in all cases only in the region $\Delta m_{21} > \Gamma_i$ when the regulator is already negligible and one is approaching the hierarchical mass limit. Also, the most accurate regulator over varying coupling strengths is the sum regulator of~\cite{Garny:2011hg}. We shall show below how the sum regulator indeed consistently emerges when we reduce the QKEs to BEs in the helicity-symmetric decoupling limit.

%%%%%%%%%%%%%%%%%%%%%%%%%%%%%%%%%%%%%%%%%%%%%%%%%%%%%%%%%%%%%%%%%%%%%%%%%%%%%%%
\paragraph{Flavour oscillation.}
%%%%%%%%%%%%%%%%%%%%%%%%%%%%%%%%%%%%%%%%%%%%%%%%%%%%%%%%%%%%%%%%%%%%%%%%%%%%%%%

\begin{figure}[t!]
    \centering
    \includegraphics{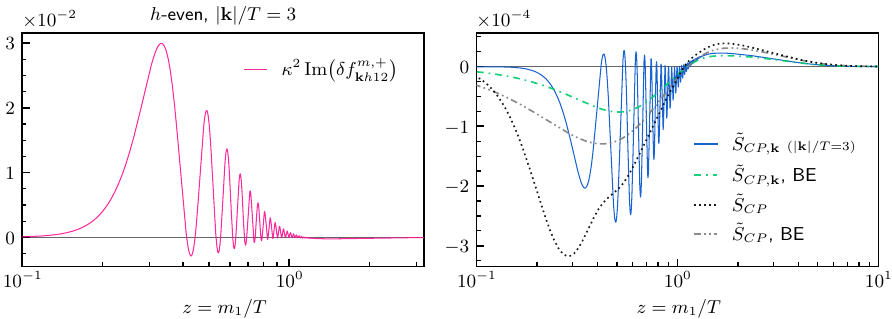}
    \caption{Flavour oscillation of $\Im(\delta f_{\evec{k},h12}^{m,+})(z)$ (helicity-even component) and $\tilde S_{CP,\evec{k}}(z)$ for a single $\evec{k}$-mode, with $\Delta m_{21} = 0.01 m_1$.
    % $(m_2^2 - m_1^2)/m_1^2 = 0.02$
    Other parameters have the benchmark values, and vacuum initial conditions were used. On the right we show the single $\evec{k}$-mode as well as the integrated lepton asymmetry source from the full QKEs and the BEs.}
    \label{fig:deltaf_and_SCP_flavour_oscillation}
\end{figure}

Let us look more closely at the role of the flavour oscillations in resonant leptogenesis. In the left panel of figure~\cref{fig:deltaf_and_SCP_flavour_oscillation} we show the evolution of the helicity-even part of the off-diagonal Majorana neutrino phase space function $\Im(\delta f_{\evec{k}h12}^{m,+})$ with $\abs{\evec{k}} = 3T$ and $\Delta m_{21} = 0.01 m_1$. This is a case with rapid flavour oscillations corresponding to $N_{12} \approx 110$. The modes of this $\delta f$-component in the range $\abs{\evec{k}}/T \sim$ 0.1--10 give the dominant contribution to the integrated lepton source term. In the right panel we plot the contribution $\tilde S_{CP,\evec{k}}$ of the same mode to the lepton source term, normalised according to $\tilde S_{CP} \equiv \int \dd^3 \evec{k}/(2 \pi)^3 \tilde S_{CP,\evec{k}}/T^3$. Both the mode and its contribution to the lepton source display a strong oscillation pattern with a quickly dying amplitude. This decay of the oscillations is precisely the reason for the emergence of the semiclassical limit (shown as the green dash-dotted line) from the QKEs. We will make this explicit in equation~\cref{eq:ImInt} below. We show also in the right panel the full integrated source term from the QKEs (dotted line) and from the Boltzmann approach with the sum regulator (dash-dot-dotted line). Even in the QKE-result all oscillations are smoothed out in the integrated source, due to the phase differences between different modes.

%------------------------------------------------------------------------------
%
%
%------------------------------------------------------------------------------

%%%%%%%%%%%%%%%%%%%%%%%%%%%%%%%%%%%%%%%%%%%%%%%%%%%%%%%%%%%%%%%%%%%%%%%%%%%%%%%
%
\section{Helicity-symmetric approximation}
\label{sec:helicity_symmetric_equations}
%
%%%%%%%%%%%%%%%%%%%%%%%%%%%%%%%%%%%%%%%%%%%%%%%%%%%%%%%%%%%%%%%%%%%%%%%%%%%%%%%

In this section we will derive a series of approximations to the QKEs~\cref{eq:delta-f-m-equation} (or equivalently~\cref{eq:final-numerical-neutrino-equation}), eventually reducing them to the semiclassical Boltzmann limit. This process also leads to a simplified source term in the asymmetry equation~\cref{eq:lepton-asymmetry-final}, which eventually reproduces the CP-asymmetry parameter with the sum regulator.

Looking more closely at equations~\cref{eq:cp-source-simplified,eq:cp-source}, one can see that the lepton asymmetry is sourced mainly by the helicity-even combination of the imaginary part of the off-diagonal function $\delta f_{\evec{k}h12}$ in the non-relativistic or mildly relativistic case. Based on this observation, we drop the tracking of the helicity asymmetry in the equations for $\delta f_{\evec{k}hij}$. We can then write a simpler set of equations for the $h$-even part, which we denote simply by $\delta f_{\evec{k}ij}$ henceforth, and a simpler form for the lepton source term $S_{CP}$ including only $\Im(\delta f_{\evec{k}12})$. We will also work with vacuum dispersion relations, setting $\Sigma^{\rm H} \equiv 0$. The resulting \emph{helicity-symmetric} equations are
\begin{align}
    \partial_t \delta f_{\evec{k}11} &= -\Gamma_{\evec{k}11} \,\delta f_{\evec{k}11}
    - \partial_t f_{{\rm ad},\evec{k}11}^{(0)} - \Gamma_{\evec{k}12} \Re(\delta f_{\evec{k}12})
    \text{,} \label{eq:diag1_equation_approx}
    \\*
    \partial_t \delta f_{\evec{k}22} &= -\Gamma_{\evec{k}22} \,\delta f_{\evec{k}22}
    - \partial_t f_{{\rm ad},\evec{k}22}^{(0)} - \Gamma_{\evec{k}21} \Re(\delta f_{\evec{k}12})
    \text{,} \label{eq:diag2_equation_approx}
    \\*
    \partial_t \delta f_{\evec{k}12} &= -\bar\Gamma_{\evec{k}12} \,\delta f_{\evec{k}12}
    - \im \Delta\omega_{\evec{k}12} \,\delta f_{\evec{k}12} - \smash{\frac{1}{2}} \bigl(
        \Gamma_{\evec{k}21} \,\delta f_{\evec{k}11} + \Gamma_{\evec{k}12} \,\delta f_{\evec{k}22}
    \bigr) \text{,} \label{eq:offd_equation_approx}
\shortintertext{and}
    S_{CP} &= -2 \,\frac{\Im(y_1^* y_2)}{\Re(y_1^* y_2)} \int \frac{\dd^3 \evec{k}}{(2\pi)^3}
    \bigl(\Gamma_{\evec{k}21} + \Gamma_{\evec{k}12}\bigr) \Im\bigl(\delta f_{\evec{k}12}\bigr)
    \text{,} \label{eq:SCP_approximation}
\end{align}
where $\bar\Gamma_{\evec{k}12} \equiv (\Gamma_{\evec{k}11} + \Gamma_{\evec{k}22})/2$ and $\Gamma_{\evec{k}il} \equiv 2 \Re(C^{+++}_{\evec{k}hilj})\rvert_{\Sigma^H \to 0}$ with $C^{+++}_{\evec{k}hilj}$ given by equation~\cref{eq:Coll_trace_function}. Note that all $\Gamma_{\evec{k}il}$ are now real so the diagonal functions $\delta f_{\evec{k}ii}$ couple directly only to $\Re(\delta f_{\evec{k}12})$. The diagonal damping rate admits the factorisation $\Gamma_{\evec{k}ii} = (m_i/\omega_{\evec{k}i}) \smash{\Gamma_i^{(0)}} \mathcal{X}_{\evec{k}i}$ where $m_i/\omega_{\evec{k}i}$ is the time dilation factor, $\smash{\Gamma_i^{(0)}} = \abs{y_i}^2 m_i/(8\pi)$ is the tree-level vacuum decay width of the Majorana neutrino and $\mathcal{X}_{\evec{k}i}$ is the thermal quantum statistical correction factor, which obeys $\mathcal{X}_{\evec{k}i} \to 1$ when $\abs{\evec{k}}/T \to \infty$ (see equation~\cref{eq:Boltzmann_damping_rate}). The damping rate $\bar\Gamma_{\evec{k}12}$ in the off-diagonal equation, given by the average of the diagonal rates, agrees with the flavour coherence damping rate found in the density matrix formalism for mixing neutrinos~\cite{Enqvist:1990ad,Kainulainen:1990ds,Enqvist:1991qj}. This is an expected result, as the two phenomena are of course closely related. Indeed, if we further assume that $\Gamma_{\evec{k}21} \simeq \Gamma_{\evec{k}12}$, we can write equations~\cref{eq:diag1_equation_approx,eq:diag2_equation_approx,eq:offd_equation_approx} in a simple density matrix form:
\begin{equation}
    \partial_t \delta f_\evec{k} = -\partial f^{(0)}_{{\rm ad},\evec{k}}
    - \im \comm{H_\evec{k}, \delta f_\evec{k}}
    - \frac{1}{2} \anticomm{\Gamma_\evec{k}, \delta f_\evec{k}}
    \text{,} \label{eq:density-matrix-form}
\end{equation}
where $H_\evec{k} \equiv \diag(\omega_{\evec{k}1}, \omega_{\evec{k}2})$. A similar equation has been used earlier to study light neutrino mixing in the early universe~\cite{Enqvist:1990ad,Barbieri:1990vx,Enqvist:1991qj,Sigl:1992fn} and in the resonant leptogenesis context it was first derived in~\cite{Garbrecht:2011aw}.%
\footnote{In the quasidegenerate limit and restricted to two neutrino flavours, it is easy to generalise equation~\cref{eq:density-matrix-form} to include helicity as in~\cite{Garbrecht:2011aw}. We do not write such an equation explicitly here, because it would be but a further special case of our general QKE~\cref{eq:delta-f-m-equation} for the flavour mixing problem.}

\begin{figure}[t!]
    \centering
    \includegraphics{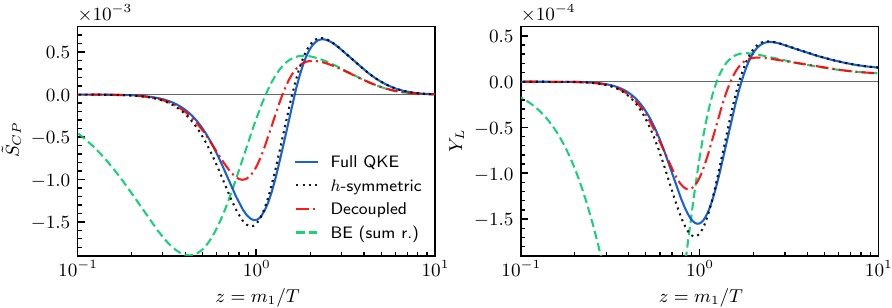}
    \caption{Comparison of the full results (solid line) to the helicity-symmetric approximation (dotted), decoupling limit (dash-dotted) and Boltzmann results with the sum regulator (dashed). We show the results for the lepton asymmetry source term (left) and the lepton asymmetry itself (right) as functions of $z$. The benchmark parameters with vacuum initial conditions were used.}
    \label{fig:approximation_comparison}
\end{figure}

The helicity-symmetric equations~\cref{eq:diag1_equation_approx,eq:diag2_equation_approx,eq:offd_equation_approx,eq:SCP_approximation} provide an excellent approximation to the full QKEs with helicity-even initial conditions, as can be seen in figure~\cref{fig:approximation_comparison} where we compare the two for our benchmark parameters. This is so because the neutrino source terms are helicity-symmetric and the helicity-asymmetry is only generated by loop effects. Also, it is evident from~\cref{eq:cp-source-simplified} that the contribution to the source from the helicity-odd $\delta f^m_{\evec{k}}$-function is thermally suppressed compared to the one coming from helicity-even part. This shows that resonant leptogenesis is dominated by the helicity-independent flavour mixing, although this conclusion partly relies on the fact that the asymmetry is mostly generated in the non-relativistic regime $T \lesssim m_1$. Helicity would play a more important role if the Majorana neutrinos were relativistic when the asymmetry is generated, like in some low scale leptogenesis scenarios~\cite{Klaric:2020lov}. Indeed, equations similar to~\cref{eq:density-matrix-form}, but keeping the helicity degree of freedom, were recently used to study leptogenesis~\cite{Drewes:2016gmt,Klaric:2020lov,Eijima:2020shs}. Our full QKEs~\cref{eq:delta-f-equation} as well as the mass-shell equations~\cref{eq:delta-f-m-equation} are more general than~\cref{eq:density-matrix-form} and the helicity dependent equations in~\cite{Klaric:2020lov,Eijima:2020shs,Klaric:2021cpi}. Note in particular, that for more than two Majorana neutrino flavours the collision terms in~\cref{eq:delta-f-m-equation} cannot in general be reduced to the canonical form for a density matrix equation.

%------------------------------------------------------------------------------
%
%
%------------------------------------------------------------------------------

%%%%%%%%%%%%%%%%%%%%%%%%%%%%%%%%%%%%%%%%%%%%%%%%%%%%%%%%%%%%%%%%%%%%%%%%%%%%%%%
\paragraph{Decoupling limit.}
%%%%%%%%%%%%%%%%%%%%%%%%%%%%%%%%%%%%%%%%%%%%%%%%%%%%%%%%%%%%%%%%%%%%%%%%%%%%%%%

Equations~\cref{eq:diag1_equation_approx,eq:diag2_equation_approx,eq:offd_equation_approx,eq:SCP_approximation} still incorporate all essential flavour mixing consistent with full resummation of the interaction terms. Now we simplify these equations further in the case where the flavour oscillations are fast ($\Delta m_{21} \gg \Gamma$). We use the same reasoning as in section~\cref{sec:mass_shell_equation} (see footnote~\cref{footnote:fc-elimation}) to argue for dropping the flavour off-diagonal terms in the diagonal equations~\cref{eq:diag1_equation_approx,eq:diag2_equation_approx}. We call this approximation the \emph{decoupling limit}. The diagonal equations, written with the expansion of the universe, are then identical to the semiclassical Boltzmann equation~\cref{eq:Boltzmann_fi_dimensionless}. The washout term can also be approximated with the Boltzmann version or even with only the $W_{\rm ad}$ contribution~\cref{eq:Boltzmann_washout_B}. Because the flavour-diagonal functions now decouple from the off-diagonal ones, they can be solved independently and their solutions can be treated as external sources to the off-diagonal function $\delta f_{\evec{k}12}$.

Assuming that initially $\delta f_{\evec{k}12}(t_0) = 0$ (a non-zero initial value could be easily added as a special solution to the homogeneous equation), the off-diagonal differential equation~\cref{eq:offd_equation_approx} can be integrated to give
\begin{equation}
    \delta f_{\evec{k}12}(t) = -\frac{1}{2} \int_{t_0}^t \dd u \,
    \bigl(\Gamma_{21} \delta f_{11} + \Gamma_{12} \delta f_{22}\bigr)_{\evec{k}}(u)
    \exp\biggl[
        -\int_u^t \dd v \,\Bigl(\bar\Gamma_{12} + \im \Delta\omega_{12}\Bigr)_{\evec{k}}(v)
    \biggr] \text{.}
\end{equation}
Substituting this into equation~\cref{eq:SCP_approximation} we then get a closed formula for $S_{CP}$ which now \emph{defines} the source term in the semiclassical Boltzmann equations; indeed comparing to the Boltzmann formula~\cref{eq:Boltzmann_SCP}, this improved form shows that the combination $\Im(y_1^* y_2) \Im(\delta f_{\evec{k}12})$ acts as an effective dynamical CP-asymmetry parameter.

In the quasidegenerate case $m_1 \simeq m_2$ (\ie~at this point we assume the weakly resonant regime $\Gamma \ll \Delta m_{21} \ll m_1$) we can further approximate $\Gamma_{\evec{k}ij} \simeq \Re(y_i^* y_j) \Gamma_{\evec{k}jj}/\abs{y_j}^2$ and $\Gamma_{\evec{k}11}/\abs{y_1}^2 \simeq \Gamma_{\evec{k}22}/\abs{y_2}^2$. Then we can write the lepton source term into an even more suggestive form:
\begin{align}
    S_{CP}(t) \approx \sum_{\substack{i = 1,2 \\ (j \neq i)}} & \Gamma_i^{(0)} g_i
    \int \frac{\dd^3 \evec{k}}{(2\pi)^3} \Bigl(\frac{m_i}{\omega_{\evec{k}i}} \mathcal{X}_{\evec{k}i}\Bigr)
    \frac{\Im\bigl[(y_1^* y_2)^2\bigr]}{\abs{y_1}^2 \abs{y_2}^2} \notag
    \\*[-0.6em]
    {}\times{}& \frac{1}{4} \int_{t_0}^t \dd u \,
    \bigl(\Gamma_{jj}(\delta f_{11} + \delta f_{22})\bigr)_{\evec{k}}(u)
    \Im\exp\biggl[
        -\int_u^t \dd v \,\Bigl(\bar\Gamma_{12} + \im \Delta\omega_{12}\Bigr)_{\evec{k}}(v)
    \biggr]\text{.}
    \label{eq:SCP-approxfminus1}
\end{align}
This result shows that the lepton asymmetry is cumulatively sourced by the non-equilibrium perturbations in the diagonal mass-shell functions $\delta f_{ii}(u)$, such that past contributions are suppressed by the flavour coherence damping rate $\bar \Gamma_{12}$. We show the approximation~\cref{eq:SCP-approxfminus1} for $S_{CP}$ (with the diagonal solutions $\delta f_{ii}$ calculated from the ordinary Boltzmann equations) and the resulting lepton asymmetry by the red dash-dotted lines in figure~\cref{fig:approximation_comparison}. Even though in the figure we used the benchmark parameters where $\Delta m_{21} \simeq \Gamma$, equation~\cref{eq:SCP-approxfminus1} is still a relatively good approximation to the full QKEs, in particular at early times $z \lesssim 1$. However, when $z \gtrsim 1$ it deviates from the full result, and at late times $z \gg 1$ the asymmetry coincides perfectly with the Boltzmann result (see below) instead, shown by the dashed line in figure~\cref{fig:approximation_comparison}.

%%%%%%%%%%%%%%%%%%%%%%%%%%%%%%%%%%%%%%%%%%%%%%%%%%%%%%%%%%%%%%%%%%%%%%%%%%%%%%%
%
\subsection{Boltzmann limit}
%
%%%%%%%%%%%%%%%%%%%%%%%%%%%%%%%%%%%%%%%%%%%%%%%%%%%%%%%%%%%%%%%%%%%%%%%%%%%%%%%

To get an even closer comparison to the existing literature, we now make stronger approximations to evaluate the time integrals in equation~\cref{eq:SCP-approxfminus1} analytically. Indeed, if one assumes that the source functions are roughly constant, one can take them outside the $u$-integral.%
\footnote{This is consistent if the damping time is much shorter than the time of variation of the diagonal $\delta f_{ii}$-functions, which is given by the Hubble time. That is, an approximation of the type of equation~\cref{eq:SCP-approxf} is valid in the strong damping limit: $\bar\Gamma_{12} \gg H$.}
If one further assumes that $\bar \Gamma_{\evec{k}12}$ and $\Delta \omega_{\evec{k}12}$ are constants (we take $\mathcal{X}_{\evec{k}i} \to 1$ in $\Gamma_{\evec{k}ii}$, assuming $\Gamma_{\evec{k}ii} \simeq (m_i/\omega_{\evec{k}i}) \smash{\Gamma_i^{(0)}}$, and neglect the Hubble expansion for the mass and momentum), one can perform the integrals to get
\begin{multline}
    \Im \int_{t_0}^t \dd u \,\exp\bigl[-\bigl(\bar\Gamma_{12} + \im \Delta\omega_{12}\bigr)(t - u)\bigr]
    \\*
    \simeq \frac{-\Delta\omega_{12}}{(\Delta \omega_{12})^2 + (\bar\Gamma_{12})^2}
    \Biggl[
        \,1 - \e^{-\bar\Gamma_{12}(t - t_0)}
        \biggl(
            \cos\bigl[\Delta\omega_{12}(t - t_0)\bigr]
            + \bar\Gamma_{12} \frac{\sin\bigl[\Delta \omega_{12}(t - t_0)\bigr]}
            {\Delta\omega_{12}}
        \biggr)
    \Biggr] \text{.}
    \label{eq:ImInt}
\end{multline}
A similar expression was also found in~\cite{DeSimone:2007gkc}, but without the exponential damping factor in the oscillating term. This is important, because now we see that taking the limit $t_0 \to -\infty$, the oscillating part is damped to zero.%
\footnote{Authors of ref.~\cite{DeSimone:2007gkc} argued that the second term in~\cref{eq:ImInt} vanishes due to averaging out over oscillations, but this is not the correct explanation; note that~\cref{eq:ImInt} is valid even in the limit $\Delta \omega_{12} \to 0$. Also, a similar expression and including the damping, albeit with a different damping factor $(\Gamma_{11} - \Gamma_{22})/2$ (giving rise to the `difference' regulator), was found in~\cite{Garbrecht:2011aw}.}
We already saw this effect in figure~\cref{fig:deltaf_and_SCP_flavour_oscillation}: while the source was there computed using the full QKE, the strongly damped rapid oscillation we observed corresponds to the second term in equation~\cref{eq:ImInt}.

Substituting~\cref{eq:ImInt} back to equation~\cref{eq:SCP-approxfminus1} and dropping the damped oscillating term, one finds
\begin{align}
    S_{CP}(t) \approx \sum_{\substack{i = 1,2 \\ (j \neq i)}} \Gamma_i^{(0)} g_i
    \int \frac{\dd^3 \evec{k}}{(2\pi)^3} & \Bigl(\frac{m_i}{\omega_{\evec{k}i}} \mathcal{X}_{\evec{k}i}\Bigr)
    \frac{1}{2} \bigl(\delta f_{11} + \delta f_{22}\bigr)_\evec{k}(t) \notag
    \\*[-1.2em]
    &{}\times{} \frac{\Im\bigl[(y_1^* y_2)^2\bigr]}{\abs{y_1}^2 \abs{y_2}^2} \frac{1}{2}
    \biggl[
        \frac{-\Delta\omega_{12} \Gamma_{jj}}{(\Delta\omega_{12})^2 + (\bar\Gamma_{12})^2}
    \biggr]_\evec{k} \text{.}
    \label{eq:SCP-approxf}
\end{align}
This form already greatly resembles the Boltzmann result~\cref{eq:Boltzmann_SCP} with the constant CP-asymmetry parameter~\cref{eq:epsilon-type-CP-asymmetry}. It is now also clear that the CP-asymmetry is regulated by the \emph{coherence damping rate} $\bar \Gamma_{12}$ in the degenerate limit $m_2 \to m_1$. We also note that dropping the damped oscillating term is the main reason why the BE approach tends to initially overestimate the asymmetry. This effect is clearly visible in figures~\cref{fig:lepton_YL_benchmark,fig:lepton_SCP_and_W_benchmark,fig:approximation_comparison}.

We can go even further and extract the CP-asymmetry parameter by using again the quasidegeneracy $m_2 \simeq m_1$, whereby $\Delta\omega_{\evec{k}12} \simeq \Delta m^2_{12}/(2\omega_{\evec{k}i})$, and taking into account the time dilation factor in the damping rate $\Gamma_{\evec{k}ii} = (m_i/\omega_{\evec{k}i}) \smash{\Gamma_i^{(0)}}$. The result~\cref{eq:SCP-approxf} can then be written as
\begin{gather}
    S_{CP}(t) \approx \sum_{i = 1,2} \Gamma_i^{(0)} g_i \,\epsilon^{CP}_{i,{\rm sum}}
    \int \frac{\dd^3 \evec{k}}{(2\pi)^3} \Bigl(\frac{m_i}{\omega_{\evec{k}i}} \mathcal{X}_{\evec{k}i}\Bigr)
    \frac{1}{2} \bigl(\delta f_{11} + \delta f_{22}\bigr)_\evec{k}(t) \text{,}
    \label{eq:SCP-approx2}
\shortintertext{where (for $j \neq i$)}
    \epsilon^{CP}_{i,{\rm sum}} =
    \frac{\Im\bigl[(y_1^* y_2)^2\bigr]}{\abs{y_1}^2 \abs{y_2}^2}
    \frac{(m_2^2 - m_1^2) m_i \Gamma_j^{(0)}}
    {(m_2^2 - m_1^2)^2 + \bigl(m_1 \Gamma_1^{(0)} + m_2 \Gamma_2^{(0)}\bigr)^2}
    \text{.} \label{eq:sum_epsilon}
\end{gather}
Equation~\cref{eq:SCP-approx2} is still different from the Boltzmann result~\cref{eq:Boltzmann_SCP} in its dependence
on the flavour-diagonal functions $\delta f_{ii}$. But using the quasidegeneracy argument once more to approximate $(m_1/\omega_{\evec{k}1}) \mathcal{X}_{\evec{k}1} \simeq (m_2/\omega_{\evec{k}2}) \mathcal{X}_{\evec{k}2}$, it can finally be written exactly in the same form as~\cref{eq:Boltzmann_SCP}.

%%%%%%%%%%%%%%%%%%%%%%%%%%%%%%%%%%%%%%%%%%%%%%%%%%%%%%%%%%%%%%%%%%%%%%%%%%%%%%%
\paragraph{Strong washout limit.}
%%%%%%%%%%%%%%%%%%%%%%%%%%%%%%%%%%%%%%%%%%%%%%%%%%%%%%%%%%%%%%%%%%%%%%%%%%%%%%%

In articles~\cite{Garbrecht:2014aga,Dev:2017wwc,Iso:2014afa} the strong washout limit was considered, where one assumes that the diagonal rate parameters are large separately. In this limit one can find an approximative late-time solution by putting all derivative terms to zero on the left-hand side of equations~\cref{eq:diag1_equation_approx,eq:diag2_equation_approx,eq:offd_equation_approx}. One can then solve all distribution functions algebraically, and eventually the CP-violating source term~\cref{eq:SCP_approximation} takes the form
\begin{align}
    S_{CP} \approx \frac{\Im(y_1^* y_2)}{\Re(y_1^* y_2)} \int \frac{\dd^3 \evec{k}}{(2\pi)^3}
    & \bigl(\Gamma_{21} + \Gamma_{12}\bigr)_\evec{k}
    \bigl(\Gamma_{21} \delta f_{11} + \Gamma_{12} \delta f_{22}\bigr)_\evec{k} \notag
    \\*
    &{}\times{} \Biggl[
        \frac{-\Delta\omega_{12}}
        {(\Delta\omega_{12})^2 + (\bar\Gamma_{12})^2\Bigl(
            1 - \frac{\Gamma_{12}\Gamma_{21}}{\Gamma_{11}\Gamma_{22}}
        \Bigr)}
    \Biggr]_\evec{k} \text{.}
    \label{eq:late_time_SCP_approx}
\end{align}
No other approximations have yet been made at this point. In the quasidegenerate limit this source then has the same form as~\cref{eq:SCP-approxf} except that the off-diagonal backreaction to the diagonal functions is taken into account by a modification of the regulator term $\bar\Gamma_{12}$.%
\footnote{The diagonal distributions in equation~\cref{eq:late_time_SCP_approx} contain no backreaction from the off-diagonals, and technically they correspond to the late-time approximations $\delta f_{\evec{k}ii}^{(\rm L)} \equiv - \partial_t f_{{\rm ad},\evec{k}ii}^{(0)}/\Gamma_{\evec{k}ii}$. However, we have improved this approximation by using here the decoupled diagonal distributions, \ie~the usual diagonal distributions $\delta f_{\evec{k}ii}$ solved from the Boltzmann equations. This gives the same late-time limit, but a more accurate early time evolution.}
This approximation can be again reduced to the Boltzmann equation with yet another effective CP-asymmetry parameter, similar to equation~\cref{eq:sum_epsilon}, but with
\begin{equation}
    \bigl(m_1 \Gamma_1^{(0)} + m_2 \Gamma_2^{(0)}\bigr)^2 \quad \longrightarrow \quad
    \bigl(m_1 \Gamma_1^{(0)} + m_2 \Gamma_2^{(0)}\bigr)^2 \sin^2\theta_{12} \text{.}
    \label{eq:effective_sum_regulator}
\end{equation}
Here we used the fact that in the single lepton flavour and quasidegeneracy limit $\Gamma_{\evec{k}12}\Gamma_{\evec{k}21} \simeq \cos^2\theta_{12} \,\Gamma_{\evec{k}11}\Gamma_{\evec{k}22}$, where the Yukawa phase $\theta_{12}$ was defined in equation~\cref{eq:Yukawa_CP_phase}. The result~\cref{eq:effective_sum_regulator} agrees with the effective sum regulator defined in~\cite{Dev:2017wwc}.

%------------------------------------------------------------------------------
%
\begin{figure}[t!]
    \centering
    \includegraphics{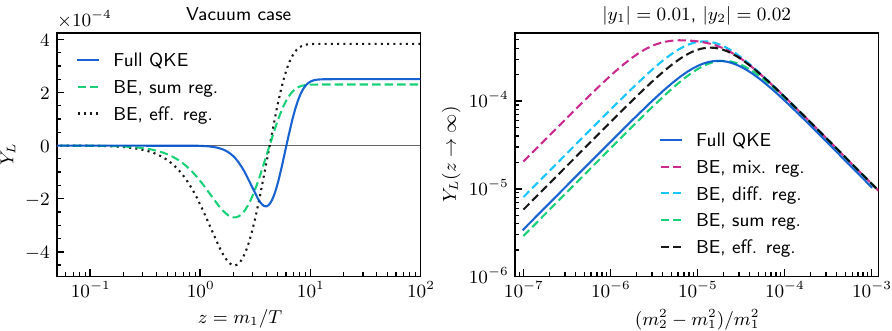}
    \caption{Comparison of the effective sum regulator with the other results in the weak washout case with $\abs{y_1} = 0.01$ and $\abs{y_1} = 0.02$. In the left panel we took $\Delta m^2_{21}/m_1^2 = (0.01^2 + 0.02^2)/(16\pi) \approx 10^{-5}$ near the resonance maximum. We used vacuum initial conditions in both panels, and other parameters have the benchmark values~\cref{eq:benchmark-parameters}.}
    \label{fig:weak_washout_with_effsumreg}
\end{figure}
%
%------------------------------------------------------------------------------

We have shown the results using this effective sum regulator in figure~\cref{fig:final_YL_vs_massdiff} as the black dashed line. This is indeed the best approximation in the strong washout limit. On the other hand, this approximation does not work well outside the strong washout case. We show in figure~\cref{fig:weak_washout_with_effsumreg} the case where both Yukawa couplings are small, corresponding to the washout strength parameters $K_1 \approx 0.28$ and $K_2 \approx 1.1$. In this case the regulator~\cref{eq:effective_sum_regulator} is worse than the simpler sum-regulator we found in the decoupling limit. These examples show that using different approximations one can accommodate the most relevant quantum effects in different parametric regions. However, no approximation remains quantitatively accurate throughout the parameter space.

%%%%%%%%%%%%%%%%%%%%%%%%%%%%%%%%%%%%%%%%%%%%%%%%%%%%%%%%%%%%%%%%%%%%%%%%%%%%%%%
\paragraph{Summary.}
%%%%%%%%%%%%%%%%%%%%%%%%%%%%%%%%%%%%%%%%%%%%%%%%%%%%%%%%%%%%%%%%%%%%%%%%%%%%%%%

We have used a series of controlled approximations to reduce the Majorana neutrino QKEs and the lepton asymmetry source term to the Boltzmann limit. The decoupling limit CP-asymmetry parameter~\cref{eq:sum_epsilon} corresponds to the sum regulator~\cref{eq:epsilon-sum-regulator} in agreement with~\cite{Garny:2011hg,Iso:2013lba,Iso:2014afa}. The previous work used some extra assumptions about the model parameters however, and their validity in the doubly degenerate limit have been questioned~\cite{Dev:2014laa}. Our derivation is very different and does not rely on these assumptions. We also find a clear origin and interpretation for the regulator, corresponding to the flavour coherence damping in the neutrino equation. Our numerical work confirmed that the sum-regulator is indeed the most accurate in the Boltzmann approach in the sense that it provided qualitatively best results in all regimes, as can be seen in figures~\cref{fig:final_YL_vs_massdiff,fig:weak_washout_with_effsumreg}, while the effective sum regulator~\cref{eq:effective_sum_regulator} is the most accurate one in the late time limit in the strong washout case. As a byproduct, we demonstrated quantitatively how the quantum oscillations are suppressed and the semiclassical limit arises in the hierarchical strongly damped limit $\Delta m_{21}\gg \bar \Gamma_{12} \gg H$. Beyond this case, the BE approach does not provide highly accurate results and QKEs are needed. However, we found that the helicity-symmetric QKEs give a very accurate final asymmetry throughout the resonant leptogenesis parameter range.

%%%%%%%%%%%%%%%%%%%%%%%%%%%%%%%%%%%%%%%%%%%%%%%%%%%%%%%%%%%%%%%%%%%%%%%%%%%%%%%
%
\section{Comparison to earlier work}
\label{sec:comparison_and_discussion}
%
%%%%%%%%%%%%%%%%%%%%%%%%%%%%%%%%%%%%%%%%%%%%%%%%%%%%%%%%%%%%%%%%%%%%%%%%%%%%%%%

Leptogenesis has been studied extensively before~\cite{Buchmuller:2000nd,Hohenegger:2008zk,Anisimov:2008dz,Garny:2009rv,Anisimov:2010aq,Beneke:2010wd,Garny:2010nz,Beneke:2010dz,Garbrecht:2010sz,Anisimov:2010dk,Garbrecht:2012qv,Drewes:2012ma,Garbrecht:2012pq,Frossard:2012pc,Garbrecht:2013gd,Garbrecht:2013urw,Frossard:2013bra,Garbrecht:2013iga,Dev:2017trv,Garbrecht:2019zaa,DeSimone:2007gkc,DeSimone:2007edo,Cirigliano:2007hb,Garny:2009qn,Garbrecht:2011aw,Garny:2011hg,Iso:2013lba,Iso:2014afa,Hohenegger:2014cpa,Garbrecht:2014aga,Dev:2014wsa,Kartavtsev:2015vto,Drewes:2016gmt,Dev:2017wwc,Garbrecht:2018mrp}, and many of the results presented here have been found in some form previously. In this section we will provide a more in-depth comparison of our results and other studies based on first-principles CTP methods, which we believe are most closely similar to ours.

In several studies of leptogenesis based on the CTP method (\eg~\cite{Garny:2011hg,Hohenegger:2014cpa,Kartavtsev:2015vto,Anisimov:2010dk,Depta:2020zmy}) the non-equilibrium part of the Majorana propagator is identified with a homogeneous transient, while the inhomogeneous part is taken to be the thermal equilibrium solution. In this approach the only non-equilibrium source is in the initial conditions, as there is no dynamical source for the asymmetry generation. Such approaches have been used to model the lepton asymmetry generation during the initial approach of the Majorana neutrinos to equilibrium~\cite{Anisimov:2010dk,Depta:2020zmy}. Our formalism contains this effect, which shows up as the initial negative $Y_L$ dip in the left-hand side panels in figures~\cref{fig:lepton_YL_benchmark,fig:lepton_YL_with_Y1_001}. However, as these figures show, a moderate washout can erase this asymmetry. This restricts the use of pure transient methods to the weak washout case.

A very careful analysis using the transient approach was given in~\cite{Garny:2011hg}. In particular off-diagonal pole propagators, which are crucial for the evolution of the non-equilibrium initial state in the two-time approach, were calculated in detail. These results are similar to our~\cref{eq:adiabatic-pole-eq}. However, in our approach where the local correlator is evolved dynamically, it is sufficient to compute the pole propagators to a leading order approximation. In~\cite{Garny:2011hg} it was also found that the sum regulator~\cref{eq:epsilon-sum-regulator} is a reasonable choice for the CP-asymmetry parameter~\cref{eq:epsilon-type-CP-asymmetry} in the Boltzmann approach. The Hubble expansion was not included in~\cite{Garny:2011hg}, but the issue was later addressed in~\cite{Iso:2013lba,Iso:2014afa}. Our derivation is more general, and does not impose restrictions on Yukawa couplings~\cite{Garny:2011hg,Iso:2013lba} or on the size of the deviation from equilibrium~\cite{Iso:2014afa}. Our derivation also reveals the physical origin of the sum regulator as corresponding to the flavour coherence damping rate.

The approach in~\cite{Garbrecht:2011aw} (see also~\cite{Garbrecht:2014aga,Drewes:2016gmt}) is similar to ours, but it relies heavily on the Wigner space representation. The neutrino correlator $S^\lt_\evec{k}$ is also expanded in a different basis, used earlier in the EWBG context~\cite{Kainulainen:2001cn,Kainulainen:2002th} and in the cQPA approach~\cite{Herranen:2008hi,Herranen:2008hu,Herranen:2008yg,Herranen:2008di,Herranen:2010mh,Fidler:2011yq,Herranen:2011zg}. In this basis the division of $S^\lt_\evec{k}$ into components with a characteristic time dependence is obscured, making it difficult to separate the particle-antiparticle and flavour coherence effects. Ref.~\cite{Garbrecht:2011aw} then used several approximations in integrating the particle-antiparticle coherences, in reduction to the spectral shell limit and a restriction to the quasidegenerate case, that we do not need to make. On the balance, the final QKE of~\cite{Garbrecht:2011aw} agrees with our equation~\cref{eq:density-matrix-form} in the quasidegenerate limit (and includes the small backreaction to the neutrino equation from the lepton chemical potential, which we omitted.) Overall our derivation is more general and displays a hierarchy of QKEs, which separate the different physical scales. We also provide detailed numerical examples and comparisons, with the Hubble expansion included.

In~\cite{Dev:2014wsa} resonant leptogenesis was studied in the interaction picture method~\cite{Millington:2012pf}, using the double momentum representation, and it was found that the lepton asymmetry source term contains two distinct contributions from mixing and oscillations.%
\footnote{Note that the `mixing contribution' in~\cite{Dev:2014wsa} contains both of the traditional $\epsilon$ and $\epsilon'$ type CP-violation sources. The oscillation part is an additional contribution. The oscillation contribution is also different from the ARS mechanism~\cite{Dev:2014wsa}.}
These results suggest that the usual Boltzmann approach, which contains only the mixing contribution, potentially captures only half of the actual late time lepton asymmetry. The findings of~\cite{Dev:2014wsa} were supported by an analysis~\cite{Kartavtsev:2015vto} performed in a simplified non-equilibrium setup in the weak-washout regime. Both articles found that the mixing contribution is mainly due to the flavour diagonal functions and the oscillation contribution mainly due to the off-diagonal functions in the Majorana neutrino correlator. Ref.~\cite{Kartavtsev:2015vto} found also another contribution resulting from the interference of the mixing and oscillation terms, which tends to cancel the other contributions in some cases. These results have been further discussed in~\cite{Dev:2017trv}.

Our results do not support these findings. Indeed, our result for the lepton asymmetry converges to the usual Boltzmann result in the weakly resonant case (see figure~\cref{fig:final_YL_vs_massdiff}) and a difference by a factor of two should be clearly visible. It is then curious to note that, similarly to refs.~\cite{Dev:2014wsa,Kartavtsev:2015vto}, our lepton asymmetry source~\cref{eq:cp-source-simplified} also contains two distinct contributions from the flavour diagonal and off-diagonal phase space functions. However, in our case the flavour-diagonal contribution is helicity suppressed as explained in section~\cref{sec:helicity_symmetric_equations}. This helicity dependence may thus be the source of the discrepancy, especially given that refs.~\cite{Dev:2014wsa,Kartavtsev:2015vto} are based on scalar toy models where the effective Majorana states do not have Dirac structure. Also, while the results of~\cite{Dev:2014wsa} were essentially reproduced with true Majorana neutrinos in~\cite{Dev:2014laa}, that analysis was based on a semiclassical approach, which again may not necessarily implement the correct helicity structure.

Finally, flavour coherent equations similar to ones used to study light neutrino mixing in the early universe in~\cite{Barbieri:1990vx,Enqvist:1990ad,Kainulainen:1990ds,Enqvist:1990ek,Barbieri:1989ti,Enqvist:1991qj,Sigl:1992fn}, were used to study leptogenesis in~\cite{Klaric:2020lov,Eijima:2020shs,Klaric:2021cpi}. We showed how these equations arise from our more general formalism. We also showed that a further reduction to even simpler, but very accurate helicity-even equations~\cref{eq:density-matrix-form} is possible in the resonant leptogenesis case. Our formalism is also self-contained giving explicit expressions for all self-energies involved.

%%%%%%%%%%%%%%%%%%%%%%%%%%%%%%%%%%%%%%%%%%%%%%%%%%%%%%%%%%%%%%%%%%%%%%%%%%%%%%%
%
\section{Conclusions and outlook}
\label{sec:conclusions}
%
%%%%%%%%%%%%%%%%%%%%%%%%%%%%%%%%%%%%%%%%%%%%%%%%%%%%%%%%%%%%%%%%%%%%%%%%%%%%%%%

We have developed a general and comprehensive formalism for problems involving quantum coherence effects in spatially homogeneous and isotropic systems with mixing fermions. Our methods are applicable to various problems ranging from vacuum particle production to neutrino physics and resonant leptogenesis, which we used as an example and a platform for the more detailed developments of the method. In particular we concentrated on a benchmark model with two Majorana neutrinos and one lepton flavour including decay and inverse decay interactions.

Our method uses the CTP formalism~\cite{Schwinger:1960qe,Keldysh:1964ud} and the 2PIEA methods~\cite{Cornwall:1974vz,Calzetta:1986cq}. An essential part in our derivation of tractable quantum kinetic equations \emph{including coherence} was finding a closed equation for the \emph{local} correlation function $S^\lt_{\evec{k}}(t,t)$. This required a method to evaluate collision integrals which contain the full correlation function $S^\lt_{\evec{k}}(t_1,t_2)$. We explained how this can be done when the system has dissipation. Two key elements in the process were the identification of proper adiabatic background solutions and the \emph{ansatz}~\cref{eq:local-approximation} to parametrise the perturbation $\delta S^\lt_{\evec{k}}(t_1,t_2)$ in terms of the local correlation function in the collision terms. However, no assumptions were made to restrict the spin or the flavour structure of the local correlator, which makes it well suited for studying dynamical mixing. Another essential element was the use of the projector basis~\cref{eq:local-correlator-parametrisation}, which provides a clean separation of physics related to different time scales, \eg~the particle-antiparticle oscillations and the flavour oscillations. Our formalism can be seen as a generalisation of the cQPA method developed in~\cite{Herranen:2008hi,Herranen:2008hu,Herranen:2008yg,Herranen:2008di,Herranen:2010mh,Fidler:2011yq,Herranen:2011zg} and further studied in~\cite{Jukkala:2019slc}.

Our main results include the quantum kinetic equation~\cref{eq:delta-f-equation} which contains complete coherence information including particle-antiparticle mixing, and its coarse-grained version~\cref{eq:delta-f-m-equation}, which completely incorporates the flavour mixing in both particle and antiparticle sectors separately. Note that these QKEs cannot in general be written as a traditional density matrix equation, if there are more than two flavours, due to the complicated flavour structure of the collision term~\cref{eq:collision_term_trace_def}. For the two-flavour case we derived even more simplified but very accurate helicity-symmetric QKEs~\cref{eq:diag1_equation_approx,eq:diag2_equation_approx,eq:offd_equation_approx,eq:SCP_approximation} and further wrote them into a density matrix form~\cref{eq:density-matrix-form} in the nearly degenerate limit $m_1\simeq m_2$. Eventually we reduced our QKEs to the diagonal Boltzmann limit, deriving the CP-asymmetry parameter $\epsilon^{CP}_i$ of leptogenesis with the `sum'-regulator~\cref{eq:sum_epsilon} directly from the quantum transport formalism. We also pointed out that the sum-regulator physically corresponds to the coherence damping rate of the Majorana neutrinos in the underlying QKEs.

The question of the correct CP-asymmetry regulator in the semiclassical approach has been under some discussion recently~\cite{Garny:2011hg,Garbrecht:2014aga,Dev:2017trv,Dev:2017wwc}. We performed careful numerical comparisons of our QKEs and the Boltzmann equations endowed with different choices for $\epsilon^{CP}_i$. We found that the sum-regulator derived here in the decoupling limit, and first found in~\cite{Garny:2011hg}, agrees best qualitatively with the QKEs throughout the parameter range of interest for resonant leptogenesis. We also implemented the integrated Boltzmann rate equations and compared them with the momentum dependent BEs and the QKEs. It turns out that the difference between the REs and BEs was always much smaller than the difference stemming from the choice of different regulators. While the sum-regulator was generically the best choice for the BE and RE approach, their results can still be wrong by a factor $\sim$ 2--4 for $\Delta m_{21} \lesssim \Gamma$. In the strong washout case the modified sum regulator~\cref{eq:effective_sum_regulator} of~\cite{Garbrecht:2014aga,Dev:2017wwc} gives even more accurate late-time results. For high accuracy results throughout the parameter range, QKEs are needed however. We found that the helicity-symmetric QKEs~\cref{eq:offd_equation_approx,eq:diag2_equation_approx,eq:offd_equation_approx,eq:SCP_approximation} give a very accurate approximation to the full flavour QKEs~\cref{eq:delta-f-m-equation} for all parameters. Finally, all of these approaches are in good agreement in the weakly resonant case, $m_i \gg \Delta m_{21} \gg \Gamma$, as expected.

Leptogenesis has been studied extensively using the CTP approach~\cite{Buchmuller:2000nd,Hohenegger:2008zk,Anisimov:2008dz,Garny:2009rv,Anisimov:2010aq,Beneke:2010wd,Garny:2010nz,Beneke:2010dz,Garbrecht:2010sz,Anisimov:2010dk,Garbrecht:2012qv,Drewes:2012ma,Garbrecht:2012pq,Frossard:2012pc,Garbrecht:2013gd,Garbrecht:2013urw,Frossard:2013bra,Garbrecht:2013iga,Dev:2017trv,Garbrecht:2019zaa,DeSimone:2007gkc,DeSimone:2007edo,Cirigliano:2007hb,Garny:2009qn,Garbrecht:2011aw,Garny:2011hg,Iso:2013lba,Iso:2014afa,Hohenegger:2014cpa,Garbrecht:2014aga,Dev:2014wsa,Kartavtsev:2015vto,Drewes:2016gmt,Dev:2017wwc,Garbrecht:2018mrp} and many of the results shown here have been found previously. We believe that our treatment stands out in displaying the most complete set of quantum kinetic equations, with a clean separation of different physics and by giving a comprehensive account of the approximations made in deriving them. Based on our results one can easily compute the effective self-energy functions to different levels of approximation in the coupling constant expansion, and including also the coherent propagators in the internal lines. In some accounts we did less than what has been done before. Definitely the phenomenological reach of our results is compromised by our not including the $2 \to 2$ scattering processes or multiple lepton flavours in our equations. We will leave these to a future work.

In this article we mainly concentrated on resonant leptogenesis, but our methods are applicable as such, or easily modifiable to other versions of the leptogenesis mechanism~\cite{Davidson:2008bu,Blanchet:2012bk}. We already briefly discussed thermal leptogenesis in the hierarchical limit in section~\cref{sec:mass_shell_equation}. In this case it would be interesting to compute corrections to the lepton asymmetry source arising from the mixed particle-antiparticle flavour correlation functions ($\delta f^{c,\pm}_{\evec{k}h12}$ in the notation of section~\cref{sec:mass_shell_equation}), starting from the full QKEs~\cref{eq:delta-f-equation} and working in the decoupling limit. Our equations can also easily accommodate dispersive corrections to the neutrino QKEs. Such corrections would generalise the vacuum Hamiltonian to include the matter effects, similar to the well known case with light neutrinos. This would replace the mass difference $\Delta m_{21}$ with an effective dynamical quantity and potentially change the quantitative predictions in resonant leptogenesis.

%%%%%%%%%%%%%%%%%%%%%%%%%%%%%%%%%%%%%%%%%%%%%%%%%%%%%%%%%%%%%%%%%%%%%%%%%%%%%%%
%
\section*{Note added}
%
%%%%%%%%%%%%%%%%%%%%%%%%%%%%%%%%%%%%%%%%%%%%%%%%%%%%%%%%%%%%%%%%%%%%%%%%%%%%%%%

The Mathematica code package that was used to compute all numerical results in this paper is publicly available at \url{https://doi.org/10.5281/zenodo.5025929}.

%%%%%%%%%%%%%%%%%%%%%%%%%%%%%%%%%%%%%%%%%%%%%%%%%%%%%%%%%%%%%%%%%%%%%%%%%%%%%%%
%
\section*{Acknowledgements}
%
%%%%%%%%%%%%%%%%%%%%%%%%%%%%%%%%%%%%%%%%%%%%%%%%%%%%%%%%%%%%%%%%%%%%%%%%%%%%%%%

This work was supported by the Academy of Finland grant 318319. HJ was in addition supported by grants from the Väisälä Fund of the Finnish Academy of Science and Letters. We wish to thank Matti Herranen for collaboration during the early stages of this work.

%%%%%%%%%%%%%%%%%%%%%%%%%%%%%%%%%%%%%%%%%%%%%%%%%%%%%%%%%%%%%%%%%%%%%%%%%%%%%%%
%
\section*{Appendices}
\addcontentsline{toc}{section}{\protect\numberline{}Appendices}
\appendix
%
%%%%%%%%%%%%%%%%%%%%%%%%%%%%%%%%%%%%%%%%%%%%%%%%%%%%%%%%%%%%%%%%%%%%%%%%%%%%%%%

%%%%%%%%%%%%%%%%%%%%%%%%%%%%%%%%%%%%%%%%%%%%%%%%%%%%%%%%%%%%%%%%%%%%%%%%%%%%%%%
%
\section{Resummation of the Schwinger--Dyson equation}
\label{sec:sd-resummation}
%
%%%%%%%%%%%%%%%%%%%%%%%%%%%%%%%%%%%%%%%%%%%%%%%%%%%%%%%%%%%%%%%%%%%%%%%%%%%%%%%

Here we present the derivation of the results~\cref{eq:formal-homog-solution,eq:formal-inhomog-solution},
starting from the Schwinger--Dyson equations~\cref{eq:Schwinger-Dyson-two-time}. We will shorten the notation explained below equations~\cref{eq:free-propagators} even further, by leaving out the momentum indices and convolution signs: \eg~$S_{0,\evec{k}}^p \nconvol \Sigma^p_\evec{k} \nconvol S^p_\evec{k} \to S^p_0 \Sigma^p S^p$. It is also important to note that the inverse free propagator $S_0^{-1}$ contains a derivative operator, and thus changing the direction of its operation generates additional surface terms.

First, the pole equation~\cref{eq:Schwinger-Dyson-two-time-A} for $p = r,a$ can be formally iterated as
\begin{align}
    S^p &= S_0^p \bigl(\idmat + \Sigma^p S^p\bigr)
    \label{eq:SD-pole-resum-1}
    \\*
    &= S_0^p \bigl(\idmat + \Sigma^p S^p_0 + \Sigma^p S^p_0 \Sigma^p S^p_0 + \cdots \bigr)
    \notag
    \\*
    &= S_0^p + S_0^p \Sigma^p S^p_0 + S_0^p \Sigma^p S^p_0 \Sigma^p S^p_0 + \cdots
    \label{eq:SD-pole-resum-3}
    \\*
    &= \bigl(\idmat + S^p_0 \Sigma^p + S^p_0 \Sigma^p S^p_0 \Sigma^p + \cdots \bigr) S_0^p
    \notag
    \\*
    &= \bigl(\idmat + S^p \Sigma^p \bigr) S_0^p \text{.}
    \label{eq:SD-pole-resum-2}
\end{align}
This is consistent with the Hermiticity properties~\cref{eq:hermiticity-two-time} of the pole propagators, which relate equation~\cref{eq:SD-pole-resum-1} for $p = r$ to equation~\cref{eq:SD-pole-resum-2} for $p = a$ and vice versa. Equation~\cref{eq:Schwinger-Dyson-two-time-B} for $S^\lt$ (similarly for $S^\gt$) can now be iterated and rearranged as follows:
\begin{align}
    S^\lt ={}& S_0^\lt + S_0^\lt \Sigma^a S^a + S_0^r \Sigma^\lt S^a + S_0^r \Sigma^r S^\lt
    \label{eq:SD-wightman-resum-start}
    \\*
    ={}& S_0^\lt + S_0^\lt \Sigma^a S^a + S_0^r \Sigma^\lt S^a
    \notag
    \\*
    &{} + S_0^r \Sigma^r \bigl(
        S_0^\lt + S_0^\lt \Sigma^a S^a + S_0^r \Sigma^\lt S^a + S_0^r \Sigma^r S^\lt
    \bigr) \notag
    \\*
    ={}& \bigl(\idmat + S_0^r \Sigma^r\bigr)\bigl(
        S_0^\lt + S_0^\lt \Sigma^a S^a + S_0^r \Sigma^\lt S^a
    \bigr) \notag
    \\*
    &{} + S_0^r \Sigma^r S_0^r \Sigma^r \bigl(
        S_0^\lt + S_0^\lt \Sigma^a S^a + S_0^r \Sigma^\lt S^a + S_0^r \Sigma^r S^\lt
    \bigr) \notag
    \\*
    \shortvdotswithin*{=}
    ={}& \bigl(\idmat + S_0^r \Sigma^r + S_0^r \Sigma^r S_0^r \Sigma^r + \cdots\bigr)\bigl(
        S_0^\lt + S_0^\lt \Sigma^a S^a + S_0^r \Sigma^\lt S^a
    \bigr) \notag
    \\*
    ={}& \bigl(\idmat + S^r \Sigma^r\bigr)\bigl(
        S_0^\lt + S_0^\lt \Sigma^a S^a + S_0^r \Sigma^\lt S^a
    \bigr) \notag
    \\*
    ={}& \bigl(\idmat + S^r \Sigma^r\bigr) S_0^\lt \bigl(\idmat +\Sigma^a S^a\bigr)
    + S^r \Sigma^\lt S^a \text{.}
    \label{eq:SD-wightman-resum-end}
\end{align}
The second and third equalities explicitly show the first and second iteration of the first equation~\cref{eq:SD-wightman-resum-start}. In the fourth step this iteration is assumed to continue indefinitely. To get the final two lines we then used equations~\cref{eq:SD-pole-resum-3,eq:SD-pole-resum-2}. Note that the final line can also be written in the form
\begin{equation}
    S^\lt = \bigl(S^r S_0^{-1}\bigr) S_0^\lt \bigl(S_0^{-1} S^a \bigr) + S^r \Sigma^\lt S^a \text{.}
\end{equation}
This result was derived also in~\cite{Greiner:1998vd} and a similar iterative solution was presented in the double momentum representation in the context of the interaction picture CTP formalism~\cite{Dev:2014wsa}.

%%%%%%%%%%%%%%%%%%%%%%%%%%%%%%%%%%%%%%%%%%%%%%%%%%%%%%%%%%%%%%%%%%%%%%%%%%%%%%%
%
\section{Majorana neutrino self-energies}
\label{sec:neutrino-self-energies}
%
%%%%%%%%%%%%%%%%%%%%%%%%%%%%%%%%%%%%%%%%%%%%%%%%%%%%%%%%%%%%%%%%%%%%%%%%%%%%%%%

The self-energy functions $\mathfrak{S}_\mu$ defined in equations~\cref{eq:neutrino-self-energy-parametrisations} are given by
\begin{subequations}
\begin{align}
    \slashed{\mathfrak{S}}_{\rm eq}^{\rm H}(k) &= \frac{1}{2} \int \frac{\dd^4 p}{(2\pi)^4}
    2\pi \sgn(p^0) \delta(p^2) \PV\biggl[\frac{1}{(k - p)^2}\biggr] \notag
    \\*
    &\hphantom{{} = \frac{1}{2} \int \frac{\dd^4 p}{(2\pi)^4}}
    {} \times \Bigl[
        \bigl(1 + 2 f_{\rm BE}(p^0)\bigr)(\slashed k - \slashed p)
        + \bigl(1 - 2 f_{\rm FD}(p^0)\bigr)\slashed p
    \Bigr] \text{,} \label{eq:Sigma-H-appendix}
    \\
    \slashed{\mathfrak{S}}_{\rm eq}^{\mathcal{A}}(k) &= \frac{1}{2} \int \frac{\dd^4 p}{(2\pi)^4}
    (2\pi)^2 \sgn(p^0) \sgn(k^0-p^0) \,\slashed p \,\delta(p^2) \delta\bigl((k - p)^2\bigr) \notag
    \\*
    &\hphantom{{} = \frac{1}{2} \int \frac{\dd^4 p}{(2\pi)^4}}
    {} \times \Bigl[f_{\rm FD}(-p^0) + f_{\rm BE}(k^0 - p^0)\Bigr]
    \text{,} \label{eq:Sigma-A-appendix}
    \\
    \delta\slashed{\mathfrak{S}}^{\lessgtr}(k) &= \int \frac{\dd^4 p}{(2\pi)^4}
    (2\pi)^2 \sgn(p^0) \sgn(k^0 - p^0) \,\slashed p \,\delta(p^2) \delta\bigl((k - p)^2\bigr) \notag
    \\*
    &\hphantom{{} = \int \frac{\dd^4 p}{(2\pi)^4}}
    {} \times f_{\rm FD}(p^0) f_{\rm FD}(-p^0) f_{\rm BE} \bigl(\pm(k^0 - p^0)\bigr)
    \text{.} \label{eq:delta-Sigma-appendix}
\end{align}
\end{subequations}
Here $\PV$ denotes the Cauchy principal value distribution, and $\pm = {+},{-}$ for $\smash{\lessgtr} = \smallless, \smallgreater$, respectively. Note that equation~\cref{eq:Sigma-H-appendix} contains also the unrenormalised vacuum part of $\mathfrak{S}_{\rm eq}^{\rm H}(k)$, which is UV-divergent. The Yukawa and chirality structure for this part are the same as in equation~\cref{eq:neutrino-self-energy-dispersive-equilibrium}. We split $\mathfrak{S}_{\rm eq}^{\rm H}$ into the vacuum and temperature dependent parts
\begin{equation}
    \slashed{\mathfrak{S}}_{\rm eq}^{\rm H}(k)
    = \slashed{\mathfrak{S}}_{\rm eq}^{{\rm H(vac)}}(k)
    + \slashed{\mathfrak{S}}_{\rm eq}^{{\rm H}(T)}(k)
    \text{,} \label{eq:Sigma-H-integral-vacuum-split}
\end{equation}
where the vacuum part was given (using dimensional regularisation) in equation~\cref{eq:Sigma-H-integral-vacuum-part} and the temperature dependent part is
\begin{equation}
    \slashed{\mathfrak{S}}_{\rm eq}^{{\rm H}(T)}(k) = 2\pi \int \frac{\dd^4 p}{(2\pi)^4}
    \delta(p^2) \PV\biggl[\frac{1}{(k - p)^2}\biggr] \Bigl(
        f_{\rm BE}\abs{p^0}(\slashed k - \slashed p) - f_{\rm FD}\abs{p^0} \,\slashed p
    \Bigr) \text{.} \label{eq:Sigma-H-integral-thermal-part}
\end{equation}
The temperature dependent integral~\cref{eq:Sigma-H-integral-thermal-part} has been worked out in~\cite{Weldon:1982bn}.

We further parametrise the integrals~\cref{eq:Sigma-A-appendix,eq:delta-Sigma-appendix,eq:Sigma-H-integral-thermal-part} with
\begin{gather}
    \slashed{\mathfrak{S}}^\alpha(k) =
    a^\alpha\bigl(k^0, \abs{\evec{k}}\bigr) \gamma^0
    + b^\alpha\bigl(k^0, \abs{\evec{k}}\bigr) \evec{\gamma} \cdot \Uevec{k}
    \text{,}
\shortintertext{and}
    a^\alpha\bigl(k^0, \abs{\evec{k}}\bigr) = T \,\tilde a^\alpha\bigl(k^0/T, \abs{\evec{k}}/T\bigr)
    \text{,}
    \\*
    b^\alpha\bigl(k^0, \abs{\evec{k}}\bigr) = T \,\tilde b^\alpha\bigl(k^0/T, \abs{\evec{k}}/T\bigr)
    \text{,}
\end{gather}
where $\tilde a^\alpha, \tilde b^\alpha$ are dimensionless functions, and $\alpha = \mathcal{A}, \mathrm{H}, \smallless, \smallgreater$ and $\mathfrak{S} = \mathfrak{S}_{\rm eq}, \delta\mathfrak{S}$. After some calculation we get the following results (with $\kappa_0 \equiv k^0/T$, $\kappa \equiv \abs{\evec{k}}/T$):
\begin{subequations}
\label{eq:ab-result-integrals}
\begin{align}
    \tilde a^{{\rm H}(T)}_{\rm eq}(\kappa_0,\kappa) ={}&
    \frac{1}{16 \pi^2 \kappa} \int_0^\infty \dd x \,\Biggl\{
        \biggl[
            2x \log\abs[\Big]{\frac{\kappa_+}{\kappa_-}} - x L_+(x)
        \biggr] \Bigl[ \tilde f_{\rm FD}(x) + \tilde f_{\rm BE}(x) \Bigr] \notag
        \\*
        &\phantom{\frac{1}{16 \pi^2 \kappa} \int_0^\infty \dd x \,\Biggl\{}
        {} - \kappa_0 L_-(x) \tilde f_{\rm BE}(x)
    \Biggr\} \text{,} \displaybreak[0]
    \\
    \tilde b^{{\rm H}(T)}_{\rm eq}(\kappa_0,\kappa) ={}&
    \frac{1}{16 \pi^2 \kappa} \int_0^\infty \dd x \,\Biggl\{
        \biggl[
            4x - 2x \frac{\kappa_0}{\kappa} \log\abs[\Big]{\frac{\kappa_+}{\kappa_-}}
            + \frac{\kappa_0^2 - \kappa^2}{2\kappa} L_-(x) + \frac{\kappa_0}{\kappa} x L_+(x)
        \biggr] \notag
        \\*
        &\phantom{\frac{1}{16 \pi^2 \kappa} \int_0^\infty \dd x \,\Biggl\{}
        {} \times \Bigl[ \tilde f_{\rm FD}(x) + \tilde f_{\rm BE}(x) \Bigr]
        + \kappa L_-(x) \tilde f_{\rm BE}(x)
    \Biggr\} \text{,} \displaybreak[0]
    \\
    \tilde a^{\mathcal{A}}_{\rm eq}(\kappa_0,\kappa) ={}&
    \frac{1}{16 \pi \kappa} \int_{\kappa_-}^{\kappa_+} \dd x \,x \Bigl[
        \tilde f_{\rm FD}(-x) + \tilde f_{\rm BE}(\kappa_0 - x)
    \Bigr] \text{,} \displaybreak[0]
    \\
    \tilde b^{\mathcal{A}}_{\rm eq}(\kappa_0,\kappa) ={}&
    \frac{1}{16 \pi \kappa} \int_{\kappa_-}^{\kappa_+} \dd x \,\biggl(
        \frac{\kappa_0^2 - \kappa^2}{2\kappa} - \frac{\kappa_0}{\kappa} x
    \biggr) \Bigl[
        \tilde f_{\rm FD}(-x) + \tilde f_{\rm BE}(\kappa_0 - x)
    \Bigr] \text{,} \displaybreak[0]
    \\
    \delta\tilde a^{\lessgtr}(\kappa_0,\kappa) ={}&
    \frac{1}{8 \pi \kappa} \int_{\kappa_-}^{\kappa_+} \dd x \,x
    \tilde f_{\rm FD}(x) \tilde f_{\rm FD}(-x) \tilde f_{\rm BE}\bigl(\pm(\kappa_0 - x)\bigr)
    \text{,} \displaybreak[0]
    \\
    \delta\tilde b^{\lessgtr}(\kappa_0,\kappa) ={}&
    \frac{1}{8 \pi \kappa} \int_{\kappa_-}^{\kappa_+} \dd x \,\biggl(
        \frac{\kappa_0^2 - \kappa^2}{2\kappa} - \frac{\kappa_0}{\kappa} x
    \biggr)
    \tilde f_{\rm FD}(x) \tilde f_{\rm FD}(-x) \tilde f_{\rm BE}\bigl(\pm(\kappa_0 - x)\bigr)
    \text{,}
\end{align}
\end{subequations}
where
\begin{subequations}
\begin{gather}
    L_{\pm} \equiv \log\abs[\Big]{\frac{x + \kappa_+}{x + \kappa_-}}
    \pm \log\abs[\Big]{\frac{x - \kappa_+}{x - \kappa_-}} \text{,} \qquad
    \kappa_{\pm} \equiv \frac{1}{2}\bigl(\kappa_0 \pm \kappa\bigr) \text{,}
\shortintertext{and}
    \tilde f_{\rm FD}(x) \equiv \frac{1}{\e^x + 1} \text{,} \qquad
    \tilde f_{\rm BE}(x) \equiv \frac{1}{\e^x - 1} \text{.}
\end{gather}
\end{subequations}
The integrals given above for $\tilde a^{\mathcal{A}}_{\rm eq}$, $\tilde b^{\mathcal{A}}_{\rm eq}$, $\delta\tilde a^{\lessgtr}$ and $\delta\tilde b^{\lessgtr}$ may be further calculated in closed form using logarithm and dilogarithm functions (like in~\cite{Beneke:2010wd}). The results given here hold only for $\abs{\kappa_0} > \kappa > 0$ (timelike four-momentum). The corresponding results for $\kappa > \abs{\kappa_0} > 0$ (spacelike four-momentum) are attained by replacing the integral operator above as follows:
\begin{equation}
    \int_{\kappa_-}^{\kappa_+} \dd x \quad \longrightarrow \quad
    -\biggl(
        \int_{\kappa_+}^\infty + \int_{-\infty}^{\kappa_-}
    \biggr) \!\dd x \text{.}
\end{equation}
Also, in the results~\cref{eq:ab-result-integrals} it was assumed that the energy parameter $\kappa_0$ is real. If these results are continued to complex values of $\kappa_0$ (as required when considering finite widths) then the implicit sign functions in the absolute values must be applied to the real parts only (\eg~$\sgn(\kappa_0) \kappa_0 = \abs{\kappa_0}$ should be replaced by $\sgn(\Re \kappa_0) \kappa_0 = \sqrt{{\kappa_0}^2}$).

Finally, the $a^\alpha$ and $b^\alpha$ functions have the following $k^0$-symmetry properties:
\begin{subequations}
\begin{alignat}{2}
    a_{\rm eq}^{\rm H}(-k^0,\abs{\evec{k}}) &=
    -a_{\rm eq}^{\rm H}(k^0,\abs{\evec{k}}) \text{,}& \qquad
    b_{\rm eq}^{\rm H}(-k^0,\abs{\evec{k}}) &=
    b_{\rm eq}^{\rm H}(k^0,\abs{\evec{k}}) \text{,}
    \label{eq:ab-H-eq-symmetries}
    \\*
    a_{\rm eq}^\mathcal{A}(-k^0,\abs{\evec{k}}) &=
    a_{\rm eq}^\mathcal{A}(k^0,\abs{\evec{k}}) \text{,}& \qquad
    b_{\rm eq}^\mathcal{A}(-k^0,\abs{\evec{k}}) &=
    -b_{\rm eq}^\mathcal{A}(k^0,\abs{\evec{k}}) \text{,}
    \label{eq:ab-A-eq-symmetries}
    \\*
    \delta a^\lt(-k^0,\abs{\evec{k}}) &=
    -\delta a^\gt(k^0,\abs{\evec{k}}) \text{,}& \qquad
    \delta b^\lt(-k^0,\abs{\evec{k}}) &=
    \delta b^\gt(k^0,\abs{\evec{k}}) \text{.}
    \label{eq:ab-Wightman-delta-symmetries}
\end{alignat}
\end{subequations}
%

%%%%%%%%%%%%%%%%%%%%%%%%%%%%%%%%%%%%%%%%%%%%%%%%%%%%%%%%%%%%%%%%%%%%%%%%%%%%%%%
%
\section{Adiabatic pole propagator inversion}
\label{sec:neutrino-pole-propagator-inversion}
%
%%%%%%%%%%%%%%%%%%%%%%%%%%%%%%%%%%%%%%%%%%%%%%%%%%%%%%%%%%%%%%%%%%%%%%%%%%%%%%%

In this paper we have used the leading order spectral approximation for the adiabatic solutions, which indeed is a good approximation in the weak coupling limit. In this appendix we show how to obtain more complete solutions for the adiabatic pole propagators $S^{r,a}_{\rm ad}(k,t)$. We start by writing equation~\cref{eq:adiabatic-pole-eq} in an equivalent form and with explicit flavour indices:
\begin{equation}
    \sum_{l=1}^{2}\Bigl(\bigl[\slashed k - m_i(t)\bigr]\delta_{il} -
    \Sigma^p_{{\rm eq},il}(k,t)\Bigr)S^p_{{\rm ad},lj}(k,t) = \delta_{ij}\idmat
    \text{.} \label{eq:pole-wigner-lo-components}
\end{equation}
Generally, the inverse of a $2 \times 2$ block matrix $A_{ij}$, satisfying $\sum_{l=1}^{2}A_{il} S_{lj} = \delta_{ij}\idmat$, is given by $S_{ij} = (A_{ji} - A_{jk}A^{-1}_{lk}A_{li})^{-1}$ with $k\neq i$ and $l\neq j$. For given $i$ and $j$ the indices $k$ and $l$ are fixed, because the block dimension is only $2$. This block-wise inversion formula can be generalised to larger block matrices, which would indeed be necessary if there were more flavours. We now solve the flavour components of the pole propagators from~\cref{eq:pole-wigner-lo-components} as follows (equivalent formulae were given in~\cite{Pilaftsis:1997jf}):
\begin{align}
    S^p_{{\rm ad},ii}(k,t) &=
    \Bigl[
        \slashed k - m_i - \Sigma^p_{{\rm eq},ii} - \Sigma^p_{{\rm eq},il}
        (\slashed k - m_l - \Sigma^p_{{\rm eq},ll})^{-1}
        \Sigma^p_{{\rm eq},li}
    \Bigr]^{-1}
    \quad (l\neq i) \text{,} \label{eq:pole-propagator-inversion-diagonal}
    \\*
    S^p_{{\rm ad},ij}(k,t) &=
    \Bigl[
        (\slashed k - m_j - \Sigma^p_{{\rm eq},jj})
        (\Sigma^p_{{\rm eq},ij})^{-1}
        (\slashed k - m_i - \Sigma^p_{{\rm eq},ii})
        - \Sigma^p_{{\rm eq},ji}
    \Bigr]^{-1}
    \quad (i\neq j) \text{.} \label{eq:pole-propagator-inversion-offdiagonal}
\end{align}
The inverses here are taken with respect to Dirac indices only and we suppressed the $(k,t)$-arguments of the self-energies. These solutions can also be obtained formally by expanding the inverse of $\slashed k - m - \Sigma^p_{\rm eq}$ as a geometric series in powers of the off-diagonal self-energy $\Sigma^p_{{\rm eq},ij}$ (with $i\neq j$) and performing a resummation of the series.

The solutions~\cref{eq:pole-propagator-inversion-diagonal,eq:pole-propagator-inversion-offdiagonal} are still general. We now use the leading order equilibrium self-energy $\Sigma_{\rm eq}$ given by equations~\cref{eq:neutrino-self-energy-absorptive-equilibrium,eq:neutrino-self-energy-dispersive-equilibrium}. We neglect here the vacuum part of $\Sigma^{\rm H}_{\rm eq}$ which acquires additional Dirac and flavour structures from the vacuum on-shell renormalisation and would make the resulting formulae below more complicated. However, near the poles the vacuum part should only give small corrections. Hence, we use here the pole self-energies
\begin{equation}
    \Sigma^p_{{\rm eq},ij}(k) =
    \cw \bigl(y_i y_j^* P_{\rm L} + y_i^* y_j P_{\rm R}\bigr) \gamma^\mu \mathfrak{S}^{p}_{{\rm eq},\mu}(k)
    \text{.} \label{eq:neutrino-self-energy-pole-equilibrium}
\end{equation}
Using this form, the Dirac matrix inverses in equations~\cref{eq:pole-propagator-inversion-diagonal,eq:pole-propagator-inversion-offdiagonal} can be written explicitly as
\begin{align}
    S^p_{{\rm ad},ii}(k,t) &=
    \frac{1}{D^p_{il}}
    \Bigl[
        D^p_l(\slashed k + m_i)
        - \widetilde{D}^p_l \Sigma^p_{{\rm eq},ii}
        - \Sigma^p_{{\rm eq},li} (\slashed k - m_l) \Sigma^p_{{\rm eq},il}
    \Bigr]
    \quad (l\neq i) \text{,} \label{eq:pole-propagator-diagonal-wigner-lo}
    \\*
    S^p_{{\rm ad},ij}(k,t) &=
    \frac{1}{D^p_{ij}}
    \Bigl[
        (\slashed k + m_i)\Sigma^p_{{\rm eq},ij}(\slashed k + m_j)
        - (\slashed k + m_i) \Sigma^p_{{\rm eq},ij} \Sigma^p_{{\rm eq},jj} \notag
        \\*
        &\hphantom{
        {}=\frac{1}{D^p_{ij}}\Bigl[
            (\slashed k + m_i)\Sigma^p_{{\rm eq},ij}(\slashed k + m_j)
        }
        - \Sigma^p_{{\rm eq},ii}\Sigma^p_{{\rm eq},ij}
        (\slashed k + m_j)
    \Bigr]
    \quad (i\neq j) \text{,} \label{eq:pole-propagator-offdiagonal-wigner-lo}
\end{align}
with the definitions
\begin{subequations}
\begin{align}
    \makebox[3em]{} D^p_{ij} &\equiv
    \begin{aligned}[t]
    \Delta_i \Delta_j \biggl[&
        1 - 2\cw (k \cdot \mathfrak{S}^p_{\rm eq})\Bigl(
            \abs{y_i}^2 \Delta_i^{-1} + \abs{y_j}^2 \Delta_j^{-1}
        \Bigr) \makebox[135pt]{}
    \\
    {} + {} & \begin{aligned}[t]
        \cw^2 (\mathfrak{S}^p_{\rm eq} \cdot \mathfrak{S}^p_{\rm eq})\Bigl(
        &\abs{y_i}^4 \Delta_i^{-1} + \abs{y_j}^4 \Delta_j^{-1}
        \\
        {} + {} 2&\abs{y_i}^2 \abs{y_j}^2 (\Delta_i \Delta_j)^{-1}
        \bigl(k^2 - \cos(2\theta_{ij}) m_i m_j\bigr)
        \Bigr)
        \biggr] \text{,}
        \end{aligned}
    \end{aligned} \raisetag{22pt} \displaybreak[0]
    \\
    D^p_l &\equiv \Delta_l - 2\cw \abs{y_l}^2 (k \cdot \mathfrak{S}^p_{\rm eq})
    + \cw^2 \abs{y_l}^4 (\mathfrak{S}^p_{\rm eq} \cdot \mathfrak{S}^p_{\rm eq})
    \text{,}
    \\*
    \widetilde D^p_l &\equiv \Delta_l - 2\cw \abs{y_l}^2 (k \cdot \mathfrak{S}^p_{\rm eq})
    \text{,}
    \\*
    \Delta_i &\equiv k^2 - m_i^2
    \text{.}
\end{align}
\end{subequations}
Note that since $D_{ij}^p = D_{ji}^p$, the denominator of all flavour components $S^p_{{\rm ad},ij}$ is the same, $D_{12}^p$. Using the formulae presented in this appendix, one can straightforwardly implement more accurate approximations for the adiabatic correlation functions~\cref{eq:two-time-Sad} and the effective self-energies~\cref{eq:two-time-Sigma-eff}.

%%%%%%%%%%%%%%%%%%%%%%%%%%%%%%%%%%%%%%%%%%%%%%%%%%%%%%%%%%%%%%%%%%%%%%%%%%%%%%%
%
\section{Independent components of the neutrino Wightman function}
\label{sec:delta_f_components}
%
%%%%%%%%%%%%%%%%%%%%%%%%%%%%%%%%%%%%%%%%%%%%%%%%%%%%%%%%%%%%%%%%%%%%%%%%%%%%%%%

The Hermiticity property~\cref{eq:wightman-hermiticity}, the sum rule~\cref{eq:sum-rule-two-time} and the Majorana condition~\cref{eq:majorana-condition-propagator} can be used to derive relations among the components of the neutrino Wightman functions. First, the Majorana condition~\cref{eq:majorana-condition-propagator} and equation~\cref{eq:two-time-def} imply that $S^\gt_\evec{k}(t_1,t_2) = -C S^\lt_{-\evec{k}}(t_2,t_1)^\transp C^{-1}$ in the two-time representation. We can then write the aforementioned three relations as
\begin{align}
    \bar S^\lt_{\evec{k}ij}(t,t) &= \bar S^\lt_{\evec{k}ji}(t,t)^\dagger \text{,}
    \\*
    \bar S^\lt_{\evec{k}ij}(t,t) &= \delta_{ij}\idmat - \bar S^\gt_{\evec{k}ij}(t,t)
    \text{,} \label{eq:sum-rule-Wightman}
    \\*
    \bar S^\gt_{\evec{k}ij}(t,t)
    &= \gamma^0 C \,\bar S^\lt_{-\evec{k},ji}(t,t)^\transp \,C^{-1} \gamma^0 \text{,}
\end{align}
where we also used $2\bar{\mathcal{A}} = \bar S^\gt + \bar S^\lt$ in the sum rule. These identities hold for the full Majorana Wightman functions $\bar S^{\lt,\gt}_{\evec{k}ij}(t,t)$. Assuming that they hold independently for the adiabatic functions $\bar S^{\lt,\gt}_{{\rm ad},\evec{k}ij}(t,t)$ implies that the non-equilibrium local correlator $\delta\bar S^\lt_{\evec{k}ij}(t,t)$ satisfies the constraints
\begin{align}
    \delta\bar S^\lt_{\evec{k}ij}(t,t) &= \delta\bar S^\lt_{\evec{k}ji}(t,t)^\dagger
    \text{,} \label{eq:hermiticity-for-noneq-correlator}
    \\*
    \delta\bar S^\lt_{\evec{k}ij}(t,t) &=
    -\gamma^0 C \,\delta\bar S^\lt_{-\evec{k},ji}(t,t)^\transp \,C^{-1} \gamma^0
    \text{.} \label{eq:majorana-condition-for-noneq-correlator}
\end{align}

Inserting the parametrisation~\cref{eq:local-correlator-parametrisation} into equations~\cref{eq:majorana-condition-for-noneq-correlator,eq:hermiticity-for-noneq-correlator} and using the identity $\gamma^0 C (\mathcal{P}_{-\evec{k},hji}^{s's})^\transp C^{-1} \gamma^0 = -\mathcal{P}_{\evec{k}hij}^{-s,-s'}$, we get constraints for the phase space functions:
\begin{equation}
    \delta f_{\evec{k}hij}^{ss'} = \bigl(\delta f_{\evec{k}hji}^{s's}\bigr)^*
    = \delta f_{\evec{k}hji}^{-s',-s} \text{.}
\end{equation}
In terms of the mass and coherence shell functions~\cref{eq:mass-shell-functions,eq:coherence-shell-functions} these read
\begin{align}
    \delta f_{\evec{k}hij}^{m,s} &= \bigl(\delta f_{\evec{k}hji}^{m,s}\bigr)^*
    = \delta f_{\evec{k}hji}^{m,-s} \text{,}
    \\*
    \delta f_{\evec{k}hij}^{c,s} &= \bigl(\delta f_{\evec{k}hji}^{c,-s}\bigr)^*
    = \delta f_{\evec{k}hji}^{c,s} \text{.}
\end{align}
In the case of two neutrino flavours ($i,j = 1,2$) these relations imply that for a given helicity $h$ the sixteen different $(s,s',i,j)$ components of $\delta f_{\evec{k}h}$ can be reduced to only six independent distributions with 10 degrees of freedom, given explicitly by the following equations:
\begin{subequations}
    \label{eq:delta-f-m-c-component-constraints}
    \begin{align}
        \delta f^{m,+}_{\evec{k}h11} &= \delta f^{m,-}_{\evec{k}h11} \quad \text{(real)} \text{,} \\*
        \delta f^{m,+}_{\evec{k}h22} &= \delta f^{m,-}_{\evec{k}h22} \quad \text{(real)} \text{,} \\*
        \delta f^{m,+}_{\evec{k}h12} &= (\delta f^{m,-}_{\evec{k}h12})^* = (\delta f^{m,+}_{\evec{k}h21})^*
        = \delta f^{m,-}_{\evec{k}h21} \quad \text{(complex)} \text{,}
\shortintertext{and}
        \delta f^{c,+}_{\evec{k}h11} &= (\delta f^{c,-}_{\evec{k}h11})^* \quad \text{(complex)} \text{,} \\*
        \delta f^{c,+}_{\evec{k}h22} &= (\delta f^{c,-}_{\evec{k}h22})^* \quad \text{(complex)} \text{,} \\*
        \delta f^{c,+}_{\evec{k}h12} &= (\delta f^{c,-}_{\evec{k}h12})^* = \delta f^{c,+}_{\evec{k}h21}
        = (\delta f^{c,-}_{\evec{k}h21})^* \quad \text{(complex)} \text{.}
    \end{align}
\end{subequations}

Note that these relations are satisfied for perturbations when the sum rule~\cref{eq:sum-rule-Wightman} is satisfied by the \emph{full} adiabatic solution. This is true \eg~for a full thermal solution and for the free particle solution~\cref{eq:adiabatic-local-wightman}. One can still use approximative forms for the adiabatic solutions in various expressions for the sources and collision terms.

%%%%%%%%%%%%%%%%%%%%%%%%%%%%%%%%%%%%%%%%%%%%%%%%%%%%%%%%%%%%%%%%%%%%%%%%%%%%%%%
%
\section{Neutrino collision term traces}
\label{sec:collision-term-traces}
%
%%%%%%%%%%%%%%%%%%%%%%%%%%%%%%%%%%%%%%%%%%%%%%%%%%%%%%%%%%%%%%%%%%%%%%%%%%%%%%%

Here we give explicit results for the collision term trace functions $C^{sss}_{\evec{k}hilj}$ which we use in the mass shell equation~\cref{eq:delta-f-m-equation} of the Majorana neutrinos. We actually need only the $s = +1$ component $C^{+++}_{\evec{k}hilj}$ because of the relations~\cref{eq:delta-f-m-c-component-constraints} and because we only solve the positive energy solutions. Using the definition~\cref{eq:collision_term_trace_def} and the leading order expansion~\cref{eq:local-Sigma-eff-LO} for the effective self-energy, we get first $C^{+++}_{\evec{k}hilj} = \im \tr\bigl[ \mathcal{P}_{\evec{k}hji}^{++} \bar\Sigma^r_{{\rm eq},il}(k_{+l}) \mathcal{P}_{\evec{k}hlj}^{++} \bigr]$. Here the self-energy is given in the Wigner representation and evaluated at the four-momentum $(k_{+l})^\mu \equiv (+\omega_{\evec{k}l}, \evec{k})$.

Next we observe that in the case with two flavours ($i,l,j = 1,2$) $\mathcal{P}^{ss}_{\evec{k}hlj} \mathcal{P}^{ss}_{\evec{k}hji} = \mathcal{P}^{ss}_{\evec{k}hli}$, which implies that $C^{+++}_{\evec{k}hilj} = \im \tr\bigl[ \mathcal{P}_{\evec{k}hli}^{++} \bar\Sigma^r_{{\rm eq},il}(k_{+l}) \bigr]$. Notably, the collision function is in this specific case independent of the last flavour index $j$. Using equations~\cref{eq:neutrino-self-energy-parametrisations,eq:renormalised-general-Sigma-p-split} for the equilibrium Majorana self-energy functions with $\bar \Sigma^r = \gamma_0 \Sigma^r$ and $\Sigma^r = \Sigma^{\rm H} - \im \Sigma^{\mathcal{A}}$, and performing the Dirac matrix trace yields the result
\begin{align}
    C^{+++}_{\evec{k}hilj} = \frac{\cw N^m_{\evec{k}il}}{2 \omega_{\evec{k}i} \omega_{\evec{k}l}} &
    \Bigl[\Re(y_i^* y_l) \bigl(m_i k_{+l} + m_l k_{+i}\bigr)^\mu
        - \im h \Im(y_i^* y_l) \bigl(m_i k^\perp_{+l} + m_l k^\perp_{+i} \bigr)^\mu
    \Bigr] \notag
    \\*
    &{}\times{} \Bigl(
        \mathfrak{S}^\mathcal{A}_{\rm eq}(k_{+l}) + \im \mathfrak{S}^{{\rm H}(T)}_{\rm eq}(k_{+l})
    \Bigr)_\mu \text{.}
    \label{eq:Coll_trace_function}
\end{align}
The self-energy four-vector functions $\mathfrak{S}_\mu$ are calculated in appendix~\cref{sec:neutrino-self-energies}. We also defined $N_{\evec{k}ij}^m \equiv N_{\evec{k}hij}^{ss}$ (see equation~\cref{eq:local-correlator-normalisation}) and $(k^\perp)^\mu \equiv (\abs{\evec{k}}, k^0 \Uevec{k})$.

%%%%%%%%%%%%%%%%%%%%%%%%%%%%%%%%%%%%%%%%%%%%%%%%%%%%%%%%%%%%%%%%%%%%%%%%%%%%%%%
%
\section{Lepton CP-source and washout terms}
\label{sec:cp-source-and-washout-appendix}
%
%%%%%%%%%%%%%%%%%%%%%%%%%%%%%%%%%%%%%%%%%%%%%%%%%%%%%%%%%%%%%%%%%%%%%%%%%%%%%%%

Explicit forms for the CP-source and washout terms were given in equations~\cref{eq:cp-source-with-distributions,eq:washout-1-with-distributions,eq:washout-2-with-distributions}. We now substitute the self-energy functions defined in~\cref{eq:neutrino-self-energy-absorptive-equilibrium,eq:neutrino-self-energy-wightman-delta} to these expressions and use the symmetry properties~\cref{eq:ab-A-eq-symmetries,eq:ab-Wightman-delta-symmetries}. We also use the lowest order result~\cref{eq:local-Sigma-eff-LO} for the effective self-energy, adapted for $\bar\Sigma^\mathcal{A}_{\rm L}\equiv P_{\rm R}\bar\Sigma^\mathcal{A}$ and the constraints~\cref{eq:delta-f-m-c-component-constraints} for the non-equilibrium distribution functions. We keep the coherence shell functions here for completeness, but split the results to separate mass and coherence shell parts, $S_{CP} = S_{CP}^m + S_{CP}^c$ and $\delta W = \delta W^m + \delta W^c$. After performing the traces, we get
\begin{subequations}
\begin{alignat}{2}
    S_{CP}^m ={}&& \sum_{i,j} \int \frac{\dd^3\evec{k}}{(2\pi)^3}
    \frac{\cw N^m_{\evec{k}ij}}{2 \omega_{\evec{k}i} \omega_{\evec{k}j}}
    \Bigl[
        &\Re(y_i^* y_j) \textstyle{\sum_h} h \Re(\delta f^{m,+}_{\evec{k}hij})
        \bigl(m_i k^\perp_{+j} + m_j k^\perp_{+i}\bigr)^\mu \notag
        \\*[-0.7em]
        && {}-{} &\Im(y_i^* y_j) \textstyle{\sum_h} \Im(\delta f^{m,+}_{\evec{k}hij})
        \bigl(m_i k_{+j} + m_j k_{+i} \bigr)^{\mu}
    \Bigr] \notag
    \\*
    && {}\times{} &\Bigl(
        \mathfrak{S}^\mathcal{A}_{\rm eq}(k_{+i}) + \mathfrak{S}^\mathcal{A}_{\rm eq}(k_{+j})
    \Bigr)_{\mu} \text{,} \label{eq:cp-source-simplified} \displaybreak[0]
    \\
    S_{CP}^c ={}&& \sum_{i,j} \int \frac{\dd^3\evec{k}}{(2\pi)^3}
    \frac{\cw N^c_{\evec{k}ij}}{2 \omega_{\evec{k}i} \omega_{\evec{k}j}}
    \Bigl[
        &{-\Re(y_i^* y_j)} \textstyle{\sum_h} h \Re(\delta f^{c,+}_{\evec{k}hij})
        \bigl(m_i k^\perp_{+j} + m_j k^\perp_{-i}\bigr)^\mu \notag
        \\*[-0.7em]
        && {}-{} &\Im(y_i^* y_j) \textstyle{\sum_h} \Im(\delta f^{c,+}_{\evec{k}hij})
        \bigl(m_i k_{+j} + m_j k_{-i} \bigr)^{\mu}
    \Bigr] \notag
    \\*
    && {}\times{} &\Bigl(
        \mathfrak{S}^\mathcal{A}_{\rm eq}(k_{-i}) + \mathfrak{S}^\mathcal{A}_{\rm eq}(k_{+j})
    \Bigr)_{\mu} \text{,}
\end{alignat}
\end{subequations}
and
\begin{subequations}
\begin{alignat}{2}
    \frac{\delta W^m}{\beta\mu_\ell} ={}&& \sum_{i,j} \int \frac{\dd^3\evec{k}}{(2\pi)^3}
    \frac{\cw N^m_{\evec{k}ij}}{2 \omega_{\evec{k}i} \omega_{\evec{k}j}}
    \Bigl[
        &\Re(y_i^* y_j) \textstyle{\sum_h} \Re(\delta f^{m,+}_{\evec{k}hij})
        \bigl(m_i k_{+j} + m_j k_{+i}\bigr)^\mu \notag
        \\*[-0.7em]
        && {}-{} &\Im(y_i^* y_j) \textstyle{\sum_h} h \Im(\delta f^{m,+}_{\evec{k}hij})
        \bigl(m_i k^\perp_{+j} + m_j k^\perp_{+i} \bigr)^\mu
    \Bigr] \notag
    \\*
    && {}\times{} &\Bigl(
        \delta\mathfrak{S}^\mathcal{A}(k_{+i}) + \delta\mathfrak{S}^\mathcal{A}(k_{+j})
    \Bigr)_\mu \text{,} \label{eq:washout-1-simplified} \displaybreak[0]
    \\
    \frac{\delta W^c}{\beta\mu_\ell} ={}&& \sum_{i,j} \int \frac{\dd^3\evec{k}}{(2\pi)^3}
    \frac{\cw N^c_{\evec{k}ij}}{2 \omega_{\evec{k}i} \omega_{\evec{k}j}}
    \Bigl[
        &{-\Re(y_i^* y_j)} \textstyle{\sum_h} \Re(\delta f^{c,+}_{\evec{k}hij})
        \bigl(m_i k_{+j} + m_j k_{-i}\bigr)^\mu \notag
        \\*[-0.7em]
        && {}-{} &\Im(y_i^* y_j) \textstyle{\sum_h} h \Im(\delta f^{c,+}_{\evec{k}hij})
        \bigl(m_i k^\perp_{+j} + m_j k^\perp_{-i} \bigr)^\mu
    \Bigr] \notag
    \\*
    && {}\times{} &\Bigl(
        \delta\mathfrak{S}^\mathcal{A}(k_{-i}) + \delta\mathfrak{S}^\mathcal{A}(k_{+j})
    \Bigr)_\mu \text{.}
\end{alignat}
\end{subequations}
Here $\mu_\ell$ is the lepton chemical potential, $\beta = 1/T$, and we further defined $(k^\perp)^\mu \equiv (\abs{\evec{k}}, k^0 \Uevec{k})$. The self-energy functions $\mathfrak{S}^{\mathcal{A}}_{{\rm eq},\mu}$ and $\delta\mathfrak{S}^{\mathcal{A}}_\mu = (\delta\mathfrak{S}^{\gt}_\mu + \delta\mathfrak{S}^{\lt}_\mu)/2$ are calculated in section~\cref{sec:neutrino-self-energies}. We also defined here $N_{\evec{k}ij}^m \equiv N_{\evec{k}hij}^{ss}$ and $N_{\evec{k}ij}^c \equiv N_{\evec{k}hij}^{s,-s}$ using equation~\cref{eq:local-correlator-normalisation}. Finally, the adiabatic washout term $W_{\rm ad}$ can be simplified to
\begin{equation}
    \frac{W_{\rm ad}}{\beta\mu_\ell} = \cw \sum_i \int \frac{\dd^3\evec{k}}{(2\pi)^3}
    \frac{2\abs{y_i}^2}{\omega_{\evec{k}i}} \Bigl[
        2f_{\rm FD}(\omega_{\evec{k}i}) \,k_{+i} \cdot \delta\mathfrak{S}^{\mathcal{A}}(k_{+i})
        - k_{+i} \cdot \delta\mathfrak{S}^{\lt}(k_{+i})
    \Bigr] \text{.}
    \label{eq:washout-2-simplified}
\end{equation}

The results~\cref{eq:cp-source-simplified,eq:washout-1-simplified} show explicitly how the helicity sums of the non-equilibrium distribution functions $\delta f_{\evec{k}h}$ enter the leading order CP-source and washout terms. In particular, from equation~\cref{eq:cp-source-simplified} we see that CP-violation is only sourced by the helicity-odd combinations $\Re(\delta f_{\evec{k},+1} - \delta f_{\evec{k},-1})$ of the real parts and helicity-even combinations $\Im(\delta f_{\evec{k},+1} + \delta f_{\evec{k},-1})$ of the imaginary parts.

%%%%%%%%%%%%%%%%%%%%%%%%%%%%%%%%%%%%%%%%%%%%%%%%%%%%%%%%%%%%%%%%%%%%%%%%%%%%%%%
%
\section{Semiclassical Boltzmann equations}
\label{sec:Boltzmann_equations}
%
%%%%%%%%%%%%%%%%%%%%%%%%%%%%%%%%%%%%%%%%%%%%%%%%%%%%%%%%%%%%%%%%%%%%%%%%%%%%%%%

For comparison with our main quantum kinetic equations, we implement and solve numerically the semiclassical Boltzmann equations (see \eg~\cite{Kolb:1979qa,Luty:1992un,Giudice:2003jh,Basboll:2006yx,Pilaftsis:2003gt}) and the momentum integrated rate equations in our model~\cref{eq:lagrangian}. We include only the decay and inverse decay contributions supplemented by a RIS-correction term required to cure the problem of spurious equilibrium source~\cite{Giudice:2003jh,Basboll:2006yx,Kolb:1979qa}. We write the equations first following ref.~\cite{Luty:1992un}, to establish the notation and to facilitate comparison with the literature. We then present the equations in a more compact form which is directly comparable to our main equations and also more suitable for a numerical implementation.

%%%%%%%%%%%%%%%%%%%%%%%%%%%%%%%%%%%%%%%%%%%%%%%%%%%%%%%%%%%%%%%%%%%%%%%%%%%%%%%
%
\subsection{Momentum-dependent equations}
\label{sec:mom-dep-equations-appendix}
%
%%%%%%%%%%%%%%%%%%%%%%%%%%%%%%%%%%%%%%%%%%%%%%%%%%%%%%%%%%%%%%%%%%%%%%%%%%%%%%%

As in the main text, we assume that SM-fields are in kinetic equilibrium and that Higgs field chemical potential vanishes, $\mu_\phi = 0$. However, we make no assumption about the form of the phase space distribution functions $f_i$ of the Majorana neutrinos $i = 1,2$, and consider full quantum statistics with Pauli blocking and stimulated emission factors. In the expanding universe, the Boltzmann equations for $f_i$ and the lepton asymmetry $n_L$ can then be written as
\begin{align}
    \frac{\partial f_i}{\partial t}
    - \abs{\evec{p}_i} H \frac{\partial f_i}{\partial \abs{\evec{p}_i}}
    ={}& \frac{1}{2\omega_i g_i} \int \dd \pi_{\ell} \dd \pi_{\phi} \,
    (2\pi)^4 \delta(\omega_i - \omega_{\ell} - \omega_{\phi})
    \delta^{(3)}(\evec{p}_i - \evec{p}_{\ell} - \evec{p}_{\phi}) \notag
    \\*
    &{} \times \abs{\mathcal{M}_i}^2 \Bigl\{
        -f_i (1 - f_{\ell}^{+}) (1 + f_{\phi}^{\rm eq})
        + f_{\ell}^{+} f_{\phi}^{\rm eq} (1 - f_i) \notag
    \\*
        &\phantom{\times \abs{\mathcal{M}_i}^2 \Bigl\{}
        {} + \epsilon_i^{CP} f_{\ell}^{-} \Bigl[
            f_i (1 + f_{\phi}^{\rm eq}) - f_{\phi}^{\rm eq} (1 - f_i)
        \Bigr]
    \Bigr\} \text{,} \label{eq:Boltzmann_fi}
    \\
    \frac{\dd n_L}{\dd t} + 3 H n_L
    ={}& \sum_i \int \dd \pi_i \dd \pi_{\ell} \dd \pi_{\phi} \,
    (2\pi)^4 \delta(\omega_i - \omega_{\ell} - \omega_{\phi})
    \delta^{(3)}(\evec{p}_i - \evec{p}_{\ell} - \evec{p}_{\phi}) \notag
    \\*
    &{} \times \abs{\mathcal{M}_i}^2 \Bigl\{
        -f_{\ell}^{-} \Bigl[ f_{\phi}^{\rm eq} (1 - f_i)
        + f_i (1 + f_{\phi}^{\rm eq}) \Bigr] \notag
    \\*
        &\phantom{\times \abs{\mathcal{M}_i}^2 \Bigl\{}
        {} + \epsilon_i^{CP} \Bigl[
            -f_{\ell}^{+} f_{\phi}^{\rm eq} (1 - f_i)
            + f_i (1 - f_L^{+}) (1 + f_{\phi}^{\rm eq})
        \Bigr]
    \Bigr\} \text{.} \label{eq:Boltzmann_nL}
\end{align}
Here $f_x(\evec{p}_x, t)$ is the phase space distribution function, $\omega_x \equiv \sqrt{\abs{\evec{p}_x}^2 + m_x^2}$ is the on-shell energy and $\dd \pi_x \equiv \dd^3 \evec{p}_x/(2\pi)^3/(2\omega_x)$ is the phase space integration element for particle species $x = i, \ell, \phi$. The RIS-subtraction has been performed as in~\cite{Giudice:2003jh,Basboll:2006yx,Kolb:1979qa}, which ensures that also the term associated with the CP-asymmetry parameter $\epsilon_i^{CP}$ in equation~\cref{eq:Boltzmann_nL} vanishes in thermal equilibrium. For the number densities and distribution functions we are using the following notations:
\begin{subequations}
\begin{alignat}{3}
    n_L &\equiv n_{\ell} - \bar{n}_{\ell} \text{,} &\qquad
    n_{\ell} &= \cw \int \frac{\dd^3 \evec{p}_{\ell}}{(2\pi)^3} \, f_{\ell} \text{,} &\qquad
    \bar{n}_{\ell} &= \cw \int \frac{\dd^3 \evec{p}_{\ell}}{(2\pi)^3} \, \bar{f}_{\ell} \text{,}
    \\*
    f_{\ell}^{\pm} &\equiv \bigl(f_{\ell} \pm \bar{f}_{\ell}\bigr)/2 \text{,} &\qquad
    f_{\ell} &= f_{\rm FD}(\omega_{\ell} - \mu_{\ell}) \text{,} &\qquad
    \bar{f}_{\ell} &= f_{\rm FD}(\omega_{\ell} + \mu_{\ell}) \text{,}
\end{alignat}
\begin{equation}
    f_{\phi} = \bar{f}_{\phi} = f_{\phi}^{\rm eq} = f_{\rm BE}(\omega_{\phi}) \text{,}
\end{equation}
\end{subequations}
where $f_{\rm FD}$ and $f_{\rm BE}$ are the Fermi--Dirac and Bose--Einstein distribution functions~\cref{eq:FD-and-BE-distributions}. The factor $\cw = 2$ is the SM doublet multiplicity and $g_i = 2$ is the number of spin (or helicity) states of the Majorana neutrino $N_i$. Finally, the tree level matrix element for the neutrino decay process, summed over the Majorana neutrino spins and the SM doublet multiplicity is:
\begin{equation}
    \abs{\mathcal{M}_i}^2 \equiv
    \abs[\big]{\mathcal{M}(N_i \to \ell \phi)}^2
    + \abs[\big]{\mathcal{M}(N_i \to \bar\ell \bar\phi)}^2
    = 2 \cw \abs{y_i}^2 m_i^2 \text{,}
\end{equation}
where $m_{\ell} = m_{\phi} = 0$ was assumed in the end. The corresponding total decay width is then $\smash{\Gamma_i^{(0)}} \equiv \abs{\mathcal{M}_i}^2/(16 \pi g_im_i) = \abs{y_i}^2 m_i/(8 \pi)$. In the following we use these tree level results for the leading approximation.

We now give equations~\cref{eq:Boltzmann_fi,eq:Boltzmann_nL} in a more compact dimensionless form similar to~\cref{eq:final-numerical-neutrino-equation,eq:final-numerical-lepton-equation}, in terms of the variable $z = m_1/T$. We also omit the lepton backreaction term (proportional to $\epsilon_i^{CP} f_{\ell}^{-}$) in equation~\cref{eq:Boltzmann_fi}, as we did in~\cref{eq:final-numerical-neutrino-equation}. Working consistently to linear order in perturbations we then find
\begin{align}
    \frac{\dd \delta f_{\evec{k}i}}{\dd z} &= -\tilde{C}_{\evec{k}i} \,\delta f_{\evec{k}i} - \frac{\dd f_{\evec{k}i}^{\rm eq}}{\dd z}
    \text{,} \label{eq:Boltzmann_fi_dimensionless}
    \\*
    \frac{\dd Y_L}{\dd z} &= \tilde{S}_{CP} + (\delta \tilde{W} + \tilde{W}_{\rm ad}) Y_L
    \text{,} \label{eq:Boltzmann_nL_dimensionless}
\end{align}
where $Y_L = n_L/s$, $s$ is the entropy density, $\delta f_{\evec{k}i} \equiv f_{\evec{k}i} - f_{\evec{k}i}^{\rm eq}$ and $f_{\evec{k}i}^{\rm eq} = f_{\rm FD}(\omega_{\evec{k}i})$. The dimensionless neutrino collision term coefficient, CP-violating lepton source term and the lepton washout term coefficients are here given by
\begin{align}
    \tilde{C}_{\evec{k}i} &= \frac{z \Gamma_i^{(0)}}{H_1} \Bigl(
        \frac{m_i}{\omega_{\evec{k}i}} \mathcal{X}_{\evec{k}i}
    \Bigr) \text{,} \label{eq:Boltzmann_damping_rate} \displaybreak[0]
    \\
    \tilde{S}_{CP} &= \sum_i \frac{z \Gamma_i^{(0)}}{H_1} \frac{g_i}{s} \,\epsilon_i^{CP}
    \int \frac{\dd^3 \evec{k}}{(2\pi)^3} \Bigl(
        \frac{m_i}{\omega_{\evec{k}i}} \mathcal{X}_{\evec{k}i}
    \Bigr) \delta f_{\evec{k}i} \text{,} \label{eq:Boltzmann_SCP} \displaybreak[0]
    \\
    \delta \tilde{W} &= \sum_i \frac{z \Gamma_i^{(0)}}{H_1} \frac{6 g_i}{\cw T^3}
    \int \frac{\dd^3 \evec{k}}{(2\pi)^3} \Bigl(
        \frac{m_i}{\omega_{\evec{k}i}} \mathcal{Y}^{(A)}_{\evec{k}i}
    \Bigr) \delta f_{\evec{k}i} \text{,} \displaybreak[0]
    \\
    \tilde{W}_{\rm ad} &= \sum_i \frac{z \Gamma_i^{(0)}}{H_1} \frac{6 g_i}{\cw T^3}
    \int \frac{\dd^3 \evec{k}}{(2\pi)^3} \Bigl(
        \frac{m_i}{\omega_{\evec{k}i}} \mathcal{Y}^{(B)}_{\evec{k}i}
    \Bigr) \text{,}
    \label{eq:Boltzmann_washout_B}
\shortintertext{with $H_1 \equiv H(m_1)$ and}
    \mathcal{X}_{\evec{k}i} &= \frac{T}{\abs{\evec{k}}} \log\biggl[
        \frac{\e^{(\omega_{\evec{k}i} + \abs{\evec{k}})/T} - 1}
        {\e^{\omega_{\evec{k}i}/T} - \e^{\abs{\evec{k}}/T}}
    \biggr] \text{,} \displaybreak[0]
    \\
    \mathcal{Y}^{(A)}_{\evec{k}i} &= \frac{T}{\abs{\evec{k}}} \biggl[
        f_{\rm FD}\Bigl(\frac{\omega_{\evec{k}i} + \abs{\evec{k}}}{2}\Bigr)
        - f_{\rm FD}\Bigl(\frac{\omega_{\evec{k}i} - \abs{\evec{k}}}{2}\Bigr)
    \biggr] \text{,} \displaybreak[0]
    \\
    \mathcal{Y}^{(B)}_{\evec{k}i} &= f_{\rm FD}(\omega_{\evec{k}i})
    \bigl(f_{\rm FD}(\omega_{\evec{k}i}) - 1\bigr) \mathcal{X}_{\evec{k}i} \text{.}
\end{align}
For the total z-derivative of $f_{\evec{k}i}^{\rm eq}$ we use $\dd f_{\evec{k}i}^{\rm eq}/\dd z = f_{\evec{k}i}^{\rm eq}(f_{\evec{k}i}^{\rm eq} - 1) m_i^2/(m_1 \omega_{\evec{k}i})$, assuming that $\abs{\evec{k}}/T$ is independent of $z$. Note also that $n_L \tilde{W}_{\rm ad} H_1/z$ is exactly equal to the $W_{\rm ad}$ defined by~\cref{eq:washout-2-simplified}, when we use $n_L \approx \cw T^2 \mu_\ell/6$.

%%%%%%%%%%%%%%%%%%%%%%%%%%%%%%%%%%%%%%%%%%%%%%%%%%%%%%%%%%%%%%%%%%%%%%%%%%%%%%%
%
\subsection{Rate equations}
%
%%%%%%%%%%%%%%%%%%%%%%%%%%%%%%%%%%%%%%%%%%%%%%%%%%%%%%%%%%%%%%%%%%%%%%%%%%%%%%%

Simplified rate equations for the number densities can be derived from the Boltzmann equations~\cref{eq:Boltzmann_fi,eq:Boltzmann_nL} with two additional assumptions: kinetic approximation for the Majorana neutrino distributions and Maxwell--Boltzmann statistics for all particle species. To this end, we use $f_i = (n_i/n_i^{\rm eq})f_i^{\rm eq}$, replace $f_{\rm FD}$ and $f_{\rm BE}$ by $f_{\rm MB}(k^0) \equiv \e^{-k^0/T}$ and remove all extra terms originating from the Pauli blocking and stimulated emission factors. Integrating equations~\cref{eq:Boltzmann_fi,eq:Boltzmann_nL} over momenta then gives:
\begin{align}
    \frac{\dd n_i}{\dd t} + 3 H n_i &= -\biggl(\frac{n_i}{n_i^{\rm eq}} - 1\biggr) \gamma_i
    - \frac{n_L}{2 n_{\ell}^{\rm eq}} \epsilon_i^{CP} \gamma_i \text{,}
    \label{eq:integrated_Boltzmann_fi}
    \\*
    \frac{\dd n_L}{\dd t} + 3 H n_L &= \sum_i \biggl[
        \biggl(\frac{n_i}{n_i^{\rm eq}} - 1\biggr) \epsilon_i^{CP} \gamma_i
        - \frac{n_L}{2 n_{\ell}^{\rm eq}} \gamma_i
    \biggr] \text{,}
    \label{eq:integrated_Boltzmann_nL}
\end{align}
where
\begin{alignat}{2}
    \gamma_i &= g_i \Gamma_i \int \frac{\dd^3 \evec{p}_i}{(2\pi)^3} \frac{m_i}{\omega_i} \e^{-\omega_i/T}
    &{}={}& n_i^{\rm eq} \frac{\mathcal{K}_1(m_i/T)}{\mathcal{K}_2(m_i/T)} \Gamma_i \text{,}
    \\*
    n_i^{\rm eq} &= g_i \int \frac{\dd^3 \evec{p}_i}{(2\pi)^3} \e^{-\omega_i/T}
    &{}={}& \frac{g_i m_i^2 T}{2\pi^2} \mathcal{K}_2(m_i/T) \text{,}
    \\*
    n_{\ell}^{\rm eq} &= \cw \int \frac{\dd^3 \evec{p}_{\ell}}{(2\pi)^3} \e^{-\abs{\evec{p}_{\ell}}/T}
    &{}={}& \frac{\cw T^3}{\pi^2} \text{.}
\end{alignat}
Here $\mathcal{K}_n$ are the modified Bessel functions of the second kind. Here we kept also the lepton backreaction term which is the last term on the right-hand side of equation~\cref{eq:integrated_Boltzmann_fi}.

Again, we can write equations~\cref{eq:integrated_Boltzmann_fi,eq:integrated_Boltzmann_nL} in a compact dimensionless form:
\begin{align}
    \frac{\dd \delta Y_i}{\dd z} &=
    -D_i \,\delta Y_i - \frac{\dd Y_i^{\rm eq}}{\dd z}
    - \frac{Y_i^{\rm eq}}{2 Y_{\ell}^{\rm eq}} \epsilon_i^{CP} D_i Y_L\text{,}
    \label{eq:integrated_Boltzmann_fi_dimensionless}
    \\*
    \frac{\dd Y_L}{\dd z} &= \sum_i \biggl[
        \epsilon_i^{CP} D_i \,\delta Y_i - \frac{Y_i^{\rm eq}}{2 Y_{\ell}^{\rm eq}} D_i Y_L
    \biggr] \text{,}
    \label{eq:integrated_Boltzmann_nL_dimensionless}
\end{align}
where $\delta Y_i \equiv Y_i - Y_i^{\rm eq}$ and
\begin{equation}
    Y_x = \frac{n_x}{s} \text{,} \qquad
    z = \frac{m_1}{T} \text{,} \qquad
    D_i = \frac{z \Gamma_i^{(0)}}{H_1}
        \frac{\mathcal{K}_1(z x_i)}{\mathcal{K}_2(z x_i)} \text{,} \qquad
    x_i = \frac{m_i}{m_1} \text{.}
\end{equation}
The lepton source term in these equations is given by $\tilde{S}_{CP} = \sum_i \epsilon_i^{CP} D_i \,\delta Y_i$ and for the washout term $\delta \tilde W = 0$ and $\tilde W_{\rm ad} = - \sum_i D_i Y_i^{\rm eq}/(2 Y_\ell^{\rm eq})$. We used these equations also to check that the lepton backreaction was indeed numerically negligible in all examples (with vanishing initial lepton asymmetry) that we studied.

%%%%%%%%%%%%%%%%%%%%%%%%%%%%%%%%%%%%%%%%%%%%%%%%%%%%%%%%%%%%%%%%%%%%%%%%%%%%%%%
%
\subsection{CP-asymmetry parameter}
\label{sec:CP_parameter}
%
%%%%%%%%%%%%%%%%%%%%%%%%%%%%%%%%%%%%%%%%%%%%%%%%%%%%%%%%%%%%%%%%%%%%%%%%%%%%%%%

The Boltzmann equations~\cref{eq:Boltzmann_fi,eq:Boltzmann_nL,eq:Boltzmann_fi_dimensionless,eq:Boltzmann_nL_dimensionless,eq:integrated_Boltzmann_fi,eq:integrated_Boltzmann_nL,eq:integrated_Boltzmann_fi_dimensionless,eq:integrated_Boltzmann_nL_dimensionless} given above do not depend on the exact form of the CP-asymmetry parameter $\epsilon_i^{CP}$. There we used the generic definition (in the unflavoured case)
\begin{equation}
    \epsilon_i^{CP} =
    \frac{\Gamma(N_i \to \ell \phi) - \Gamma(N_i \to \bar\ell \bar\phi)}
    {\Gamma(N_i \to \ell \phi) + \Gamma(N_i \to \bar\ell \bar\phi)} \text{,}
    \label{eq:CP-asymmetry-def}
\end{equation}
where $\Gamma(N_i \to \ell \phi) + \Gamma(N_i \to \bar\ell \bar\phi) = \Gamma_i$ is the total (vacuum) decay width of the Majorana neutrino to the lepton and Higgs doublets. The CP-asymmetry parameter vanishes at tree-level and is calculated from the interference of tree level and higher order amplitudes. This is highly nontrivial, as the calculation breaks down when using ordinary perturbation theory in the degenerate limit $m_2 \to m_1$.

We only consider the self-energy contribution (also called indirect or $\epsilon$-type CP-violation), neglecting the vertex correction (\ie~the direct $\epsilon'$-type CP-violation), and use the generic form
\begin{equation}
    \epsilon^{CP}_{i, x} =
    \frac{\Im\bigl[(y_1^* y_2)^2\bigr]}{\abs{y_1}^2 \abs{y_2}^2}
    \frac{(m_2^2 - m_1^2) m_i \Gamma_j^{(0)}}{(m_2^2 - m_1^2)^2 + (R_{ij,x})^2}
    \text{.} \qquad \text{($j \neq i$)}
    \label{eq:epsilon-type-CP-asymmetry}
\end{equation}
This result is specific to the case of two Majorana neutrinos and one lepton flavour. Here $R_{ij,x}$ is the regulator which removes the singularity that would occur when $m_2 \to m_1$. We consider four alternative regulators~\cite{Buchmuller:1997yu,Pilaftsis:1997jf,Pilaftsis:2003gt,Garny:2011hg,Dev:2017wwc} (the subscript $x$): the `mixed' regulator, the `difference' regulator, the `sum' regulator and the `effective' sum regulator given by
\begin{align}
    R_{ij,{\rm mix}} &= m_i \Gamma_j^{(0)} \text{,}
    \label{eq:epsilon-mixed-regulator} \displaybreak[0]
    \\
    R_{ij,{\rm diff}} &= m_i \Gamma_i^{(0)} - m_j \Gamma_j^{(0)} \text{,}
    \label{eq:epsilon-difference-regulator} \displaybreak[0]
    \\
    R_{ij,{\rm sum}} &= m_i \Gamma_i^{(0)} + m_j \Gamma_j^{(0)} \text{,}
    \label{eq:epsilon-sum-regulator}
    \\
    R_{ij,{\rm eff}} &= (m_i \Gamma_i^{(0)} + m_j \Gamma_j^{(0)})
    \abs{\sin\theta_{ij}} \text{.} \label{eq:epsilon-eff-regulator}
\end{align}
The relative phase $\theta_{ij}$ of the Yukawa couplings was defined in equation~\cref{eq:Yukawa_CP_phase}. Note that $\smash{\epsilon^{CP}_{i, {\rm diff}}}$ with the difference regulator is still singular if $\abs{y_1} = \abs{y_2}$. This means that it is unbound and leads to unphysical results when approaching the doubly degenerate limit $\abs{y_2} \to \abs{y_1}$ and $m_2 \to m_1$. For more discussion about the validity of different regulators, see~\cite{Dev:2014laa}.

%%%%%%%%%%%%%%%%%%%%%%%%%%%%%%%%%%%%%%%%%%%%%%%%%%%%%%%%%%%%%%%%%%%%%%%%%%%%%%%

\pagebreak % first entry of References was alone on the last page
\bibliographystyle{JHEP}
\bibliography{leptogen}

\providecommand{\href}[2]{#2}\begingroup\raggedright\begin{thebibliography}{100}

\bibitem{Bilenky:1987ty}
S.M.~Bilenky and S.T.~Petcov, \emph{{Massive Neutrinos and Neutrino
  Oscillations}}, \href{https://doi.org/10.1103/RevModPhys.59.671}{\emph{Rev.
  Mod. Phys.} {\bfseries 59} (1987) 671}.

\bibitem{Barbieri:1989ti}
R.~Barbieri and A.~Dolgov, \emph{{Bounds on Sterile-neutrinos from
  Nucleosynthesis}},
  \href{https://doi.org/10.1016/0370-2693(90)91203-N}{\emph{Phys. Lett. B}
  {\bfseries 237} (1990) 440}.

\bibitem{Enqvist:1990ad}
K.~Enqvist, K.~Kainulainen and J.~Maalampi, \emph{{Refraction and Oscillations
  of Neutrinos in the Early Universe}},
  \href{https://doi.org/10.1016/0550-3213(91)90397-G}{\emph{Nucl. Phys. B}
  {\bfseries 349} (1991) 754}.

\bibitem{Kainulainen:1990ds}
K.~Kainulainen, \emph{{Light Singlet Neutrinos and the Primordial
  Nucleosynthesis}},
  \href{https://doi.org/10.1016/0370-2693(90)90054-A}{\emph{Phys. Lett. B}
  {\bfseries 244} (1990) 191}.

\bibitem{Enqvist:1990ek}
K.~Enqvist, K.~Kainulainen and J.~Maalampi, \emph{{Resonant neutrino
  transitions and nucleosynthesis}},
  \href{https://doi.org/10.1016/0370-2693(90)91030-F}{\emph{Phys. Lett. B}
  {\bfseries 249} (1990) 531}.

\bibitem{Barbieri:1990vx}
R.~Barbieri and A.~Dolgov, \emph{{Neutrino oscillations in the early
  universe}}, \href{https://doi.org/10.1016/0550-3213(91)90396-F}{\emph{Nucl.
  Phys. B} {\bfseries 349} (1991) 743}.

\bibitem{Enqvist:1991qj}
K.~Enqvist, K.~Kainulainen and M.J.~Thomson, \emph{{Stringent cosmological
  bounds on inert neutrino mixing}},
  \href{https://doi.org/10.1016/0550-3213(92)90442-E}{\emph{Nucl. Phys. B}
  {\bfseries 373} (1992) 498}.

\bibitem{Sigl:1992fn}
G.~Sigl and G.~Raffelt, \emph{{General kinetic description of relativistic
  mixed neutrinos}},
  \href{https://doi.org/10.1016/0550-3213(93)90175-O}{\emph{Nucl. Phys. B}
  {\bfseries 406} (1993) 423}.

\bibitem{Vlasenko:2013fja}
A.~Vlasenko, G.M.~Fuller and V.~Cirigliano, \emph{{Neutrino Quantum Kinetics}},
  \href{https://doi.org/10.1103/PhysRevD.89.105004}{\emph{Phys. Rev. D}
  {\bfseries 89} (2014) 105004}
  [\href{https://arxiv.org/abs/1309.2628}{{\ttfamily 1309.2628}}].

\bibitem{Joyce:1994zn}
M.~Joyce, T.~Prokopec and N.~Turok, \emph{{Nonlocal electroweak baryogenesis.
  Part 1: Thin wall regime}},
  \href{https://doi.org/10.1103/PhysRevD.53.2930}{\emph{Phys. Rev.} {\bfseries
  D53} (1996) 2930} [\href{https://arxiv.org/abs/hep-ph/9410281}{{\ttfamily
  hep-ph/9410281}}].

\bibitem{Cline:1997vk}
J.M.~Cline, M.~Joyce and K.~Kainulainen, \emph{{Supersymmetric electroweak
  baryogenesis in the WKB approximation}},
  \href{https://doi.org/10.1016/S0370-2693(99)00033-7,
  10.1016/S0370-2693(97)01361-0}{\emph{Phys. Lett.} {\bfseries B417} (1998) 79}
  [\href{https://arxiv.org/abs/hep-ph/9708393}{{\ttfamily hep-ph/9708393}}].

\bibitem{Kainulainen:2001cn}
K.~Kainulainen, T.~Prokopec, M.G.~Schmidt and S.~Weinstock, \emph{{First
  principle derivation of semiclassical force for electroweak baryogenesis}},
  \href{https://doi.org/10.1088/1126-6708/2001/06/031}{\emph{JHEP} {\bfseries
  06} (2001) 031} [\href{https://arxiv.org/abs/hep-ph/0105295}{{\ttfamily
  hep-ph/0105295}}].

\bibitem{Kainulainen:2002th}
K.~Kainulainen, T.~Prokopec, M.G.~Schmidt and S.~Weinstock,
  \emph{{Semiclassical force for electroweak baryogenesis: Three-dimensional
  derivation}}, \href{https://doi.org/10.1103/PhysRevD.66.043502}{\emph{Phys.
  Rev.} {\bfseries D66} (2002) 043502}
  [\href{https://arxiv.org/abs/hep-ph/0202177}{{\ttfamily hep-ph/0202177}}].

\bibitem{Cline:2020jre}
J.M.~Cline and K.~Kainulainen, \emph{{Electroweak baryogenesis at high wall
  velocities}},  \href{https://arxiv.org/abs/2001.00568}{{\ttfamily
  2001.00568}}.

\bibitem{Nelson:1991ab}
A.E.~Nelson, D.B.~Kaplan and A.G.~Cohen, \emph{{Why there is something rather
  than nothing: Matter from weak interactions}},
  \href{https://doi.org/10.1016/0550-3213(92)90440-M}{\emph{Nucl. Phys.}
  {\bfseries B373} (1992) 453}.

\bibitem{Huet:1995mm}
P.~Huet and A.E.~Nelson, \emph{{CP violation and electroweak baryogenesis in
  extensions of the standard model}},
  \href{https://doi.org/10.1016/0370-2693(95)00674-A}{\emph{Phys. Lett.}
  {\bfseries B355} (1995) 229}
  [\href{https://arxiv.org/abs/hep-ph/9504427}{{\ttfamily hep-ph/9504427}}].

\bibitem{Riotto:1997gu}
A.~Riotto, \emph{{More about electroweak baryogenesis in the minimal
  supersymmetric standard model}},
  \href{https://doi.org/10.1142/S0218271898000541}{\emph{Int. J. Mod. Phys.}
  {\bfseries D7} (1998) 815}
  [\href{https://arxiv.org/abs/hep-ph/9709286}{{\ttfamily hep-ph/9709286}}].

\bibitem{Postma:2019scv}
M.~Postma and J.~Van De~Vis, \emph{{Source terms for electroweak baryogenesis
  in the vev-insertion approximation beyond leading order}},
  \href{https://arxiv.org/abs/1910.11794}{{\ttfamily 1910.11794}}.

\bibitem{Fukugita:1986hr}
M.~Fukugita and T.~Yanagida, \emph{{Baryogenesis Without Grand Unification}},
  \href{https://doi.org/10.1016/0370-2693(86)91126-3}{\emph{Phys. Lett. B}
  {\bfseries 174} (1986) 45}.

\bibitem{Davidson:2008bu}
S.~Davidson, E.~Nardi and Y.~Nir, \emph{{Leptogenesis}},
  \href{https://doi.org/10.1016/j.physrep.2008.06.002}{\emph{Phys. Rept.}
  {\bfseries 466} (2008) 105}
  [\href{https://arxiv.org/abs/0802.2962}{{\ttfamily 0802.2962}}].

\bibitem{Blanchet:2012bk}
S.~Blanchet and P.~Di~Bari, \emph{{The minimal scenario of leptogenesis}},
  \href{https://doi.org/10.1088/1367-2630/14/12/125012}{\emph{New J. Phys.}
  {\bfseries 14} (2012) 125012}
  [\href{https://arxiv.org/abs/1211.0512}{{\ttfamily 1211.0512}}].

\bibitem{Chung:1999ve}
D.J.H.~Chung, E.W.~Kolb, A.~Riotto and I.I.~Tkachev, \emph{{Probing Planckian
  physics: Resonant production of particles during inflation and features in
  the primordial power spectrum}},
  \href{https://doi.org/10.1103/PhysRevD.62.043508}{\emph{Phys. Rev. D}
  {\bfseries 62} (2000) 043508}
  [\href{https://arxiv.org/abs/hep-ph/9910437}{{\ttfamily hep-ph/9910437}}].

\bibitem{Kofman:1986wm}
L.A.~Kofman and A.D.~Linde, \emph{{Generation of Density Perturbations in the
  Inflationary Cosmology}},
  \href{https://doi.org/10.1016/0550-3213(87)90698-5}{\emph{Nucl. Phys. B}
  {\bfseries 282} (1987) 555}.

\bibitem{Fairbairn:2018bsw}
M.~Fairbairn, K.~Kainulainen, T.~Markkanen and S.~Nurmi, \emph{{Despicable Dark
  Relics: generated by gravity with unconstrained masses}},
  \href{https://doi.org/10.1088/1475-7516/2019/04/005}{\emph{JCAP} {\bfseries
  04} (2019) 005} [\href{https://arxiv.org/abs/1808.08236}{{\ttfamily
  1808.08236}}].

\bibitem{Kuzmin:1985mm}
V.A.~Kuzmin, V.A.~Rubakov and M.E.~Shaposhnikov, \emph{{On the Anomalous
  Electroweak Baryon Number Nonconservation in the Early Universe}},
  \href{https://doi.org/10.1016/0370-2693(85)91028-7}{\emph{Phys. Lett. B}
  {\bfseries 155} (1985) 36}.

\bibitem{tHooft:1976snw}
G.~'t~Hooft, \emph{{Computation of the Quantum Effects Due to a
  Four-Dimensional Pseudoparticle}},
  \href{https://doi.org/10.1103/PhysRevD.14.3432}{\emph{Phys. Rev. D}
  {\bfseries 14} (1976) 3432}.

\bibitem{Klinkhamer:1984di}
F.R.~Klinkhamer and N.S.~Manton, \emph{{A Saddle Point Solution in the
  Weinberg-Salam Theory}},
  \href{https://doi.org/10.1103/PhysRevD.30.2212}{\emph{Phys. Rev. D}
  {\bfseries 30} (1984) 2212}.

\bibitem{Arnold:1987mh}
P.B.~Arnold and L.D.~McLerran, \emph{{Sphalerons, Small Fluctuations and Baryon
  Number Violation in Electroweak Theory}},
  \href{https://doi.org/10.1103/PhysRevD.36.581}{\emph{Phys. Rev. D} {\bfseries
  36} (1987) 581}.

\bibitem{Arnold:1987zg}
P.B.~Arnold and L.D.~McLerran, \emph{{The Sphaleron Strikes Back}},
  \href{https://doi.org/10.1103/PhysRevD.37.1020}{\emph{Phys. Rev. D}
  {\bfseries 37} (1988) 1020}.

\bibitem{Schwinger:1960qe}
J.S.~Schwinger, \emph{{Brownian motion of a quantum oscillator}},
  \href{https://doi.org/10.1063/1.1703727}{\emph{J. Math. Phys.} {\bfseries 2}
  (1961) 407}.

\bibitem{Keldysh:1964ud}
L.V.~Keldysh, \emph{{Diagram technique for nonequilibrium processes}},
  {\emph{Zh. Eksp. Teor. Fiz.} {\bfseries 47} (1964) 1515}.

\bibitem{Baym:1961zz}
G.~Baym and L.P.~Kadanoff, \emph{{Conservation Laws and Correlation
  Functions}}, \href{https://doi.org/10.1103/PhysRev.124.287}{\emph{Phys. Rev.}
  {\bfseries 124} (1961) 287}.

\bibitem{Kadanoff:1962book}
L.P.~Kadanoff and G.~Baym, \emph{{Quantum Statistical Mechanics}}, Benjamin,
  New York (1962).

\bibitem{Cornwall:1974vz}
J.M.~Cornwall, R.~Jackiw and E.~Tomboulis, \emph{{Effective Action for
  Composite Operators}},
  \href{https://doi.org/10.1103/PhysRevD.10.2428}{\emph{Phys. Rev.} {\bfseries
  D10} (1974) 2428}.

\bibitem{Calzetta:1986cq}
E.~Calzetta and B.L.~Hu, \emph{{Nonequilibrium Quantum Fields: Closed Time Path
  Effective Action, Wigner Function and Boltzmann Equation}},
  \href{https://doi.org/10.1103/PhysRevD.37.2878}{\emph{Phys. Rev.} {\bfseries
  D37} (1988) 2878}.

\bibitem{Greiner:1998vd}
C.~Greiner and S.~Leupold, \emph{{Stochastic interpretation of Kadanoff-Baym
  equations and their relation to Langevin processes}},
  \href{https://doi.org/10.1006/aphy.1998.5849}{\emph{Annals Phys.} {\bfseries
  270} (1998) 328} [\href{https://arxiv.org/abs/hep-ph/9802312}{{\ttfamily
  hep-ph/9802312}}].

\bibitem{Buchmuller:2000nd}
W.~Buchmüller and S.~Fredenhagen, \emph{{Quantum mechanics of baryogenesis}},
  \href{https://doi.org/10.1016/S0370-2693(00)00573-6}{\emph{Phys. Lett.}
  {\bfseries B483} (2000) 217}
  [\href{https://arxiv.org/abs/hep-ph/0004145}{{\ttfamily hep-ph/0004145}}].

\bibitem{Hohenegger:2008zk}
A.~Hohenegger, A.~Kartavtsev and M.~Lindner, \emph{{Deriving Boltzmann
  Equations from Kadanoff-Baym Equations in Curved Space-Time}},
  \href{https://doi.org/10.1103/PhysRevD.78.085027}{\emph{Phys. Rev.}
  {\bfseries D78} (2008) 085027}
  [\href{https://arxiv.org/abs/0807.4551}{{\ttfamily 0807.4551}}].

\bibitem{Anisimov:2008dz}
A.~Anisimov, W.~Buchmüller, M.~Drewes and S.~Mendizabal, \emph{{Nonequilibrium
  Dynamics of Scalar Fields in a Thermal Bath}},
  \href{https://doi.org/10.1016/j.aop.2009.01.001}{\emph{Annals Phys.}
  {\bfseries 324} (2009) 1234}
  [\href{https://arxiv.org/abs/0812.1934}{{\ttfamily 0812.1934}}].

\bibitem{Garny:2009rv}
M.~Garny, A.~Hohenegger, A.~Kartavtsev and M.~Lindner, \emph{{Systematic
  approach to leptogenesis in nonequilibrium QFT: Vertex contribution to the
  $CP$-violating parameter}},
  \href{https://doi.org/10.1103/PhysRevD.80.125027}{\emph{Phys. Rev.}
  {\bfseries D80} (2009) 125027}
  [\href{https://arxiv.org/abs/0909.1559}{{\ttfamily 0909.1559}}].

\bibitem{Anisimov:2010aq}
A.~Anisimov, W.~Buchmüller, M.~Drewes and S.~Mendizabal, \emph{{Leptogenesis
  from Quantum Interference in a Thermal Bath}},
  \href{https://doi.org/10.1103/PhysRevLett.104.121102}{\emph{Phys. Rev. Lett.}
  {\bfseries 104} (2010) 121102}
  [\href{https://arxiv.org/abs/1001.3856}{{\ttfamily 1001.3856}}].

\bibitem{Beneke:2010wd}
M.~Beneke, B.~Garbrecht, M.~Herranen and P.~Schwaller, \emph{{Finite Number
  Density Corrections to Leptogenesis}},
  \href{https://doi.org/10.1016/j.nuclphysb.2010.05.003}{\emph{Nucl. Phys.}
  {\bfseries B838} (2010) 1} [\href{https://arxiv.org/abs/1002.1326}{{\ttfamily
  1002.1326}}].

\bibitem{Garny:2010nz}
M.~Garny, A.~Hohenegger and A.~Kartavtsev, \emph{{Quantum corrections to
  leptogenesis from the gradient expansion}},
  \href{https://arxiv.org/abs/1005.5385}{{\ttfamily 1005.5385}}.

\bibitem{Beneke:2010dz}
M.~Beneke, B.~Garbrecht, C.~Fidler, M.~Herranen and P.~Schwaller,
  \emph{{Flavoured Leptogenesis in the CTP Formalism}},
  \href{https://doi.org/10.1016/j.nuclphysb.2010.10.001}{\emph{Nucl. Phys.}
  {\bfseries B843} (2011) 177}
  [\href{https://arxiv.org/abs/1007.4783}{{\ttfamily 1007.4783}}].

\bibitem{Garbrecht:2010sz}
B.~Garbrecht, \emph{{Leptogenesis: The Other Cuts}},
  \href{https://doi.org/10.1016/j.nuclphysb.2011.01.033}{\emph{Nucl. Phys.}
  {\bfseries B847} (2011) 350}
  [\href{https://arxiv.org/abs/1011.3122}{{\ttfamily 1011.3122}}].

\bibitem{Anisimov:2010dk}
A.~Anisimov, W.~Buchmüller, M.~Drewes and S.~Mendizabal, \emph{{Quantum
  Leptogenesis I}}, \href{https://doi.org/10.1016/j.aop.2011.02.002,
  10.1016/j.aop.2013.05.00}{\emph{Annals Phys.} {\bfseries 326} (2011) 1998}
  [\href{https://arxiv.org/abs/1012.5821}{{\ttfamily 1012.5821}}].

\bibitem{Garbrecht:2012qv}
B.~Garbrecht, \emph{{Leptogenesis from Additional Higgs Doublets}},
  \href{https://doi.org/10.1103/PhysRevD.85.123509}{\emph{Phys. Rev.}
  {\bfseries D85} (2012) 123509}
  [\href{https://arxiv.org/abs/1201.5126}{{\ttfamily 1201.5126}}].

\bibitem{Drewes:2012ma}
M.~Drewes and B.~Garbrecht, \emph{{Leptogenesis from a GeV Seesaw without Mass
  Degeneracy}}, \href{https://doi.org/10.1007/JHEP03(2013)096}{\emph{JHEP}
  {\bfseries 03} (2013) 096} [\href{https://arxiv.org/abs/1206.5537}{{\ttfamily
  1206.5537}}].

\bibitem{Garbrecht:2012pq}
B.~Garbrecht, \emph{{Baryogenesis from Mixing of Lepton Doublets}},
  \href{https://doi.org/10.1016/j.nuclphysb.2012.11.021}{\emph{Nucl. Phys.}
  {\bfseries B868} (2013) 557}
  [\href{https://arxiv.org/abs/1210.0553}{{\ttfamily 1210.0553}}].

\bibitem{Frossard:2012pc}
T.~Frossard, M.~Garny, A.~Hohenegger, A.~Kartavtsev and D.~Mitrouskas,
  \emph{{Systematic approach to thermal leptogenesis}},
  \href{https://doi.org/10.1103/PhysRevD.87.085009}{\emph{Phys. Rev.}
  {\bfseries D87} (2013) 085009}
  [\href{https://arxiv.org/abs/1211.2140}{{\ttfamily 1211.2140}}].

\bibitem{Garbrecht:2013gd}
B.~Garbrecht, F.~Glowna and M.~Herranen, \emph{{Right-Handed Neutrino
  Production at Finite Temperature: Radiative Corrections, Soft and Collinear
  Divergences}}, \href{https://doi.org/10.1007/JHEP04(2013)099}{\emph{JHEP}
  {\bfseries 04} (2013) 099} [\href{https://arxiv.org/abs/1302.0743}{{\ttfamily
  1302.0743}}].

\bibitem{Garbrecht:2013urw}
B.~Garbrecht, F.~Glowna and P.~Schwaller, \emph{{Scattering Rates For
  Leptogenesis: Damping of Lepton Flavour Coherence and Production of Singlet
  Neutrinos}},
  \href{https://doi.org/10.1016/j.nuclphysb.2013.08.020}{\emph{Nucl. Phys.}
  {\bfseries B877} (2013) 1} [\href{https://arxiv.org/abs/1303.5498}{{\ttfamily
  1303.5498}}].

\bibitem{Frossard:2013bra}
T.~Frossard, A.~Kartavtsev and D.~Mitrouskas, \emph{{Systematic approach to
  $\Delta L=1$ processes in thermal leptogenesis}},
  \href{https://doi.org/10.1103/PhysRevD.87.125006}{\emph{Phys. Rev.}
  {\bfseries D87} (2013) 125006}
  [\href{https://arxiv.org/abs/1304.1719}{{\ttfamily 1304.1719}}].

\bibitem{Garbrecht:2013iga}
B.~Garbrecht and M.J.~Ramsey-Musolf, \emph{{Cuts, Cancellations and the Closed
  Time Path: The Soft Leptogenesis Example}},
  \href{https://doi.org/10.1016/j.nuclphysb.2014.02.012}{\emph{Nucl. Phys.}
  {\bfseries B882} (2014) 145}
  [\href{https://arxiv.org/abs/1307.0524}{{\ttfamily 1307.0524}}].

\bibitem{Dev:2017trv}
P.S.B.~Dev, P.~Di~Bari, B.~Garbrecht, S.~Lavignac, P.~Millington and D.~Teresi,
  \emph{{Flavor effects in leptogenesis}},
  \href{https://doi.org/10.1142/S0217751X18420010}{\emph{Int. J. Mod. Phys.}
  {\bfseries A33} (2018) 1842001}
  [\href{https://arxiv.org/abs/1711.02861}{{\ttfamily 1711.02861}}].

\bibitem{Garbrecht:2019zaa}
B.~Garbrecht, P.~Klose and C.~Tamarit, \emph{{Relativistic and spectator
  effects in leptogenesis with heavy sterile neutrinos}},
  \href{https://doi.org/10.1007/JHEP02(2020)117}{\emph{JHEP} {\bfseries 02}
  (2020) 117} [\href{https://arxiv.org/abs/1904.09956}{{\ttfamily
  1904.09956}}].

\bibitem{DeSimone:2007gkc}
A.~De~Simone and A.~Riotto, \emph{{Quantum Boltzmann Equations and
  Leptogenesis}},
  \href{https://doi.org/10.1088/1475-7516/2007/08/002}{\emph{JCAP} {\bfseries
  0708} (2007) 002} [\href{https://arxiv.org/abs/hep-ph/0703175}{{\ttfamily
  hep-ph/0703175}}].

\bibitem{DeSimone:2007edo}
A.~De~Simone and A.~Riotto, \emph{{On Resonant Leptogenesis}},
  \href{https://doi.org/10.1088/1475-7516/2007/08/013}{\emph{JCAP} {\bfseries
  0708} (2007) 013} [\href{https://arxiv.org/abs/0705.2183}{{\ttfamily
  0705.2183}}].

\bibitem{Cirigliano:2007hb}
V.~Cirigliano, A.~De~Simone, G.~Isidori, I.~Masina and A.~Riotto,
  \emph{{Quantum Resonant Leptogenesis and Minimal Lepton Flavour Violation}},
  \href{https://doi.org/10.1088/1475-7516/2008/01/004}{\emph{JCAP} {\bfseries
  0801} (2008) 004} [\href{https://arxiv.org/abs/0711.0778}{{\ttfamily
  0711.0778}}].

\bibitem{Garny:2009qn}
M.~Garny, A.~Hohenegger, A.~Kartavtsev and M.~Lindner, \emph{{Systematic
  approach to leptogenesis in nonequilibrium QFT: Self-energy contribution to
  the $CP$-violating parameter}},
  \href{https://doi.org/10.1103/PhysRevD.81.085027}{\emph{Phys. Rev.}
  {\bfseries D81} (2010) 085027}
  [\href{https://arxiv.org/abs/0911.4122}{{\ttfamily 0911.4122}}].

\bibitem{Garbrecht:2011aw}
B.~Garbrecht and M.~Herranen, \emph{{Effective Theory of Resonant Leptogenesis
  in the Closed-Time-Path Approach}},
  \href{https://doi.org/10.1016/j.nuclphysb.2012.03.009}{\emph{Nucl. Phys.}
  {\bfseries B861} (2012) 17}
  [\href{https://arxiv.org/abs/1112.5954}{{\ttfamily 1112.5954}}].

\bibitem{Garny:2011hg}
M.~Garny, A.~Kartavtsev and A.~Hohenegger, \emph{{Leptogenesis from first
  principles in the resonant regime}},
  \href{https://doi.org/10.1016/j.aop.2012.10.007}{\emph{Annals Phys.}
  {\bfseries 328} (2013) 26} [\href{https://arxiv.org/abs/1112.6428}{{\ttfamily
  1112.6428}}].

\bibitem{Iso:2013lba}
S.~Iso, K.~Shimada and M.~Yamanaka, \emph{{Kadanoff-Baym approach to the
  thermal resonant leptogenesis}},
  \href{https://doi.org/10.1007/JHEP04(2014)062}{\emph{JHEP} {\bfseries 04}
  (2014) 062} [\href{https://arxiv.org/abs/1312.7680}{{\ttfamily 1312.7680}}].

\bibitem{Iso:2014afa}
S.~Iso and K.~Shimada, \emph{{Coherent Flavour Oscillation and $CP$ Violating
  Parameter in Thermal Resonant Leptogenesis}},
  \href{https://doi.org/10.1007/JHEP08(2014)043}{\emph{JHEP} {\bfseries 08}
  (2014) 043} [\href{https://arxiv.org/abs/1404.4816}{{\ttfamily 1404.4816}}].

\bibitem{Hohenegger:2014cpa}
A.~Hohenegger and A.~Kartavtsev, \emph{{Leptogenesis in crossing and runaway
  regimes}}, \href{https://doi.org/10.1007/JHEP07(2014)130}{\emph{JHEP}
  {\bfseries 07} (2014) 130} [\href{https://arxiv.org/abs/1404.5309}{{\ttfamily
  1404.5309}}].

\bibitem{Garbrecht:2014aga}
B.~Garbrecht, F.~Gautier and J.~Klarić, \emph{{Strong Washout Approximation to
  Resonant Leptogenesis}},
  \href{https://doi.org/10.1088/1475-7516/2014/09/033}{\emph{JCAP} {\bfseries
  1409} (2014) 033} [\href{https://arxiv.org/abs/1406.4190}{{\ttfamily
  1406.4190}}].

\bibitem{Dev:2014wsa}
P.S.B.~Dev, P.~Millington, A.~Pilaftsis and D.~Teresi, \emph{{Kadanoff–Baym
  approach to flavour mixing and oscillations in resonant leptogenesis}},
  \href{https://doi.org/10.1016/j.nuclphysb.2014.12.003}{\emph{Nucl. Phys.}
  {\bfseries B891} (2015) 128}
  [\href{https://arxiv.org/abs/1410.6434}{{\ttfamily 1410.6434}}].

\bibitem{Kartavtsev:2015vto}
A.~Kartavtsev, P.~Millington and H.~Vogel, \emph{{Lepton asymmetry from mixing
  and oscillations}},
  \href{https://doi.org/10.1007/JHEP06(2016)066}{\emph{JHEP} {\bfseries 06}
  (2016) 066} [\href{https://arxiv.org/abs/1601.03086}{{\ttfamily
  1601.03086}}].

\bibitem{Drewes:2016gmt}
M.~Drewes, B.~Garbrecht, D.~Gueter and J.~Klarić, \emph{{Leptogenesis from
  Oscillations of Heavy Neutrinos with Large Mixing Angles}},
  \href{https://doi.org/10.1007/JHEP12(2016)150}{\emph{JHEP} {\bfseries 12}
  (2016) 150} [\href{https://arxiv.org/abs/1606.06690}{{\ttfamily
  1606.06690}}].

\bibitem{Dev:2017wwc}
P.S.B.~Dev, M.~Garny, J.~Klarić, P.~Millington and D.~Teresi, \emph{{Resonant
  enhancement in leptogenesis}},
  \href{https://doi.org/10.1142/S0217751X18420034}{\emph{Int. J. Mod. Phys.}
  {\bfseries A33} (2018) 1842003}
  [\href{https://arxiv.org/abs/1711.02863}{{\ttfamily 1711.02863}}].

\bibitem{Garbrecht:2018mrp}
B.~Garbrecht, \emph{{Why is there more matter than antimatter? Calculational
  methods for leptogenesis and electroweak baryogenesis}},
  \href{https://doi.org/10.1016/j.ppnp.2019.103727}{\emph{Prog. Part. Nucl.
  Phys.} {\bfseries 110} (2020) 103727}
  [\href{https://arxiv.org/abs/1812.02651}{{\ttfamily 1812.02651}}].

\bibitem{Herranen:2008hi}
M.~Herranen, K.~Kainulainen and P.M.~Rahkila, \emph{{Towards a kinetic theory
  for fermions with quantum coherence}},
  \href{https://doi.org/10.1016/j.nuclphysb.2008.09.032}{\emph{Nucl. Phys. B}
  {\bfseries 810} (2009) 389}
  [\href{https://arxiv.org/abs/0807.1415}{{\ttfamily 0807.1415}}].

\bibitem{Herranen:2008hu}
M.~Herranen, K.~Kainulainen and P.M.~Rahkila, \emph{{Quantum kinetic theory for
  fermions in temporally varying backgrounds}},
  \href{https://doi.org/10.1088/1126-6708/2008/09/032}{\emph{JHEP} {\bfseries
  09} (2008) 032} [\href{https://arxiv.org/abs/0807.1435}{{\ttfamily
  0807.1435}}].

\bibitem{Herranen:2008yg}
M.~Herranen, K.~Kainulainen and P.M.~Rahkila, \emph{{Kinetic transport theory
  with quantum coherence}},
  \href{https://doi.org/10.1016/j.nuclphysa.2009.01.050}{\emph{Nucl. Phys. A}
  {\bfseries 820} (2009) 203C}
  [\href{https://arxiv.org/abs/0811.0936}{{\ttfamily 0811.0936}}].

\bibitem{Herranen:2008di}
M.~Herranen, K.~Kainulainen and P.M.~Rahkila, \emph{{Kinetic theory for scalar
  fields with nonlocal quantum coherence}},
  \href{https://doi.org/10.1088/1126-6708/2009/05/119}{\emph{JHEP} {\bfseries
  05} (2009) 119} [\href{https://arxiv.org/abs/0812.4029}{{\ttfamily
  0812.4029}}].

\bibitem{Herranen:2010mh}
M.~Herranen, K.~Kainulainen and P.M.~Rahkila, \emph{{Coherent quantum Boltzmann
  equations from cQPA}},
  \href{https://doi.org/10.1007/JHEP12(2010)072}{\emph{JHEP} {\bfseries 12}
  (2010) 072} [\href{https://arxiv.org/abs/1006.1929}{{\ttfamily 1006.1929}}].

\bibitem{Fidler:2011yq}
C.~Fidler, M.~Herranen, K.~Kainulainen and P.M.~Rahkila, \emph{{Flavoured
  quantum Boltzmann equations from cQPA}},
  \href{https://doi.org/10.1007/JHEP02(2012)065}{\emph{JHEP} {\bfseries 02}
  (2012) 065} [\href{https://arxiv.org/abs/1108.2309}{{\ttfamily 1108.2309}}].

\bibitem{Herranen:2011zg}
M.~Herranen, K.~Kainulainen and P.M.~Rahkila, \emph{{Flavour-coherent
  propagators and Feynman rules: Covariant cQPA formulation}},
  \href{https://doi.org/10.1007/JHEP02(2012)080}{\emph{JHEP} {\bfseries 02}
  (2012) 080} [\href{https://arxiv.org/abs/1108.2371}{{\ttfamily 1108.2371}}].

\bibitem{Jukkala:2019slc}
H.~Jukkala, K.~Kainulainen and O.~Koskivaara, \emph{{Quantum transport and the
  phase space structure of the Wightman functions}},
  \href{https://doi.org/10.1007/JHEP01(2020)012}{\emph{JHEP} {\bfseries 01}
  (2020) 012} [\href{https://arxiv.org/abs/1910.10979}{{\ttfamily
  1910.10979}}].

\bibitem{Berges:2004yj}
J.~Berges, \emph{{Introduction to nonequilibrium quantum field theory}},
  \href{https://doi.org/10.1063/1.1843591}{\emph{AIP Conf. Proc.} {\bfseries
  739} (2004) 3} [\href{https://arxiv.org/abs/hep-ph/0409233}{{\ttfamily
  hep-ph/0409233}}].

\bibitem{Garny:2015oza}
M.~Garny and U.~Reinosa, \emph{{Renormalization out of equilibrium in a
  superrenormalizable theory}},
  \href{https://doi.org/10.1103/PhysRevD.94.045012}{\emph{Phys. Rev. D}
  {\bfseries 94} (2016) 045012}
  [\href{https://arxiv.org/abs/1504.06643}{{\ttfamily 1504.06643}}].

\bibitem{Mahan:1987251}
G.~Mahan, \emph{{Quantum transport equation for electric and magnetic fields}},
  \href{https://doi.org/10.1016/0370-1573(87)90004-4}{\emph{Phys. Rep.}
  {\bfseries 145} (1987) 251}.

\bibitem{Pilaftsis:1997jf}
A.~Pilaftsis, \emph{{CP violation and baryogenesis due to heavy Majorana
  neutrinos}}, \href{https://doi.org/10.1103/PhysRevD.56.5431}{\emph{Phys. Rev.
  D} {\bfseries 56} (1997) 5431}
  [\href{https://arxiv.org/abs/hep-ph/9707235}{{\ttfamily hep-ph/9707235}}].

\bibitem{Pilaftsis:2003gt}
A.~Pilaftsis and T.E.~Underwood, \emph{{Resonant leptogenesis}},
  \href{https://doi.org/10.1016/j.nuclphysb.2004.05.029}{\emph{Nucl. Phys. B}
  {\bfseries 692} (2004) 303}
  [\href{https://arxiv.org/abs/hep-ph/0309342}{{\ttfamily hep-ph/0309342}}].

\bibitem{Akhmedov:1998qx}
E.K.~Akhmedov, V.A.~Rubakov and A.Y.~Smirnov, \emph{{Baryogenesis via neutrino
  oscillations}},
  \href{https://doi.org/10.1103/PhysRevLett.81.1359}{\emph{Phys. Rev. Lett.}
  {\bfseries 81} (1998) 1359}
  [\href{https://arxiv.org/abs/hep-ph/9803255}{{\ttfamily hep-ph/9803255}}].

\bibitem{Klaric:2020lov}
J.~Klari\'c, M.~Shaposhnikov and I.~Timiryasov, \emph{{Uniting low-scale
  leptogeneses}},  \href{https://arxiv.org/abs/2008.13771}{{\ttfamily
  2008.13771}}.

\bibitem{Granelli:2020ysj}
A.~Granelli, K.~Moffat and S.T.~Petcov, \emph{{Flavoured Resonant Leptogenesis
  at Sub-TeV Scales}},  \href{https://arxiv.org/abs/2009.03166}{{\ttfamily
  2009.03166}}.

\bibitem{Kolb:1979qa}
E.W.~Kolb and S.~Wolfram, \emph{{Baryon Number Generation in the Early
  Universe}}, \href{https://doi.org/10.1016/0550-3213(82)90012-8}{\emph{Nucl.
  Phys. B} {\bfseries 172} (1980) 224}.

\bibitem{Kniehl:1996bd}
B.A.~Kniehl and A.~Pilaftsis, \emph{{Mixing renormalization in Majorana
  neutrino theories}},
  \href{https://doi.org/10.1016/0550-3213(96)00280-5}{\emph{Nucl. Phys.}
  {\bfseries B474} (1996) 286}
  [\href{https://arxiv.org/abs/hep-ph/9601390}{{\ttfamily hep-ph/9601390}}].

\bibitem{Kniehl:2014dra}
B.A.~Kniehl, \emph{{All-order renormalization of propagator matrix for unstable
  Dirac fermions}},
  \href{https://doi.org/10.1103/PhysRevD.89.096005}{\emph{Phys. Rev. D}
  {\bfseries 89} (2014) 096005}.

\bibitem{Fuchs:2016swt}
E.~Fuchs and G.~Weiglein, \emph{{Breit-Wigner approximation for propagators of
  mixed unstable states}},
  \href{https://doi.org/10.1007/JHEP09(2017)079}{\emph{JHEP} {\bfseries 09}
  (2017) 079} [\href{https://arxiv.org/abs/1610.06193}{{\ttfamily
  1610.06193}}].

\bibitem{Aoki:1982ed}
K.I.~Aoki, Z.~Hioki, M.~Konuma, R.~Kawabe and T.~Muta, \emph{{Electroweak
  Theory. Framework of On-Shell Renormalization and Study of Higher Order
  Effects}}, \href{https://doi.org/10.1143/PTPS.73.1}{\emph{Prog. Theor. Phys.
  Suppl.} {\bfseries 73} (1982) 1}.

\bibitem{Espriu:2000fq}
D.~Espriu and J.~Manzano, \emph{{$CP$ violation and family mixing in the
  effective electroweak Lagrangian}},
  \href{https://doi.org/10.1103/PhysRevD.63.073008}{\emph{Phys. Rev.}
  {\bfseries D63} (2001) 073008}
  [\href{https://arxiv.org/abs/hep-ph/0011036}{{\ttfamily hep-ph/0011036}}].

\bibitem{JKP_in_progress}
H.~Jukkala, K.~Kainulainen and H.~Parkkinen, ``{work in progress}.''.

\bibitem{Anisimov:2005hr}
A.~Anisimov, A.~Broncano and M.~Plumacher, \emph{{The CP-asymmetry in resonant
  leptogenesis}},
  \href{https://doi.org/10.1016/j.nuclphysb.2006.01.003}{\emph{Nucl. Phys. B}
  {\bfseries 737} (2006) 176}
  [\href{https://arxiv.org/abs/hep-ph/0511248}{{\ttfamily hep-ph/0511248}}].

\bibitem{Dev:2014laa}
P.~Bhupal~Dev, P.~Millington, A.~Pilaftsis and D.~Teresi, \emph{{Flavour
  Covariant Transport Equations: an Application to Resonant Leptogenesis}},
  \href{https://doi.org/10.1016/j.nuclphysb.2014.06.020}{\emph{Nucl. Phys. B}
  {\bfseries 886} (2014) 569}
  [\href{https://arxiv.org/abs/1404.1003}{{\ttfamily 1404.1003}}].

\bibitem{Basboll:2006yx}
A.~Basboll and S.~Hannestad, \emph{{Decay of heavy Majorana neutrinos using the
  full Boltzmann equation including its implications for leptogenesis}},
  \href{https://doi.org/10.1088/1475-7516/2007/01/003}{\emph{JCAP} {\bfseries
  01} (2007) 003} [\href{https://arxiv.org/abs/hep-ph/0609025}{{\ttfamily
  hep-ph/0609025}}].

\bibitem{Eijima:2020shs}
S.~Eijima, M.~Shaposhnikov and I.~Timiryasov, \emph{{Freeze-in generation of
  lepton asymmetries after baryogenesis in the $\nu$MSM}},
  \href{https://arxiv.org/abs/2011.12637}{{\ttfamily 2011.12637}}.

\bibitem{Klaric:2021cpi}
J.~Klaric, M.~Shaposhnikov and I.~Timiryasov, \emph{{Reconciling resonant
  leptogenesis and baryogenesis via neutrino oscillations}},
  \href{https://arxiv.org/abs/2103.16545}{{\ttfamily 2103.16545}}.

\bibitem{Depta:2020zmy}
P.F.~Depta, A.~Halsch, J.~H\"utig, S.~Mendizabal and O.~Philipsen,
  \emph{{Complete leading-order standard model corrections to quantum
  leptogenesis}}, \href{https://doi.org/10.1007/JHEP09(2020)036}{\emph{JHEP}
  {\bfseries 09} (2020) 036}
  [\href{https://arxiv.org/abs/2005.01728}{{\ttfamily 2005.01728}}].

\bibitem{Millington:2012pf}
P.~Millington and A.~Pilaftsis, \emph{{Perturbative nonequilibrium thermal
  field theory}}, \href{https://doi.org/10.1103/PhysRevD.88.085009}{\emph{Phys.
  Rev. D} {\bfseries 88} (2013) 085009}
  [\href{https://arxiv.org/abs/1211.3152}{{\ttfamily 1211.3152}}].

\bibitem{Weldon:1982bn}
H.A.~Weldon, \emph{{Effective Fermion Masses of Order gT in High Temperature
  Gauge Theories with Exact Chiral Invariance}},
  \href{https://doi.org/10.1103/PhysRevD.26.2789}{\emph{Phys. Rev.} {\bfseries
  D26} (1982) 2789}.

\bibitem{Luty:1992un}
M.~Luty, \emph{{Baryogenesis via leptogenesis}},
  \href{https://doi.org/10.1103/PhysRevD.45.455}{\emph{Phys. Rev. D} {\bfseries
  45} (1992) 455}.

\bibitem{Giudice:2003jh}
G.~Giudice, A.~Notari, M.~Raidal, A.~Riotto and A.~Strumia, \emph{{Towards a
  complete theory of thermal leptogenesis in the SM and MSSM}},
  \href{https://doi.org/10.1016/j.nuclphysb.2004.02.019}{\emph{Nucl. Phys. B}
  {\bfseries 685} (2004) 89}
  [\href{https://arxiv.org/abs/hep-ph/0310123}{{\ttfamily hep-ph/0310123}}].

\bibitem{Buchmuller:1997yu}
W.~Buchmuller and M.~Plumacher, \emph{{CP asymmetry in Majorana neutrino
  decays}}, \href{https://doi.org/10.1016/S0370-2693(97)01548-7}{\emph{Phys.
  Lett. B} {\bfseries 431} (1998) 354}
  [\href{https://arxiv.org/abs/hep-ph/9710460}{{\ttfamily hep-ph/9710460}}].

\end{thebibliography}\endgroup

\end{document}